\RequirePackage{lineno} 

\documentclass[twocolumn,preprintnumbers,amsmath,amssymb,nofootinbib]{revtex4}

\usepackage{graphicx}
\usepackage{dcolumn}
\usepackage{bm}
 
\usepackage{epstopdf}

\def\bea{\begin{eqnarray}}
\def\eea{\end{eqnarray}}

\def\pp{\mbox{$p$-$p$}}

\def\pa{\mbox{$p$-A}}
\def\da{\mbox{$d$-A}}
\def\auau{\mbox{Au-Au}}

\def\pbpb{\mbox{Pb-Pb}}
\def\ppb{\mbox{$p$-Pb}}

\def\pn{\mbox{$p$-N}}
\def\aa{\mbox{A-A}}
\def\nn{\mbox{N-N}}
\def\ss{\mbox{S-S}}

\def\ee{\mbox{$e^+$-$e^-$}}
\def\ppbar{\mbox{$p$-$\bar p$}}
\def\qqbar{\mbox{$q$-$\bar q$}}
\def\pt{$p_t$}
\def\pz{$p_z$}
\def\mt{$m_t$}
\def\yt{$y_t$}
\def\yz{$y_z$}

\def\nch{$n_{ch}$}
\def\mmpt{$\bar p_t$}

\begin{document} 

\setlength{\pdfpagewidth}{8.5in}
\setlength{\pdfpageheight}{11in}

\setpagewiselinenumbers
\modulolinenumbers[5]

\preprint{version 1.8}

\title{Evaluating the blast-wave model as a description of 5 TeV p-Pb $\bf p_t$ spectra
}

\author{Thomas A.\ Trainor}\affiliation{University of Washington, Seattle, WA 98195}


\date{\today}

\begin{abstract}

The blast-wave (BW) spectrum model is interpreted to reveal relativistic motion (collective flow) of the hadron emission system relative to the center-of-momentum (CM) frame in high-energy A-B collisions. In essence, any spectrum deviation in the CM frame from a reference distribution (e.g.\ Boltzmann distribution on transverse mass \mt) is interpreted to reveal a flowing particle source. The ALICE collaboration has applied the BW model to identified hadron (PID) spectra for four hadron species from 5 TeV $p$-Pb collisions.  From model fits BW parameters $T_{kin}$ (freeze-out temperature) and $\langle \beta_t \rangle$ (transverse speed) are inferred that suggest strong radial expansion in more-central $p$-Pb collisions. Such results from the small $p$-Pb collision system are counterintuitive given that strong radial expansion should be driven by large density gradients. The present study is intended to address that problem. Several methods are employed to evaluate the quality of the BW model data description, including logarithmic derivatives and the Z-score statistic. The stability of the BW model definition across several applications to data is investigated. The BW model data description is compare to that of the two-component (soft+hard) model (TCM) that has been previously applied to the same $p$-Pb PID spectra. The general conclusion is that the BW model is falsified by $p$-Pb PID spectrum data according to standard statistical measures and that the fitted parameter values do not convey the intended meaning. Statistically acceptable data descriptions provided by the TCM indicate that other collision mechanisms (projectile-nucleon dissociation, dijet production), that are consistent with conventional QCD, are more likely responsible for observed spectrum characteristics.
\end{abstract}


\maketitle

 \section{Introduction}

The blast-wave (BW) spectrum model has been widely applied to identified-hadron (PID) spectra from A-B collisions in the context of the high-energy heavy ion program whose central goal has been formation of a quark-gluon plasma (QGP). The basic premise is that BW spectrum analysis might confirm the presence of {\em radial flow} -- radial expansion of a conjectured fluid assumed to be the dominant source of final-state particle emission. Initial application of the BW model to SPS \ss\ collisions (with $\sqrt{s_{NN}} \approx 19$ GeV) proved inconclusive~\cite{ssflow}. However, early results from the relativistic heavy ion collider (RHIC) ($\sqrt{s_{NN}} \approx 130$ and 200 GeV) were interpreted to indicate the presence of significant radial flow (and hence QGP?), at least in more-central \auau\ collisions~\cite{phenradflow}. 

In motivating the RHIC experimental program it had been assumed that only for more-central \aa\ collisions involving the largest nuclei might the requisite energy and matter densities be generated to form a QGP. However, application of the BW model to smaller collision systems (e.g.\ \pa, \da\ and even \pp) has lead to claims that significant radial flow appears in all A-B collisions~\cite{starradflow}. A recent example of the latter is BW analysis of PID spectra from 5 TeV \ppb\ collisions~\cite{aliceppbpid}. Given the novelty and significance of such claims several questions emerge regarding BW model fits to PID spectra:

How well-defined is the theoretical basis for the BW model? That question goes beyond the basic Cooper-Frye formulation~\cite{cooper} relating particle emission in a relativistic moving frame to observed particle momenta in the center-of-momentum (CM) or laboratory (lab) frame. It concerns basic assumptions about the particle emission process -- what does ``particle emission'' mean -- and what constraints does the emission environment (presumably high-density matter of uncertain nature) impose? Also, how consistent is the BW model formulation from one application to another? Can different research groups apply the {\em same} model to similar data and get the same result, or is each application unique and therefore unverifiable?

Regarding BW model fits to data what are the standards for concluding that {\em a} BW model (which one?) describes spectrum data with sufficient accuracy to {\em confirm} a flowing particle-source scenario? Does a {\em well-defined} BW model describe all available relevant data within {\em statistical} uncertainties. Is there a competing spectrum model that describes spectrum data qualitatively more accurately? Within the context of {\em prevailing practice} for applying  the model is it possible to {\em falsify} the BW model?

To address those questions and other related considerations the following strategy is adopted: A reference model for comparison with BW results is introduced in the form of a two-component (soft + hard) model (TCM) applied to PID spectrum data from 5 TeV \ppb\ collisions~\cite{pidpart1,pidpart2}. A BW model description of the same PID data, as reported in Ref.~\cite{aliceppbpid}, is reviewed. Several model-independent measures are employed to compare BW and TCM data descriptions. Limiting cases of the BW model are compared with a Boltzmann distribution (assuming thermal emission from a stationary particle source) and the TCM soft component (a limiting case corresponding to zero particle density and no jet contribution to spectra). Model fit quality is assessed with Z-scores based on published statistical and systematic data uncertainties.

To provide fuller context for model comparisons the concept of ``two cultures'' is introduced relating an approach based on perturbative QCD and parton cascades or showers on the one hand and an approach based on fluid-dynamic descriptions emerging from preQCD collision models on the other. Finally, the concept of ``elementary collisions'' as a reference  is briefly considered with reference to particle data from \pp\ and \ee\ collisions. Appendix~\ref{derive} defines relativistic quantities appearing in BW derivations and reviews several BW model versions.

The goals of this study are: (a) understand the assumptions and formulations of the BW model as manifested by several versions, (b) assess whether underlying BW model assumptions are physically reasonable given present knowledge of QCD theory and improved data quality, (c) determine whether the BW model describes available data within uncertainties (statistical rather than systematic) or is falsified by data, and (d) consider an alternative spectrum model that appears superior as a data model and more physically interpretable.

This article is arranged as follows:
Section~\ref{spectrumtcm} reviews a TCM description of PID spectrum data from 5 TeV \ppb\ collisions reported in Ref.~\cite{aliceppbpid}.
Section~\ref{bwmodel} briefly presents the BW model as it is applied to the same PID spectrum data.
Section~\ref{compare} compares the shapes of the two spectrum models via several model-independent shape measures.
Section~\ref{quality} evaluates data fit quality for BW and TCM models using the Z-score statistical measure.
Section~\ref{bwmodelx} provides a survey of BW model evolution over several decades.
Section~\ref{sys}  reviews systematic uncertainties.
Sections~\ref{disc} and~\ref{summ} present discussion and summary. 
Appendix~\ref{derive} presents a detailed review of relativistic quantities and several BW model derivations.

\section{TCM $\bf vs$ $\bf p$-$\bf Pb$ PID Spectrum data} \label{spectrumtcm}

The two-component model (TCM) of PID hadron spectra described here provides a simple reference based on conventional soft and hard QCD processes that may furnish a better understanding of the BW model and its relation to spectrum data and collision dynamics. The PID spectrum data for seven charge-multiplicity \nch\ classes from 5 TeV \ppb\ collisions are as reported in Ref.~\cite{aliceppbpid}. Previous TCM analysis of those data is reported in Refs.~\cite{ppbpid,pidpart1,pidpart2}. A version of the PID TCM in which two parameters are varied with \nch\ to accommodate spectrum data describes all data within their statistical uncertainties as demonstrated in Ref.~\cite{pidpart2}.

\subsection{p-Pb spectrum TCM for identified hadrons} \label{pidspecc}

The \yt\ spectrum TCM for {\em unidentified} hadrons is~\cite{ppbpid}
\bea \label{pidspectcm}
\bar \rho_{0}(y_t,n_s) &\approx& \frac{d^2 n_{ch}}{y_t dy_t d\eta}
\\ \nonumber
&=& S(y_t,n_s)+ H(y_t,n_s)
\\ \nonumber
&\approx& \bar \rho_{s}(n_s) \hat S_{0}(y_t) +  \bar \rho_{h}(n_s) \hat H_{0}(y_t),
\eea
where $y_t \equiv \ln[(m_t + p_t)/m_0]$ is transverse rapidity with  $m_0 \rightarrow m_\pi$ assumed. $ \hat S_{0}(y_t)$ and $\hat H_{0}(y_t)$ are unit-integral fixed model functions (independent of A-B centrality or charge multiplicity \nch), and $\bar \rho_{s}(n_s)$ and $ \bar \rho_{h}(n_s)$ are soft and hard charge densities with $\bar \rho_x \equiv n_x / \Delta \eta$ and $\bar \rho_{0} = \bar \rho_{s} + \bar \rho_{h}$. Soft yield $n_s$ serves as a centrality index. A-B charge densities are defined in terms of \nn\ quantities by $\bar \rho_s = (N_{part}/2) \bar \rho_{sNN}$ and $\bar \rho_h = N_{bin} \bar \rho_{hNN}$ with $x \equiv \rho_{hNN} / \rho_{sNN}$ and $\nu \equiv 2N_{bin} / N_{part}$ and with $\bar \rho_h / \bar \rho_s = x(n_s)\nu(n_s)$. For linear superposition of \nn\ collisions in A-B collisions  $\bar \rho_{hNN} \approx \alpha(\sqrt{s_{NN}}) \bar \rho_{sNN}^2$~\cite{ppprd,ppbpid}.

Given the A-B spectrum TCM for unidentified hadron spectra a corresponding TCM for identified hadrons can be generated by assuming that each hadron species $i$ comprises certain {\em fractions} of soft and hard TCM components denoted by $z_{si}(n_s)$ and $z_{hi}(n_s)$ assumed {independent of \yt}. The PID spectrum TCM is then expressed as~\cite{pidpart1,pidpart2}
\bea \label{pidspectcm}
\bar \rho_{0i}(y_t,n_s) &=& S_i(y_t,n_s)+ H_i(y_t,n_s)
\\ \nonumber
&\approx& z_{si}(n_s)  \bar \rho_{s} \hat S_{0i}(y_t) +  z_{hi}(n_s)  \bar \rho_{h} \hat H_{0i}(y_t,n_s) 
\nonumber \\ \label{eq4}
\frac{\bar \rho_{0i}(y_t,n_s)}{ z_{si}(n_s)  \bar \rho_{s}} &\equiv&   X_i(y_t,n_s)
\\ \nonumber
&\approx & \hat S_{0i}(y_t) +  \tilde z_{i}(n_s)x(n_s)\nu(n_s) \hat H_{0i}(y_t),
\eea
where $\tilde z_{i}(n_s) \equiv z_{hi}(n_s) / z_{si}(n_s)$ and unit-integral model functions $\hat S_{0i}(y_t)$ and $\hat H_{0i}(y_t)$ depend on hadron species $i$. Soft fraction $z_{si}(n_s)$ is expressed within the TCM by
\bea \label{rhosi}
z_{si}(n_s)&=& \left[\frac{1 + x(n_s) \nu(n_s)}{1 + \tilde z_{i}(n_s) x(n_s) \nu(n_s)} \right]  {z_{0i}},
\eea
where $z_{hi}(n_s) = \tilde z_{i}(n_s) z_{si}(n_s)$
and $\bar \rho_{0i} \equiv z_{0i} \bar \rho_0$ defines $z_{0i}$. Thus, if $\tilde z_{i}(n_s)$ and $z_{0i}$ are specified for relevant hadron species then all of the PID TCM is determined. Model functions $\hat S_{0i}(y_t)$ are defined on proper $m_{ti}$ for a given hadron species $i$ and then transformed to $y_{t\pi}$. $\hat H_{0i}(y_t)$ are always defined on $y_{t\pi}$. The basis for the $\hat S_{0i}(y_t)$ model definitions is the limit of normalized spectra $X_i(y_t,n_s) $ as $n_{ch} \rightarrow 0$ ({\em zero particle density}). Soft-component models are thus always derived from data spectra. Inferred {\em data} spectrum hard components are then by definition the complement of {\em model} soft components so defined.

Properties of TCM soft-component models $\hat S_{0i}(m_t)$ are consistent with nuclear transparency in \pa\ collisions~\cite{busza} and A-B collisions~\cite{stopping} and with the wounded-nucleon model~\cite{bialas}. Properties of TCM hard-component models $\hat H_{0i}(y_t)$  are consistent with measured properties of minimum-bias jets within A-B collisions if they consist of {\em linear superpositions} of \nn\ collisions~\cite{hardspec,fragevo,eeprd,jetspec2,mbdijets}.

\subsection{5 TeV $\bf p$-Pb TCM model parameters}

Table~\ref{rppbdata} presents TCM geometry parameters for 5 TeV \ppb\ collisions inferred from the analysis in Ref.~\cite{tomglauber}. Those geometry parameters, derived from \ppb\ \pt\ spectrum and ensemble \mmpt\ data for unidentified hadrons~\cite{alicempt,tommpt,tomglauber}, are assumed valid for each identified-hadron species (confirmed in Ref.~\cite{ppbpid}). 
$\bar \rho_0 = n_{ch} / \Delta \eta$ charge densities are measured quantities inferred from Fig.~16 of Ref.~\cite{aliceglauber}. Relations $N_{bin} = N_{part} + 1$ and $\nu = 2 N_{bin} / N_{part}$ involve number of nucleon N participants and \nn\ binary collisions. $\bar \rho_{sNN}$ is the mean soft-component charge density per participant pair averaged over all pairs. $x \equiv \bar \rho_{hNN}/\bar \rho_{sNN} \approx \alpha \bar \rho_{sNN}$ is the \nn\ hard/soft density ratio. Column $\sigma' / \sigma_0$ presents  nominal centralities (bin centers) reported by Ref.~\cite{aliceppbpid} in connection with measured charge densities $\bar \rho_0$ whereas column $\sigma / \sigma_0$ presents values inferred in Ref.~\cite{tomglauber}. The remaining values in the table are results of the latter analysis. 

\begin{table}[h]
\caption{TCM fractional cross sections $\sigma / \sigma_0$ (bin centers) and  geometry parameters, midrapidity charge density $\bar \rho_0$, TCM \nn\ soft component $\bar \rho_{sNN}$ and \nn\ hard/soft ratio $x(n_s)$ used for 5 TeV \ppb\ PID spectrum analysis~\cite{tomglauber}. Centrality parameters are from Ref.~\cite{tomglauber}. $\sigma' / \sigma_0$ values are from Table~1 of Ref.~\cite{aliceppbpid} as determined in Ref.~\cite{aliceglauber}.
	}
	\label{rppbdata}
	\begin{center}
		\begin{tabular}{|c|c|c|c|c|c|c|c|} \hline
			$n$ &   $\sigma' / \sigma_0$ &   $\sigma / \sigma_0$     & $N_{bin}$  & $\nu$ & $\bar \rho_0$ & $\bar \rho_{sNN}$ & $x(n_s)$ \\ \hline
			1	   &      0.025   & 0.15   & 3.20   & 1.52 & 44.6 & 16.6  & 0.188 \\ \hline
			2	 &  0.075  & 0.24    & 2.59   & 1.43 & 35.9 &15.9  & 0.180 \\ \hline
			3	 &  0.15  & 0.37 & 2.16  &  1.37 & 30.0  & 15.2  & 0.172 \\ \hline
			4	 &  0.30 & 0.58  & 1.70   & 1.26  & 23.0  & 14.1  & 0.159  \\ \hline
			5	 &  0.50   &0.80    & 1.31   & 1.13 & 15.8 &   12.1 & 0.137  \\ \hline
			6	 &  0.70  & 0.95   & 1.07   & 1.03  & 9.7  &  8.7 & 0.098 \\ \hline
			7	 & 0.90  & 0.99  & 1.00  & 1.00  &  4.4  & 4.2 &0.047  \\ \hline
		\end{tabular}
	\end{center}
\end{table}

Table~\ref{pidparamsxx} presents hard-component model parameters based on optimized descriptions of spectrum hard components for the {\em most-central} ($n = 1$) \ppb\ event class as described in Ref.~\cite{pidpart1}. TCM soft-component parameters remain unchanged from those reported in Table II of Ref.~\cite{pidpart1}, with $(T,n) =$ (145 MeV,8.5), (200 MeV,14) and (210 MeV,14) for pions, kaons and baryons respectively. More-massive hadrons are less sensitive to exponent $n$. Those parameters and these in Tables~\ref{rppbdata} and \ref{pidparamsxx} define a {\em fixed} PID TCM established as a reference. In Ref.~\cite{pidpart2} a {\em variable} TCM is defined wherein certain hard-component parameters (Gaussian widths $\sigma_{y_t}$ below {\em or} above the hard-component mode or centroids $\bar y_t$) are varied linearly with centrality measure $x\nu$ to accommodate data (see Fig.~4 of Ref.~\cite{pidpart2}). The variable TCM describes all spectra within statistical uncertainties as in Fig.~\ref{tcmdata}.
 
 \begin{table}[h]
 	\caption{Revised PID TCM hard-component model parameters $(\bar y_t,\sigma_{y_t},q)$ for identified hadrons from 5 TeV \ppb\ collisions derived from the differential spectrum analysis in Rev.~\cite{pidpart1}.  Parameter $z_{0i}$ values, inferred as limiting values of $z_{si}(n_s)$ centrality trends (Fig.~9 of Ref.~\cite{pidpart1}), are also included. Uncertainties are determined as one half the parameter change that would produce an obvious variation in $z_{hi}(y_t,n_s)$ ratios (e.g.\ Fig.~3 of Ref.~\cite{pidpart2}). Values with no uncertainties are duplicated from a related particle type.
 	}
 	\label{pidparamsxx}
 	\begin{center}
 		\begin{tabular}{|c|c|c|c|c|} \hline
 			&  $\bar y_t$ & $\sigma_{y_t}$ & $q$ & 	$z_{0i}$  \\ \hline
 			$ \pi^\pm $     
 			&	   $2.46\pm0.005$ & $0.575\pm0.005$ & $4.1\pm0.5$  &  0.82 $\pm$0.01 \\ \hline
 			$K^\pm$    
 			&	  $2.655$  & $0.568$ & $4.1$  & 0.128 $\pm$0.002 \\ \hline
 			$K_\text{S}^0$          
 			& 	   $2.655\pm0.005$ & $0.568\pm0.003$ & $4.1\pm0.1$ & 0.064 $\pm$0.002  \\ \hline
 			$p$        
 			& 	  $2.99\pm0.005$  & $0.47\pm0.005$ & $5.0$ & 0.065 $\pm$0.002  \\ \hline
 			$\Lambda$       
 			& 	  $2.99\pm0.005$  & $0.47\pm0.005$ & $5.0\pm0.05$  & 0.034 $\pm$0.002 \\ \hline	
 		\end{tabular}
 	\end{center}
 \end{table}
 
 \subsection{5 TeV $\bf p$-Pb PID  spectrum data} \label{alicedata}
 
Identified-hadron spectrum data obtained from Ref.~\cite{aliceppbpid} for the present analysis were produced by the ALICE collaboration at the LHC.  
Collision events were divided into seven charge-multiplicity \nch\ or \ppb\ centrality classes based on yields in a VZERO-A (V0A) counter subtending $2.8 < \eta_{lab} < 5.1$ in the Pb direction. Hadron species include charged pions $\pi^\pm$, charged kaons $K^\pm$, neutral kaons $K^0_\text{S}$, protons $p,~\bar p$ and Lambdas $\Lambda,~\bar \Lambda$. 
 
\begin{figure}[h]
 	\includegraphics[width=1.65in]{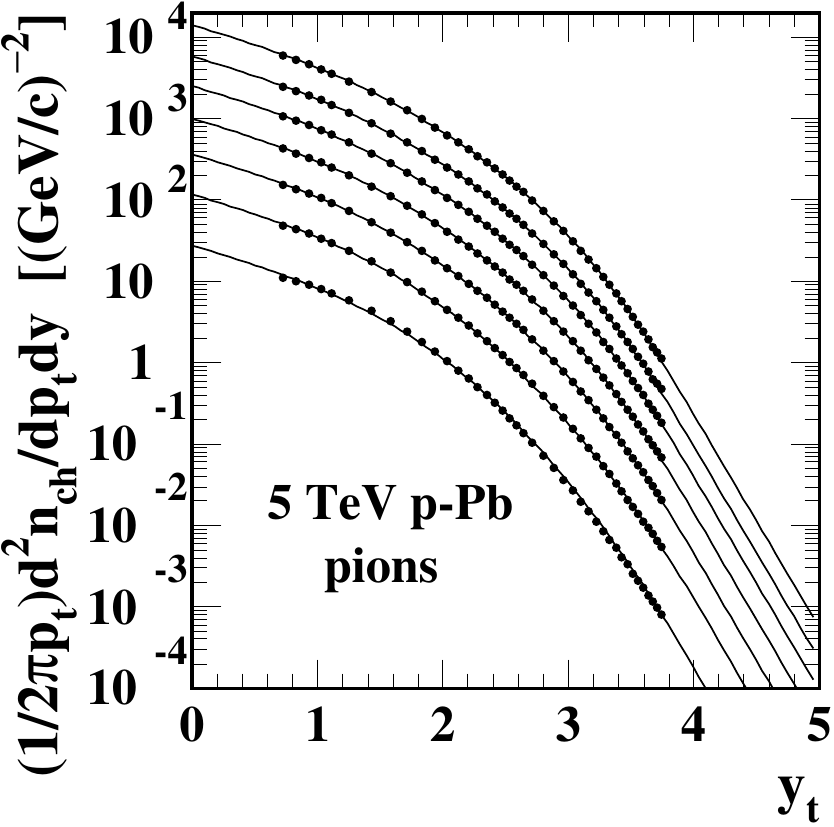}
 	\includegraphics[width=1.65in]{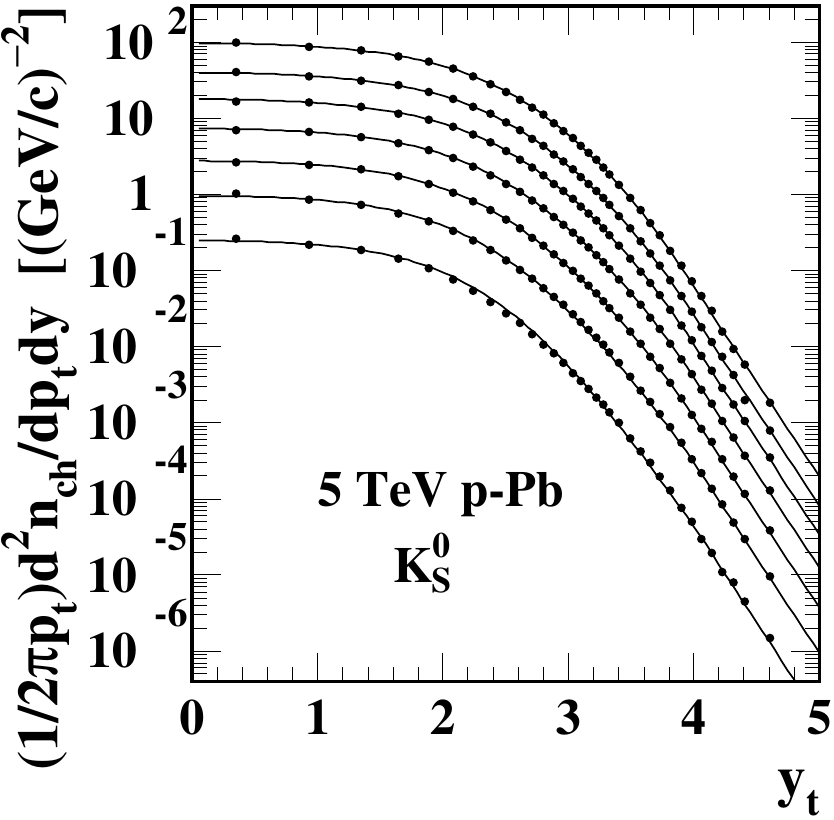}
 	\put(-142,105) {\bf (a)}
 	\put(-23,105) {\bf (b)}\\
 	\includegraphics[width=1.65in]{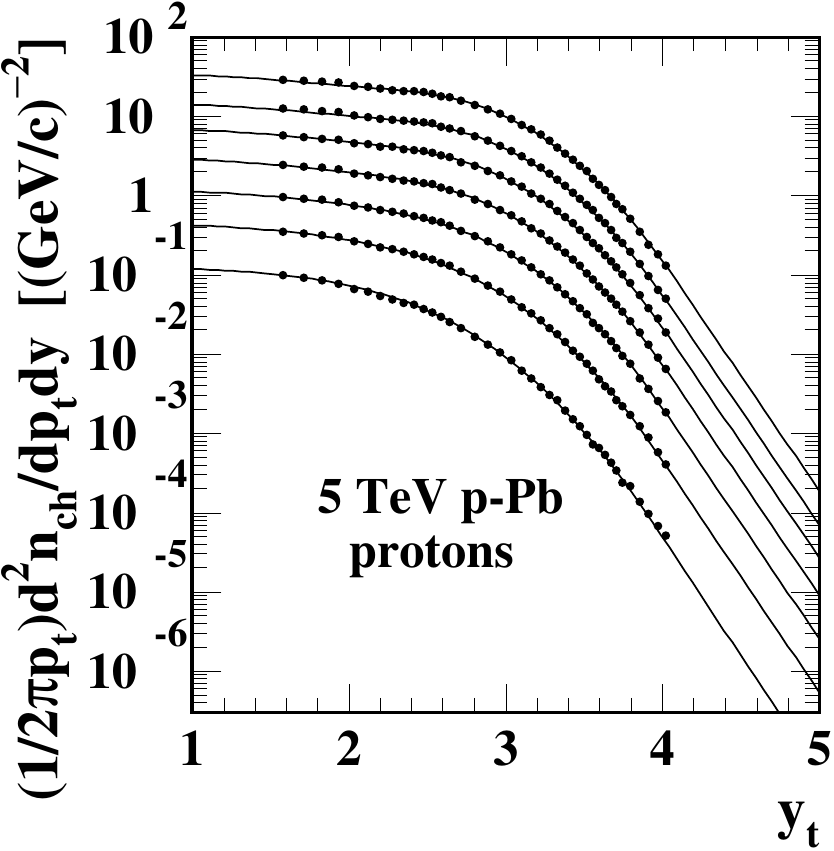}
 	\includegraphics[width=1.65in]{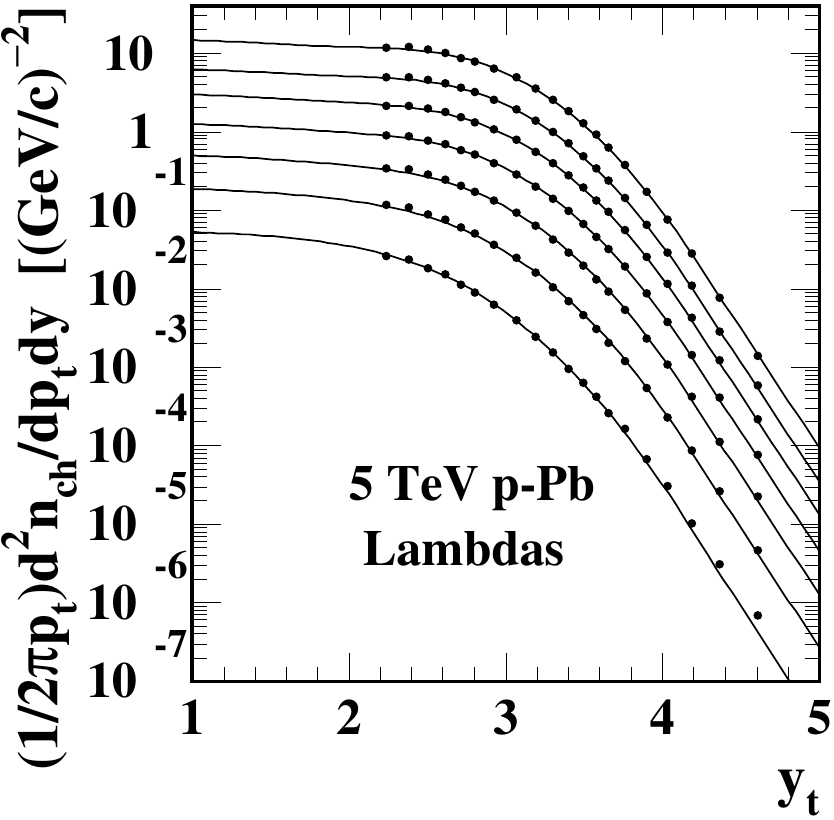}
 	\put(-142,105) {\bf (c)}
 	\put(-23,105) {\bf (d)}\\
 	\caption{\label{tcmdata}
 		\pt\ spectra for identified hadrons from 5 TeV \ppb\ collisions~\cite{aliceppbpid} plotted vs pion transverse rapidity \yt\ (default) for:
 		(a) pions,
 		(b) neutral kaons,
 		(c) corrected protons,
 		(d) Lambdas.
 		Solid curves represent the PID spectrum TCM from Ref.~\cite{pidpart2}. Proton inefficiency corrections are described in Ref.~\cite{pidpart1}. The statistical uncertainties (0.1 to 1\% of data values, e.g.\ see Fig.~\ref{bwsyser1}) are all smaller than the point size.
 	}  
 \end{figure}

Figure~\ref{tcmdata} shows PID spectrum data (densities on \pt) from Ref.~\cite{aliceppbpid} (points) plotted vs logarithmic variable \yt. That format (log-log with respect to \pt) provides detailed access to low-\pt\ structure (where most jet fragments appear) and clearly shows power-law trends at higher \pt. The curves are TCM parametrizations as reported above and demonstrated in Ref.~\cite{pidpart2}  to describe spectrum data within their statistical uncertainties.  As in Ref.~\cite{aliceppbpid} the spectra are scaled up by powers of 2 according to $2^{n-1}$ where $n \in [1,7]$ is the centrality class index and $n = 1$ is {\em least} central (following the usage in Ref.~\cite{aliceppbpid}). In this paper $n=1$ denotes the {\em most} central data as in Table~\ref{rppbdata}. 
   
Note that in the format of Fig.~\ref{tcmdata} a Boltzmann exponential on transverse mass \mt\ appears as $A - m_i\cosh(y_t) / T_i$, where $m_i$ is the mass for hadron species $i$ and $A$ is a constant,  and a power law on \mt\ appears as a straight line. The exception is for pion spectra including a substantial resonance contribution for $y_t < 2$ ($p_t < 0.5$ GeV/c)~\cite{pidpart1}.

\section{BW model $\bf vs$ $\bf p$-$\bf Pb$ PID spectrum data} \label{bwmodel}

Reference~\cite{aliceppbpid}  applies a blast-wave model to \pt\ spectrum data for 5 TeV V0A \ppb\ spectra, with fit parameters reported in its Fig.~6.  The BW fit parameters are mean radial speed $\langle \beta_T \rangle \rightarrow \bar \beta_t$, ``kinetic freezeout'' temperature $T_{kin}$ and radial-speed $\beta_t(r)$ profile parameter $n$ as in Table~5 of Ref.~\cite{aliceppbpid} (and see Table~\ref{flowparams} below).

Application of certain fit models (e.g.\ BW models) to spectrum data for small collision systems has been invoked in a number of studies to claim presence of collectivity (flows) in such systems. For example, in Ref.~\cite{cleymans} (reporting BW analysis of unidentified hadrons from \pp\ collisions) the BW model is said to be a ``more standard description'' and is ``based on collective flow in small systems.''  The BW model is said to be ``quite good in explaining the bulk part of the system, however it fails at low-$p_T$ {below 0.5 GeV/c} which could possibly be due to the decays of hadronic resonances.'' ``The applicability of [the BW model] is verified by fitting the transverse momentum spectra of the bulk part ($\sim$ 2.5 GeV/c)....'' 

\begin{table}[h]
	\caption{Blast-wave parameters for simultaneous fits of pion, charged-kaon, neutral kaon, proton and Lambda spectra from 5 TeV \ppb\ collisions~\cite{aliceppbpid}. The $\sigma' / \sigma_0$ values are the cross-section fractions (bin centers) reported in Ref.~\cite{aliceppbpid}. Table~\ref{rppbdata} shows alternative $\sigma / \sigma_0$ values determined in Ref.~\cite{tomglauber}.
}	\label{flowparams}
	\begin{center}
		\begin{tabular}{|c|c|c|c|c|c|} \hline
			$n$ &   $\sigma' / \sigma_0$ &   $\langle \beta_T \rangle$     & $T_{kin}$ (GeV)  & $n$ & $\chi^2$/ndf  \\ \hline
			1	   &      0.025   & 0.547   & 0.143   & 1.07 & 0.27  \\ \hline
			2	 &  0.075  & 0.531    & 0.147   & 1.14 & 0.33  \\ \hline
			3	 &  0.15  & 0.511 & 0.151  &  1.24 & 0.36   \\ \hline
			4	 &  0.30 & 0.478  & 0.157   & 1.41  & 0.35    \\ \hline
			5	 &  0.50   &0.428    & 0.164   & 1.73 & 0.43   \\ \hline
			6	 &  0.70  & 0.36   & 0.169   & 2.4  & 0.54  \\ \hline
			7	 & 0.90  & 0.26  & 0.166  & 3.9  &  0.84   \\ \hline
		\end{tabular}
	\end{center}
\end{table}

\vskip .2in

The argument presented in Ref.~\cite{aliceppbpid} for applying the BW model to PID spectrum data and interpreting the result in a hydrodynamic context proceeds as follows:

(a) Low-\pt\ regions of hadron spectra convey important information: ``The $p_T$ distributions and yields of particles of different mass at low and intermediate momenta of less than a few GeV/c (where the vast majority of particles is produced) can provide important information about the system created in high-energy hadron reactions.''

(b) Hydrodynamics-based spectrum models (e.g.\ blast-wave model) provide useful data descriptions: ``The measured $p_T$ distributions are compared to...hydrodynamic models.'' ``Several [spectrum] parametrizations have been tested, among which the blast-wave function...gives the best description of the data over the full $p_T$ range... [model applied {\em individually} to different hadron species].''

(c) Spectrum trends for heavy-ion collisions appear to buttress validity of a blast-wave spectrum model:  ``In heavy-ion collisions, the flattening of transverse momentum distribution and its mass ordering find their natural explanation in the collective radial expansion of the system. This picture can be tested in a blast-wave framework with a simultaneous fit to all particles for each multiplicity bin. This parameterization assumes a locally thermalized medium, expanding collectively with a common velocity field and undergoing an instantaneous common freeze-out.'' ``This [collective hydrodynamic flow] results in a characteristic dependence of the [spectrum] shape which can be described with a common kinetic freeze-out temperature parameter $T_{kin}$ and a collective average expansion velocity $\langle \beta_t \rangle$~[citing Ref.~\cite{ssflow}].''

In referring to {\em simultaneous} BW fits to PID spectra  Ref.~\cite{aliceppbpid} presents no BW-data fit residuals to support its very low $\chi^2$ values (as reproduced in Table~\ref{flowparams}). Claimed good agreement between {\em individual} BW fits and data appears to extend (for neutral kaons and Lambdas) up to 7 GeV/c where one might expect parton fragmentation to jets to produce essentially {\em all} detected hadrons. Why should the BW model have any relation to that \pt\ region?

The BW model used in Ref.~\cite{aliceppbpid} to fit spectrum data is  adopted from Ref.~\cite{ssflow} that introduced a hydrodynamics-based formula to describe pion spectra from 200 GeV fixed-target ($\sqrt{s_{NN}} \approx 19$ GeV) \mbox{S-S} collisions at the CERN SPS. The relevant formula is Eq.~(7) of Ref.~\cite{ssflow}
\bea \label{bweq}
\frac{dn}{m_t dm_t} \hspace{-.05in} &\propto & \hspace{-.05in} m_t \hspace{-.05in} \int_0^R \hspace{-.1in} r dr I_0\left[\frac{p_t \sinh(\rho)}{T}\right] K_1\left[\frac{m_t \cosh(\rho)}{T}\right],~~~~
\eea
with boost $\rho = \tanh^{-1}(\beta_t)$ and transverse speed $\beta_t(r) = \beta_s(r/R)^n$, where $n = 1$ corresponds to Hubble expansion, $\beta_s$ is the expansion speed at the emitter surface and $I_0$ and $K_1$ are modified Bessel functions. The mean transverse speed is $\langle \beta_t \rangle \rightarrow \bar \beta_t = 2 \beta_s/(n+2)$. Equation~(\ref{bweq}) then in effect represents a thermal (Boltzmann) energy spectrum in the boost (comoving) frame convoluted with a source boost (speed) distribution on source radius to describe the particle spectrum measured in the lab frame.  
 
\begin{figure}[h]
	\includegraphics[width=1.65in]{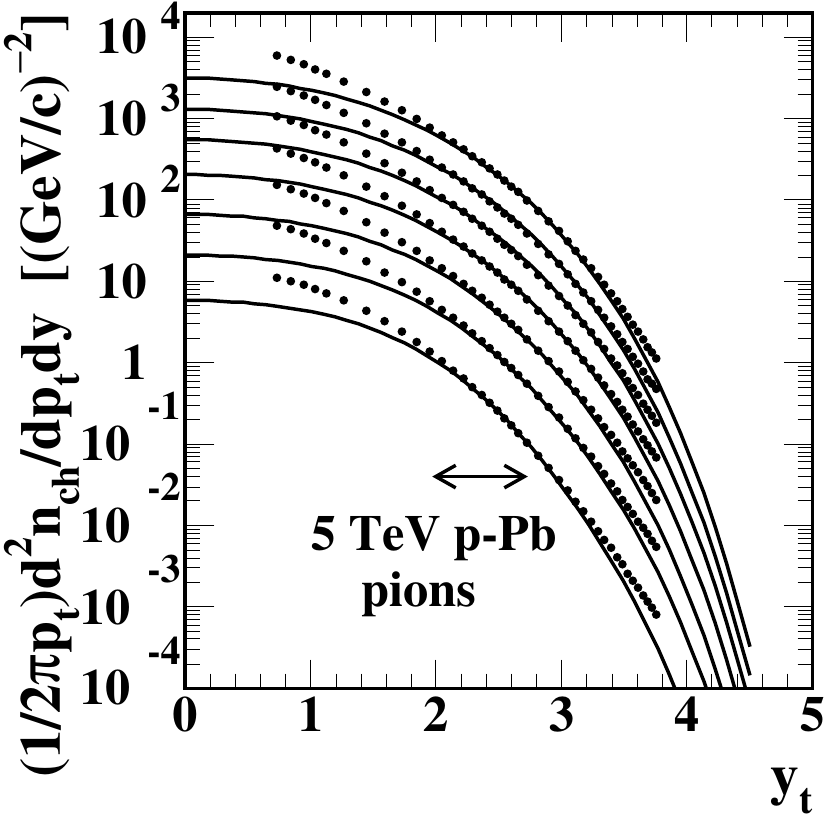}
	\includegraphics[width=1.65in]{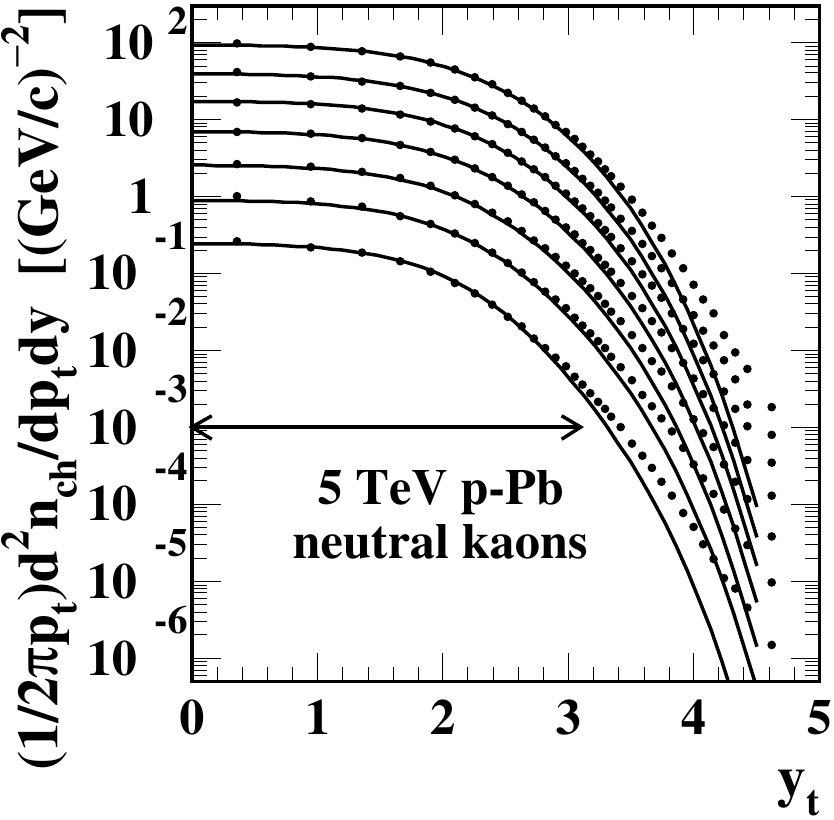}
	\put(-142,105) {\bf (a)}
	\put(-23,105) {\bf (b)}\\
	\includegraphics[width=1.65in]{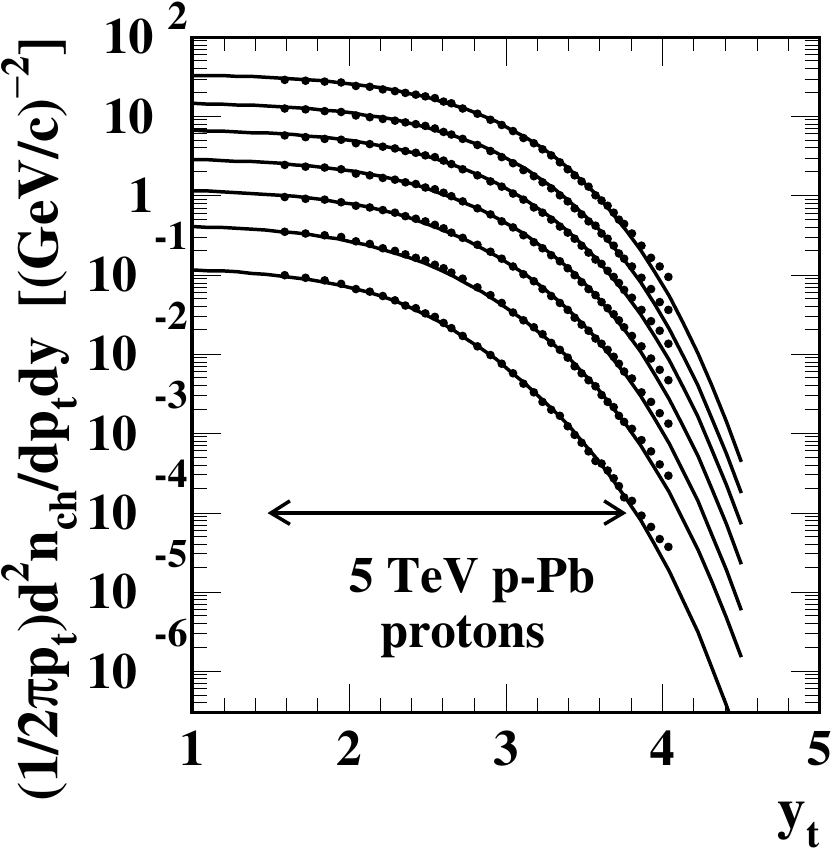}
	\includegraphics[width=1.65in]{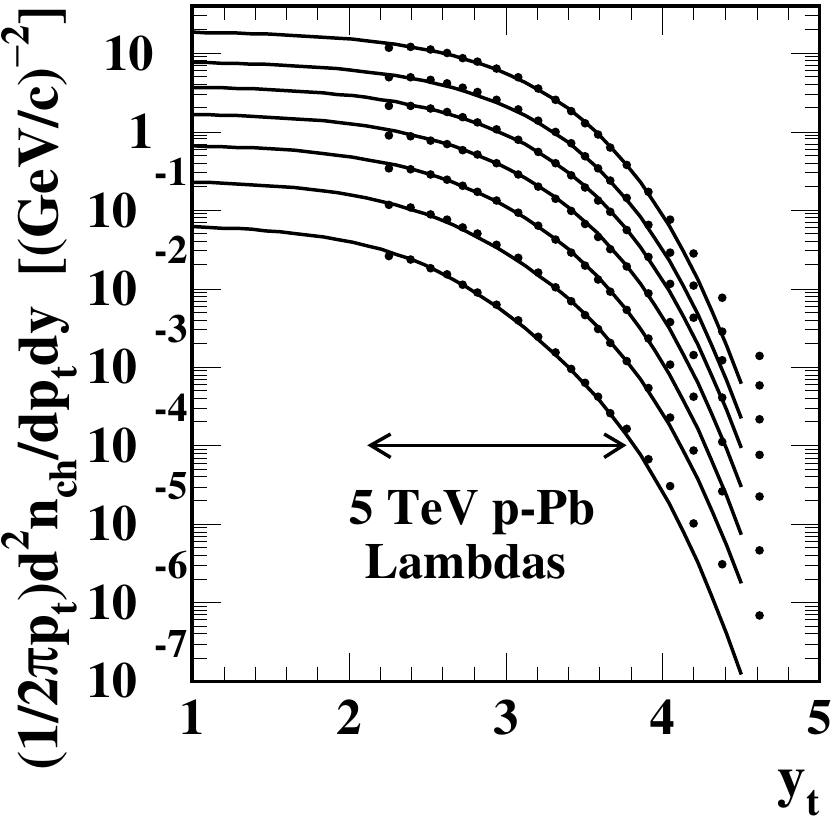}
	\put(-142,105) {\bf (c)}
	\put(-23,105) {\bf (d)}\\
	\caption{\label{bwdata}
		\pt\ spectra for identified hadrons from 5 TeV \ppb\ collisions~\cite{aliceppbpid} plotted vs pion transverse rapidity \yt\ (default) for:
		(a) pions,
		(b) neutral kaons,
		(c) corrected protons,
		(d) Lambdas.
		Solid curves represent blast-wave model fits from Ref.~\cite{aliceppbpid} with amplitudes adjusted here to best describe data within reported BW fit intervals (arrows).
	} 
\end{figure}

Figure~\ref{bwdata} shows the same PID spectrum data from 5 TeV \ppb\ collisions appearing in Fig.~\ref{tcmdata} (points) accompanied in  this case by {\em simultaneous} BW fits to four hadron species (solid) with fit values corresponding to Table~\ref{flowparams}. The various \nch\ classes are scaled up by powers of 2 just as in  Fig.~\ref{tcmdata}. The BW model spectra have been scaled in this case to best agree with data spectra within the indicated BW fit intervals (arrows). Cursory inspection suggests that the BW description of pion spectra is not relevant to interpretation. The BW descriptions of spectra for more-massive hadrons appear to accommodate  data reasonably well below \yt\ = 3-3.6 ($p_t \approx $ 1.4-2.6 GeV/c), but see Sec.~\ref{quality} for detailed assessment of fit quality.

\section{Spectrum model comparisons} \label{compare}

The TCM provides absolute predictions for hadron yields as  well as spectrum shapes. TCM predictions span \nch\ or centrality variation within a given A-B collision system as well as variations from system to system (e.g.\ from \pp\ to \ppb\ to \pbpb). In contrast, the BW model is not expected to provide such absolute predictions: ``...the normalization of the spectrum...we will always adjust for a best fit to the data, because we are {\em only interested in the shape} of the spectra to reveal the dynamics of the collision zone at freeze-out [emphasis added]''~\cite{ssflow}.
Given those limitations any direct model comparisons must be based on shape measures that are not model dependent so as to provide unbiased results. Setting aside absolute yields how best can one demonstrate the evolution of spectrum shape, thus how best test model validity?

For what follows groups of spectra under comparison (e.g.\ \nch\ variation for given hadron species) are rescaled so as to agree in amplitude near \pt\ = 0. The specific common value is not relevant. Successive differences between spectra are considered in relation to TCM hard-component model shapes. Integrals of high-\pt\ and low-\pt\ intervals (absolute yields) for variable-TCM representations of data spectra are examined. Given that TCM context, {\em ratios} of high-\pt\ to low-\pt\ intervals for both TCM and BW model are considered as a function of \nch\ or centrality. Logarithmic derivatives (curvature measures) are applied to TCM and BW models to provide differential spectrum shape information. Spectrum shape variations for each BW model parameter (with others held fixed) are examined in turn. Finally, the BW model is compared with a Boltzmann exponential and with TCM $\hat S_0(m_t)$.

\subsection{Spectrum differences}

\begin{figure}[h]
	\includegraphics[width=1.65in]{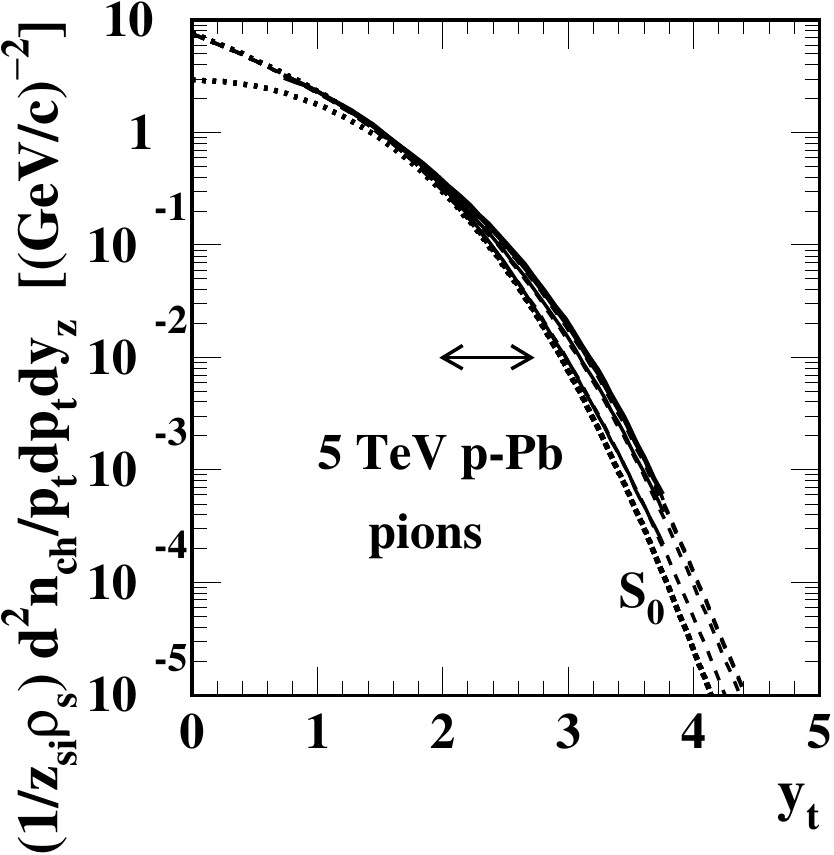}
	\includegraphics[width=1.65in]{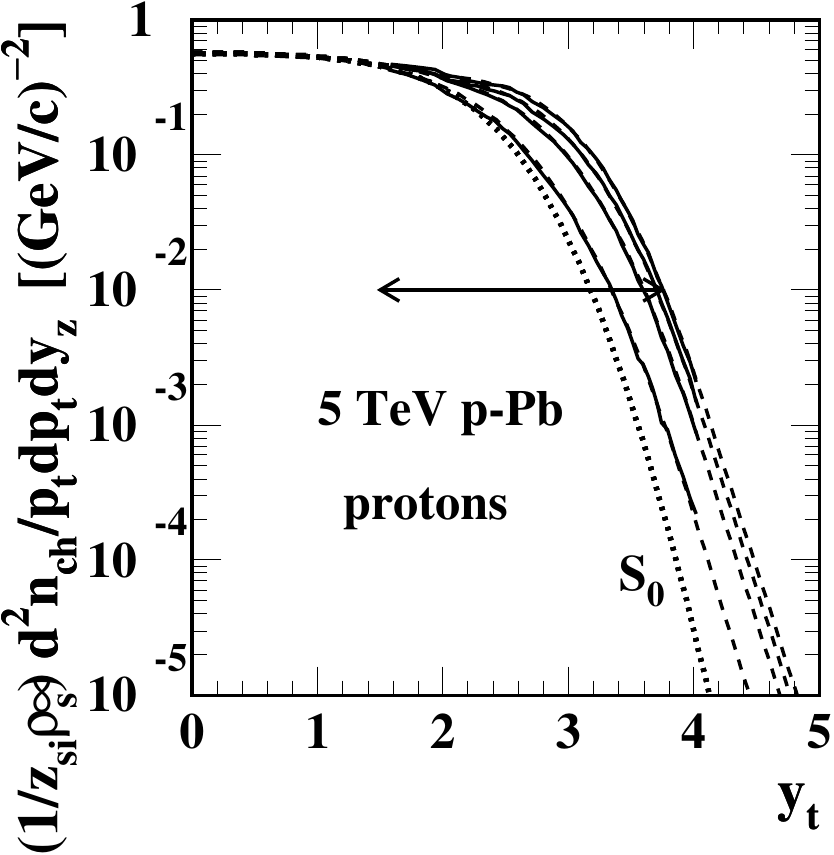}
	\caption{\label{piddatax}
	Data spectra (densities on \pt) for identified pions and protons from 5 TeV \ppb\ collisions~\cite{aliceppbpid} (solid curves) and corresponding TCM curves (dashed) for four centrality classes plotted on logarithmic transverse rapidity \yt. The spectra are rescaled to quantity $X_i(y_t)$ as defined in Eq.~(\ref{eq4}). For clarity only curves for centralities $n = $ 1, 3, 5, 7 are plotted. The lower dotted $\hat S_0(y_t)$ curve for pions is without resonances.
	}
\end{figure}

Figure~\ref{piddatax} shows data spectra (solid) for pions (left) and protons (right). The \pt\ acceptances are rather limited and thus not pursued further in this section. The TCM (dashed) describes data within their statistical uncertainties as demonstrated in Ref.~\cite{pidpart2} Sec.~III E. TCM soft components $\hat S_0(y_t)$ (dotted) represent the limiting case for spectrum data with charge density $\bar \rho_0 \rightarrow 0$ (i.e.~zero particle density). For pions (left) a second $\hat S_0(y_t)$ curve below the data at low \yt\ includes no correction for conjectured resonance contributions (see Sec.~III A of Ref.~\cite{pidpart1}). Arrows in these and other panels indicate \yt\ intervals used for BW fits as reported in Ref.~\cite{aliceppbpid}.

\begin{figure}[h]
	\includegraphics[width=3.3in]{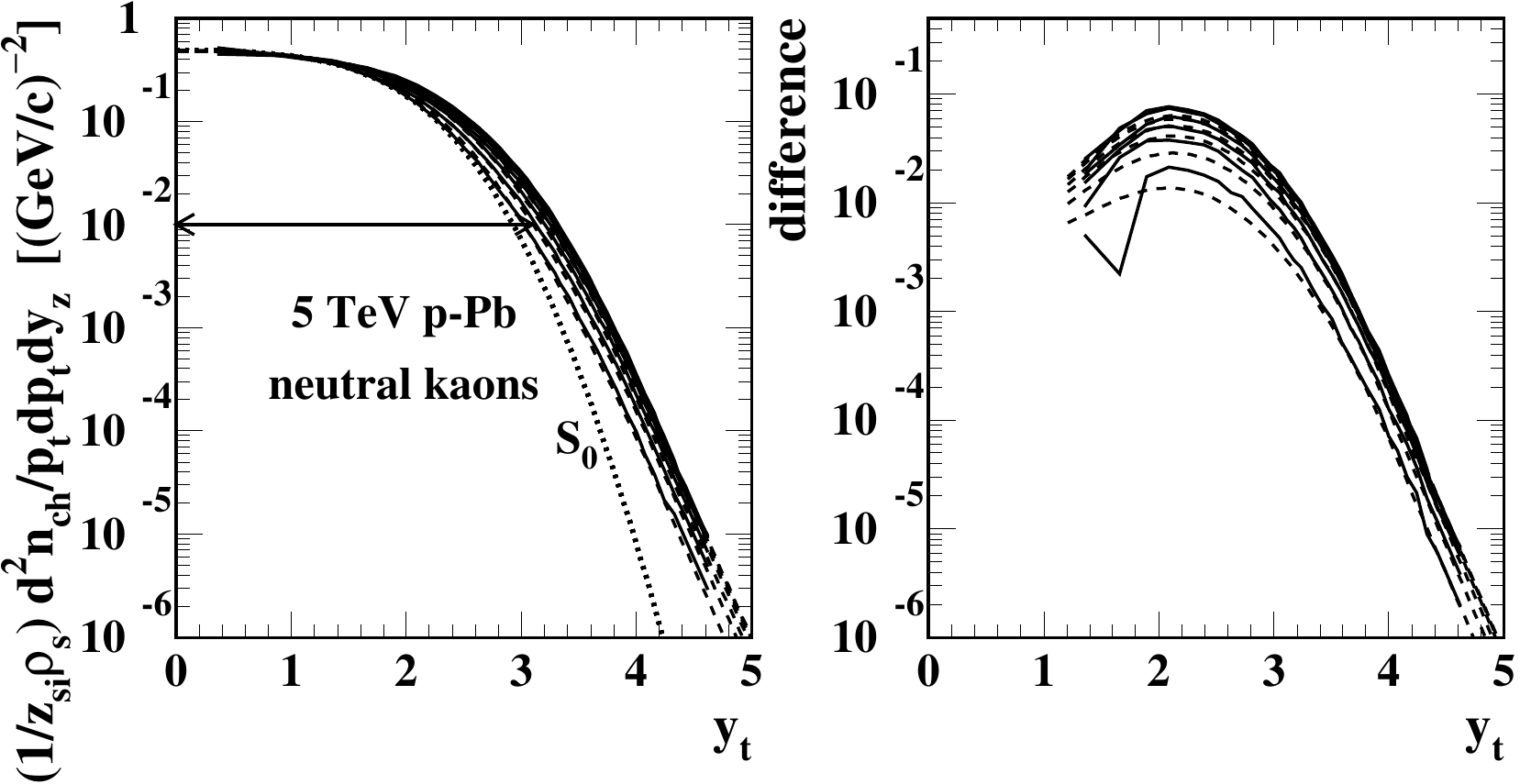}
	\put(-142,105) {\bf (a)}
	\put(-23,105) {\bf (b)}\\
	\includegraphics[width=3.3in]{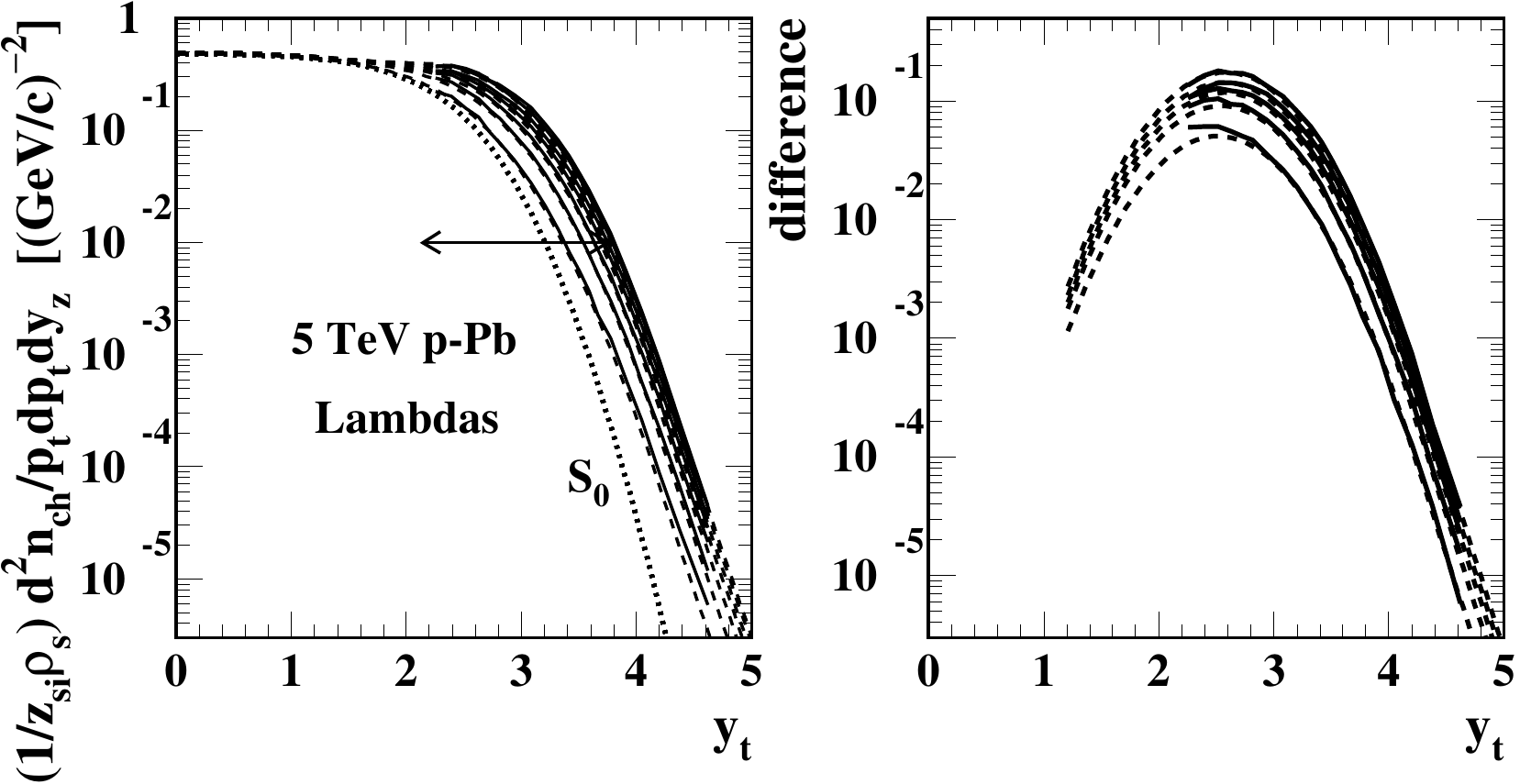}
	\put(-142,105) {\bf (c)}
	\put(-23,105) {\bf (d)}
	\caption{\label{kalam}
		Data \pt\ spectra for identified neutral kaons (a,b) and Lambdas (c,d) from 5 TeV \ppb\ collisions~\cite{aliceppbpid} (solid curves) and corresponding TCM curves (dashed) for four centrality classes. Panels (a,c) compare rescaled data spectra (solid) to TCM (dashed). Panels (b,d) show differences between spectra for $n = 1-6$ and peripheral spectrum $n = 7$.
	} 
\end{figure}

Figure~\ref{kalam} shows spectrum data for neutral kaons (a) and Lambdas (c) with the same line styles as described above. Again the TCM describes data within their statistical uncertainties. The \pt\ acceptances are large in both cases, providing meaningful model tests below.

Figure~\ref{kalam} (c,d) show differences between spectra for a given \nch\ class ($n = $ 1-6) and spectra for the lowest \nch\ class ($n = 7$) for each of neutral kaons (b) and Lambdas (d), and for data (solid) and TCM (dashed). The point of this exercise is to isolate spectrum ``hard components'' without resorting to {\em a priori} physical assumptions, thereby repeating the empirical procedure employed in Ref.~\cite{ppprd} first used to establish the TCM for 200 GeV \pp\ collisions. As noted in that \pp\ analysis  data structures so isolated are approximately independent of \nch\ (e.g.\ \ppb\ centrality),  thus demonstrating that full spectra can be represented as superpositions of two fixed model functions. The amplitude of one (hard) varies as the square of the other (soft) for each \nn\ collision.

More specifically, modes of the ``hard components'' move from lower to higher \yt\ with increasing hadron mass such that spectra for various hadron species tend to coincide at high \yt\ (e.g.\ near \yt\ = 5 or $p_t \approx 10$ GeV/c). Baryon hard components also tend to fall more rapidly below the mode than mesons. Those trends are consistent with fragmentation functions for identified hadrons from \ee\ collisions~\cite{eeprd}. Systematic uncertainties below the mode for difference spectra in panels (b,d) are substantially greater than reported in Refs.~\cite{pidpart1,pidpart2} because of the simpler method invoked here to reduce model dependence. Similar results are obtained by subtracting $\hat S_0(y_t)$ model functions (dotted) in the left panels.

\begin{figure}[h]
	\includegraphics[width=1.65in]{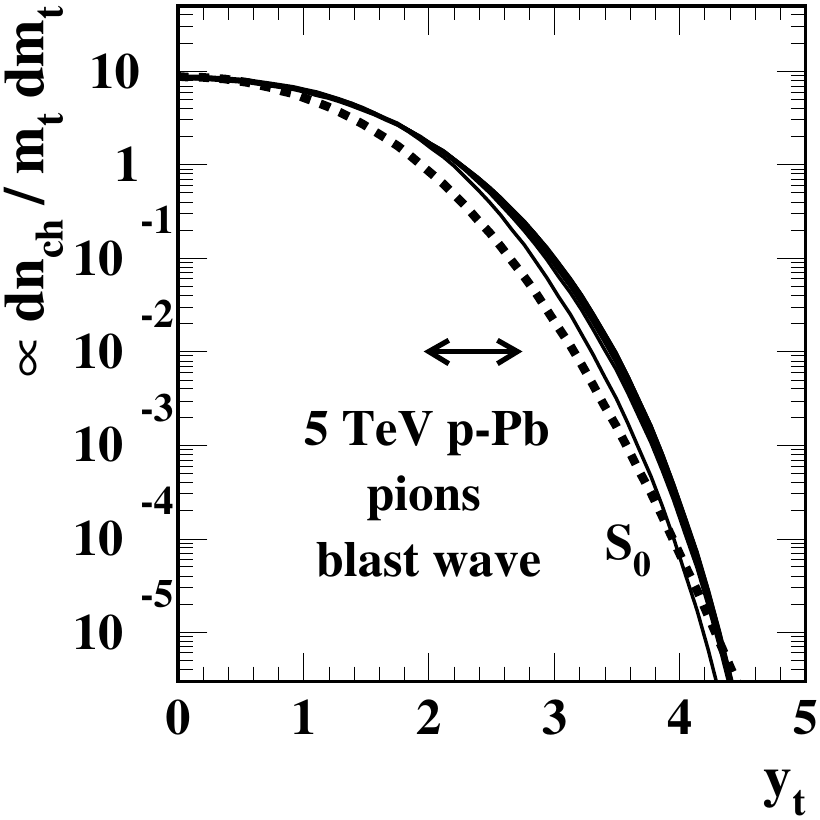}
	\includegraphics[width=1.65in]{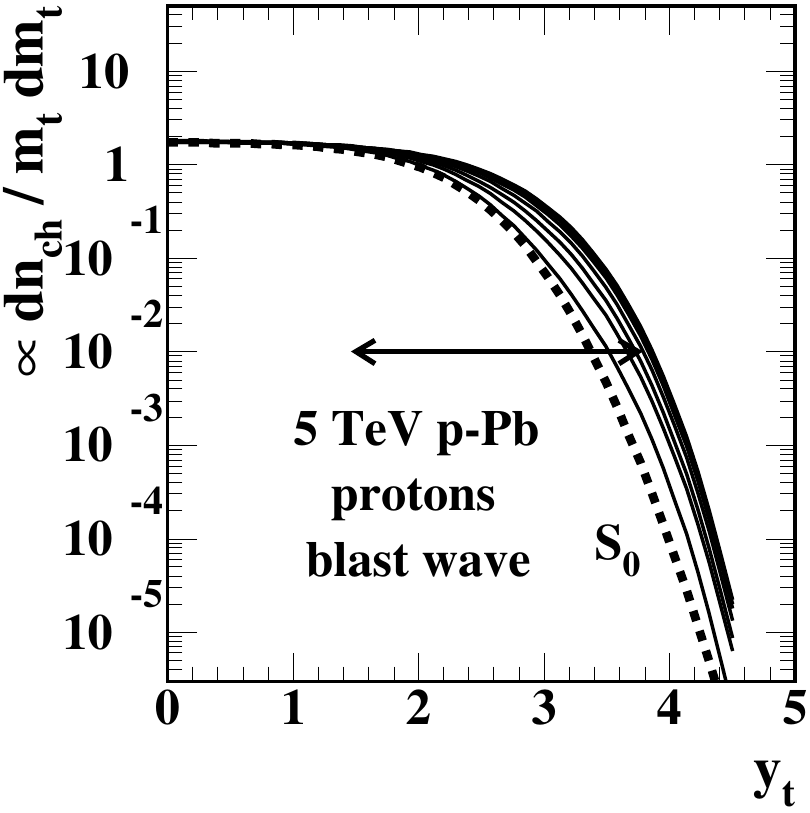}
	\caption{\label{hydropip}
	Blast-wave model \mt\ spectra for identified pions and protons from 5 TeV \ppb\ collisions~\cite{aliceppbpid} for seven centrality classes. The bold dashed curves are TCM soft-component model $\hat S_0(y_t)$. The curves are scaled to coincide at \yt\ = 0.
	} 
\end{figure}

Figure~\ref{hydropip} shows pion (left) and proton (right) spectra from the BW model in Eq.~(\ref{bweq}) using fitted parameter values in Table~\ref{flowparams}, both as reported in Ref.~\cite{aliceppbpid}. Soft-component model $\hat S_0(y_t)$ (dashed) at left does not include the conjectured resonance contribution that appears in Fig.~\ref{piddatax} (left, upper dotted) and describes pion data. The BW curves (solid) have been rescaled to coincide near \yt\ = 0 with the $\hat S_0(y_t)$ soft-component curves (dashed).

\begin{figure}[h]
	\includegraphics[width=3.3in]{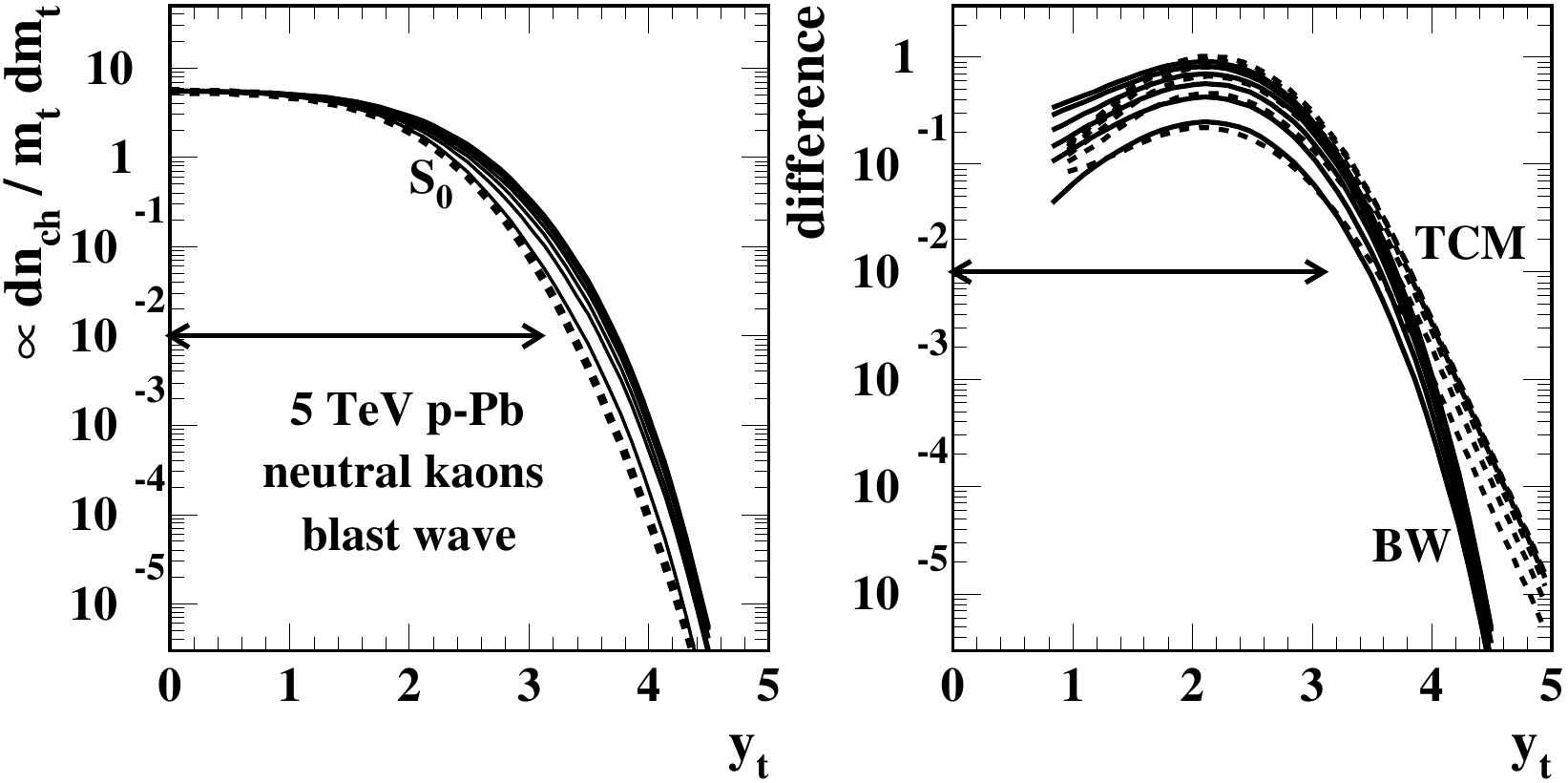}
	\put(-142,105) {\bf (a)}
	\put(-23,105) {\bf (b)}\\
	\includegraphics[width=3.3in]{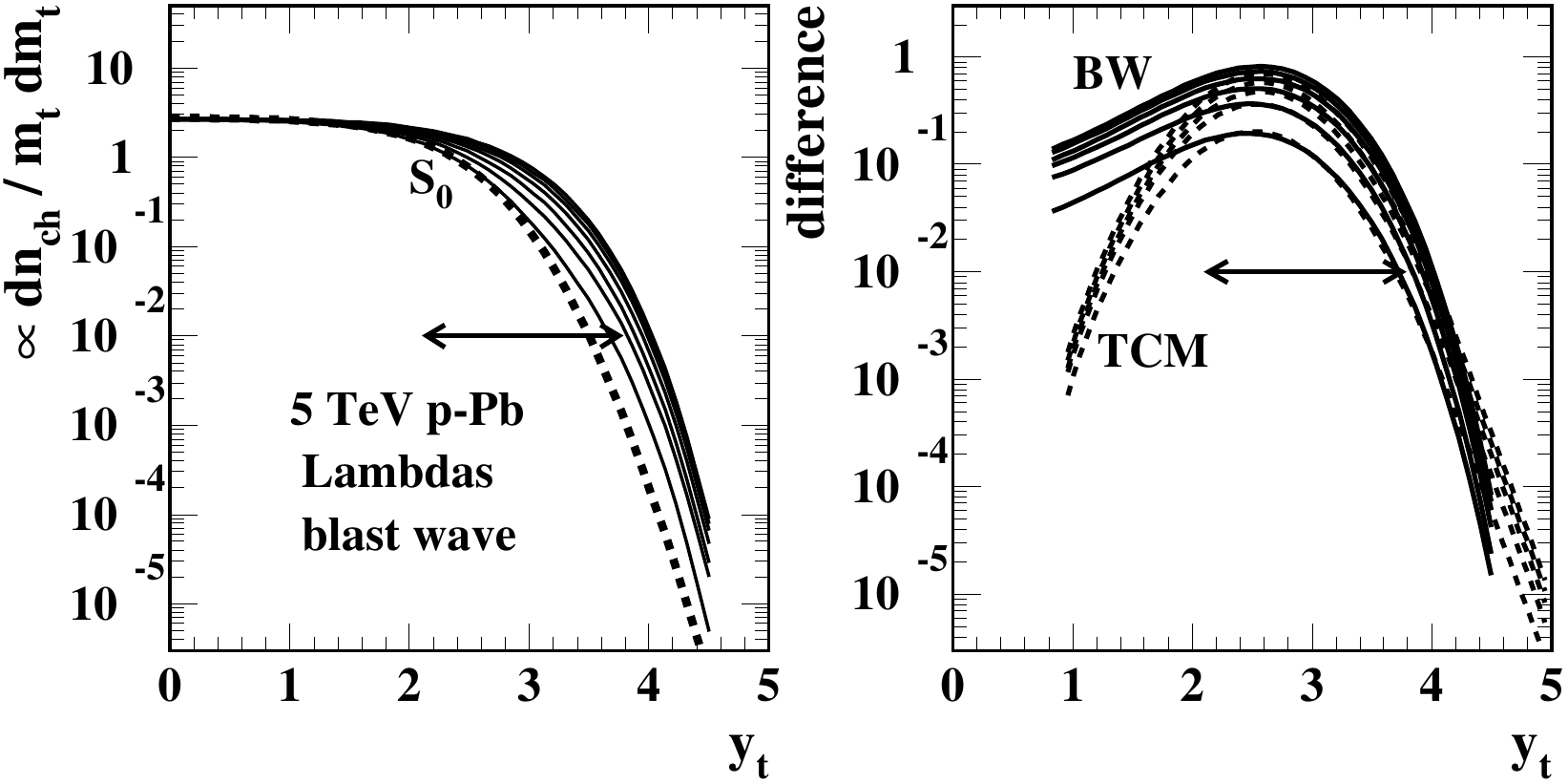}
	\put(-142,105) {\bf (c)}
	\put(-23,105) {\bf (d)}
	\caption{\label{hydrokalam}
	Blast-wave model \mt\ spectra (solid) for identified neutral kaons (a,b) and Lambdas (c,d) from 5 TeV \ppb\ collisions~\cite{aliceppbpid}. The bold dashed curves in (a,c) are TCM soft-component model $\hat S_0(y_t)$. Curves in (a,c) are scaled to coincide at \yt\ = 0.  Panels (b,d) show differences (solid) between BW spectra for $n = 1-6$ and peripheral spectrum $n = 7$. Dashed curves in (b,d) are TCM dashed curves from Fig.~\ref{kalam} (b,d) that represent the data spectrum differences there. The TCM curves are all rescaled by the same factor to best correspond with BW curves within fit intervals (arrows).
	}  
\end{figure}

Figure~\ref{hydrokalam} (a,c) shows neutral kaon (a) and Lambda (c) spectra from the BW model in Eq.~(\ref{bweq}) using fitted parameter values in Table~\ref{flowparams}, both as reported in Ref.~\cite{aliceppbpid}.

Figure~\ref{hydrokalam} (b,d) shows  differences between BW spectra for a given \nch\ class ($n = $ 1-6) and  spectra for the lowest \nch\ class ($n = 7$) for each of neutral kaons (b) and Lambdas (d). The dashed curves are TCM differences repeated from panels (b,d) of Fig.~\ref{kalam} to provide a reference. In either case the dashed curves have all been rescaled by the {\em same factor} to  accommodate the solid BW curves.

This comparison highlights two issues: (a) The BW curves below the mode remain high while the TCM data representations fall more rapidly. That is especially obvious for the Lambda spectra since baryon hard components have modes at higher \yt\ and are narrower compared to meson hard components~\cite{pidpart2}. (b) The BW curves above the mode fall off rapidly while the TCM data representations follow a power-law trend that is more gradual. That is especially obvious for the kaon spectra for the same reasons given above. The contrast is amplified by the different BW fit intervals used for the two cases.

\subsection{High/low ratio centrality trends}

\begin{figure}[h]
	\includegraphics[width=3.3in]{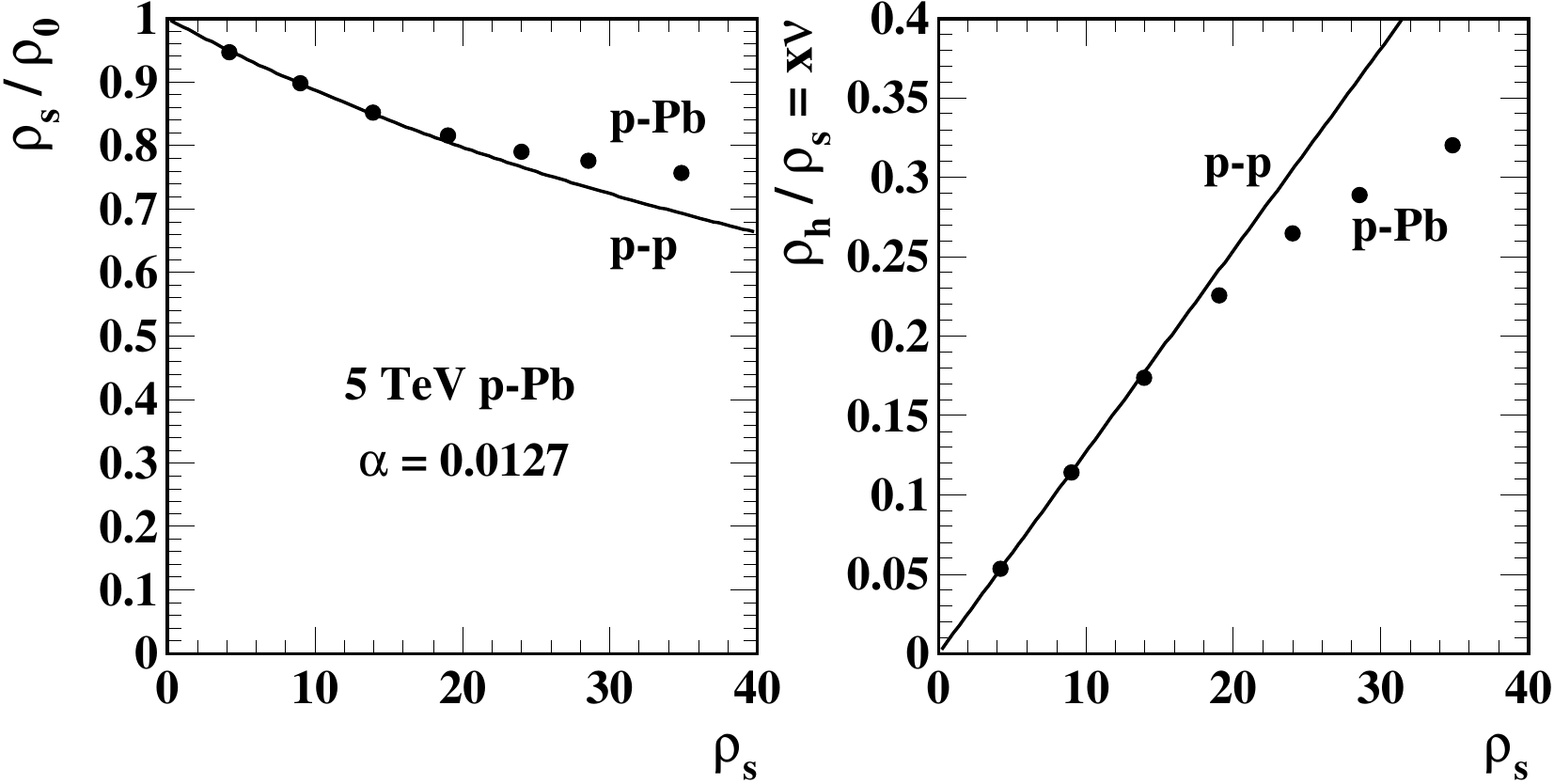}
	\caption{\label{geomsum}
Soft fraction $\bar \rho_s / \bar \rho_0$ (left) and hard/soft ratio $\bar \rho_h / \bar \rho_s \equiv x(n_s)\nu(n_s)$ (right) vs charge-density soft component $\bar \rho_s$.  The straight line at right represents the relation $\bar \rho_h \approx \alpha \bar \rho_s^2$ (with $\alpha = 0.013$) that is observed for single \nn\ collisions~\cite{ppprd}.
	}  
\end{figure}

Certain centrality trends are more interpretable when presented in the context of 5 TeV \ppb\ centrality variation as reported in Ref.~\cite{tomglauber} based on accurate TCM description of ensemble \mmpt\ data for that collision system. That centrality model then provided the basis for PID spectrum studies reported in Refs.~\cite{ppbpid}, \cite{pidpart1} and \cite{pidpart2},

Figure~\ref{geomsum} [Fig.~17 of Ref.~\cite{pidpart1} in relation to its Eqs.~(26) and (27)] shows soft fraction $\bar \rho_s / \bar \rho_0$ (left) and hard/soft ratio $\bar \rho_h / \bar \rho_s \equiv x(n_s)\nu(n_s)$ (right) vs charge-density soft component $\bar \rho_s$, with $\bar \rho_x \equiv n_x / \Delta \eta$ and $\bar \rho_0 = \bar \rho_s + \bar \rho_h$. The straight line at right represents the relation $\bar \rho_h \approx \alpha \bar \rho_s^2$ (as first reported in Ref.~\cite{ppprd}) with $\alpha \approx 0.013$ for 5 TeV \nn\ collisions~\cite{alicetomspec,pidpart1}. It is notable that (a) the lowest four of seven \ppb\ \nch\ classes are effectively equivalent to single peripheral \pn\ collisions and (b) for higher \nch\ classes \ppb\ centrality {\em does} increase significantly but jet production (measured by $\bar \rho_h / \bar \rho_s$) increases less rapidly.

The comparison of TCM and BW spectrum variation with centrality in the previous subsection, while model-independent, is not quantitative and is therefore  not conclusive. One can alternatively compare yields at low and high \pt\ for data (via variable TCM) and the BW model. 

\begin{figure}[h]
	\includegraphics[width=1.65in]{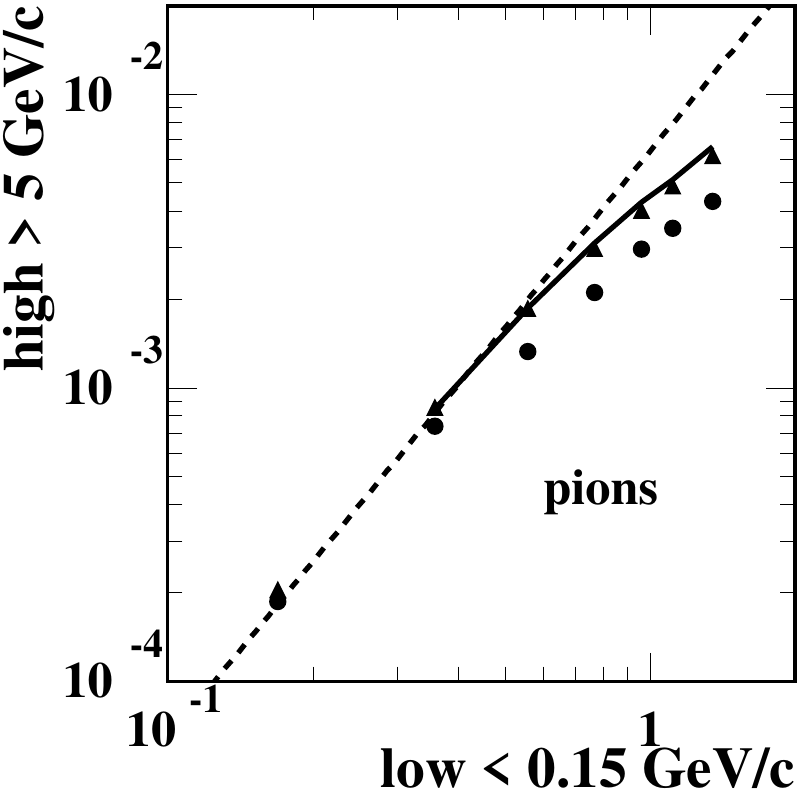}
	\includegraphics[width=1.65in]{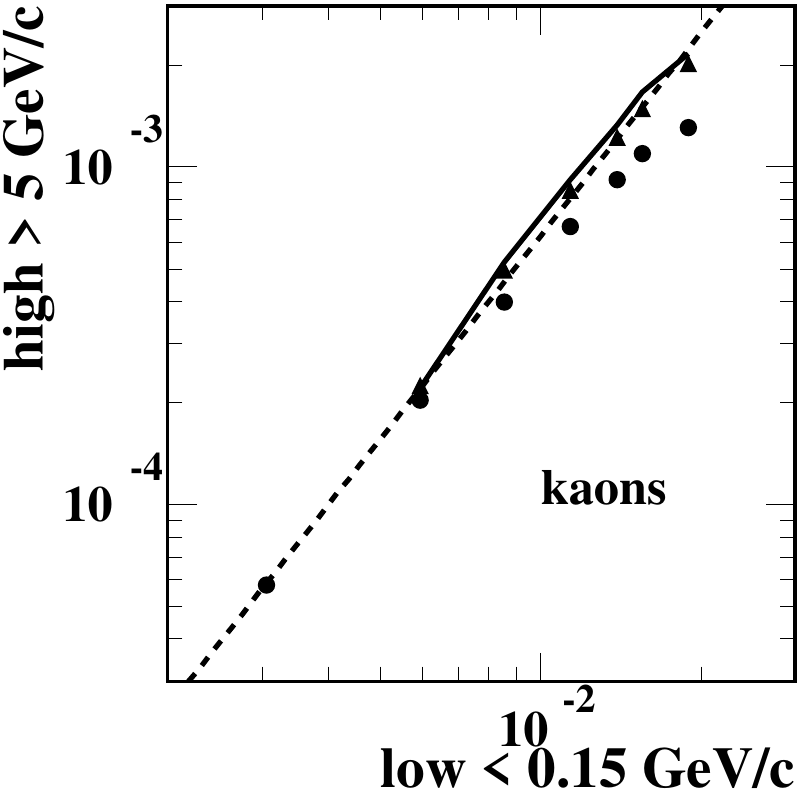}
	\put(-162,105) {\bf (a)}
	\put(-43,105) {\bf (b)}\\
	\includegraphics[width=1.65in]{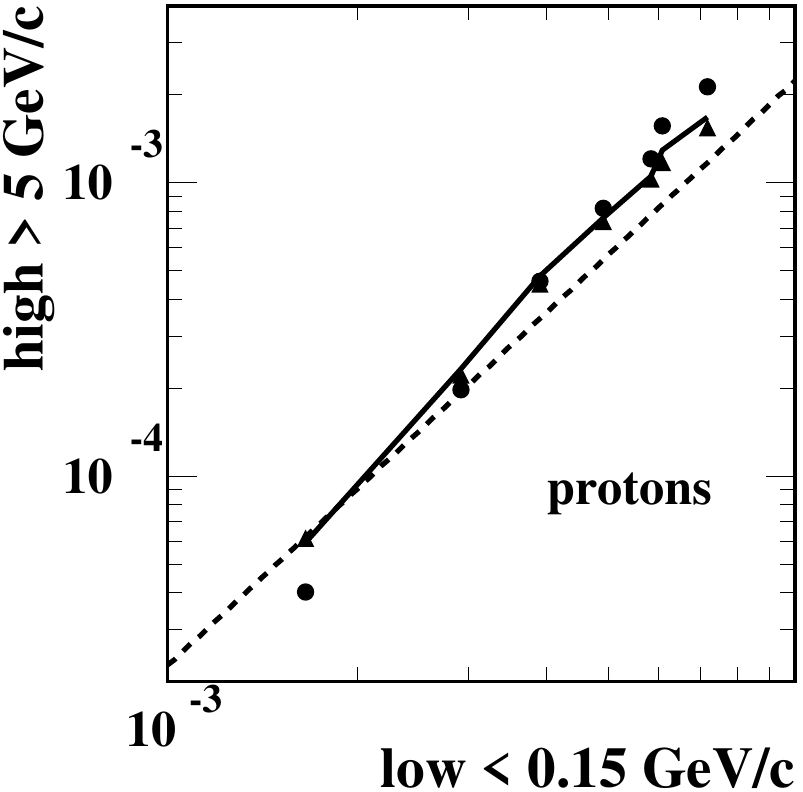}
	\includegraphics[width=1.65in]{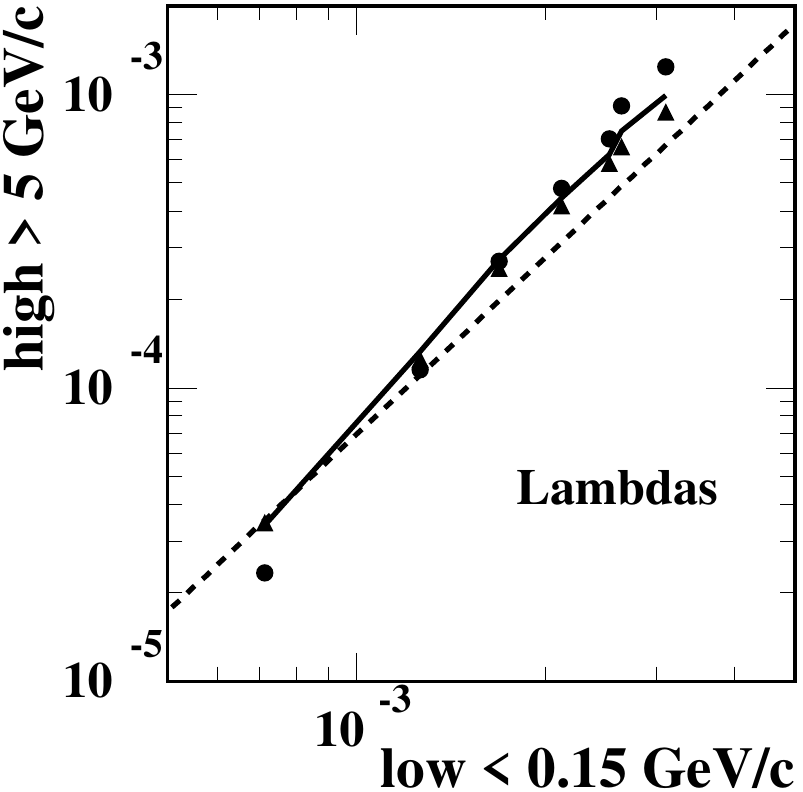}
	\put(-162,105) {\bf (c)}
	\put(-43,105) {\bf (d)}
	\caption{\label{hilowtrends} High vs low spectrum-integral trends (solid dots) for PID data spectra from 5 TeV \ppb\ collisions and for
		(a) pions,
		(b) neutral kaons,
		(c) corrected protons,
		(d) Lambdas. The \yt\ integrals have the form $(1/2\pi)dn_{ch}/dy_z$. The dashed lines represent high $\propto$ low$^2$ reference trends (expected for \nn\ collisions). The solid triangles are $x\nu \times$low trends. The solid curves are $\tilde z x \nu \times$low trends. Refer to text for further details.
	}  
\end{figure}

Figure~\ref{hilowtrends} compares yields within \pt\ intervals for $p_t < 0.15$ GeV/c (low) to those for $p_t > 5$ GeV/c (high). The solid points are derived from data  extrapolations defined by the {\em variable} TCM (which represents PID spectrum data accurately as established in Ref.~\cite{pidpart2}). A reference high-vs-low trend may be derived from the {\em fixed} PID TCM from Ref.~\cite{pidpart1} as follows. Since~\cite{pidpart1}
\bea
\bar \rho_{si} &\approx &  z_{si}(n_s) \bar \rho_s~~\text{and}
\\ \nonumber
\bar \rho_{hi} &\approx& z_{hi}(n_s) \bar \rho_h \approx \tilde z_{i}  x(n_s)\nu(n_s) \bar \rho_{si}
\eea
for hadron species $i$ then $\text{high} \approx \tilde z_{i}  x(n_s)\nu(n_s) \times \text{low}$
assuming fixed TCM model functions (solid curves). Values for $x(n_s)$ and $\nu(n_s)$ are taken from Table~I and for ratios $\tilde z_i$ are derived from Tables~IV and V of Ref.~\cite{pidpart1}. Note that for those tables $z_{si}$ are determined near 0.15 GeV/c ($y_t \approx 1$) while $z_{hi}$ are obtained near 1 GeV/c ($y_t \approx 2.7$).

The solid dots represent accurate variable-TCM representations of data spectra integrated over stated \pt\ intervals. The dashed lines are high $\approx$ low$^2$ $\rightarrow$ $x(n_s) \times$low (reference trends) that would be expected for single peripheral \pn\ collisions and fixed TCM model functions. The triangles are $x(n_s) \nu(n_s) \times$low scaled to agree with the \pn\ (dashed) trend for peripheral collisions. The relation of triangles to dashed lines for pions may be compare with Fig.~~\ref{geomsum} (right). The solid curves are $\tilde z_i x(n_s) \nu(n_s) \times$low scaled to agree with dashed trends for peripheral collisions again assuming fixed TCM model functions.

Deviations of data trends (solid dots) from $\tilde z_i x(n_s) \nu(n_s) \times$low trends (solid curves) indicate the effect of hard-component model variations with \nch\ as reported in Sec.~III of Ref.~\cite{pidpart2}. Meson trends fall well below fixed-TCM trends because meson hard components above the mode are transported to lower \yt\ with increasing \ppb\ \nch. While baryon hard components also vary substantially with increasing \nch\ (modes transported to {\em higher \yt}) the variation is mainly below the mode. Baryon solid-dot and solid-curve high vs low trends in panels (c,d) are thus different but much closer.

Those result rely on the TCM to provide accurate descriptions of {\em absolute} data yields and spectra. As typically utilized, the BW model is not expected to predict absolute yields: ``...the normalization of the spectrum...we will always adjust for a best fit to the data...''~\cite{ssflow}. However, spectrum high/low {\em ratios} can be compared between TCM and BW models.

\begin{figure}[h]
	\includegraphics[width=3.3in]{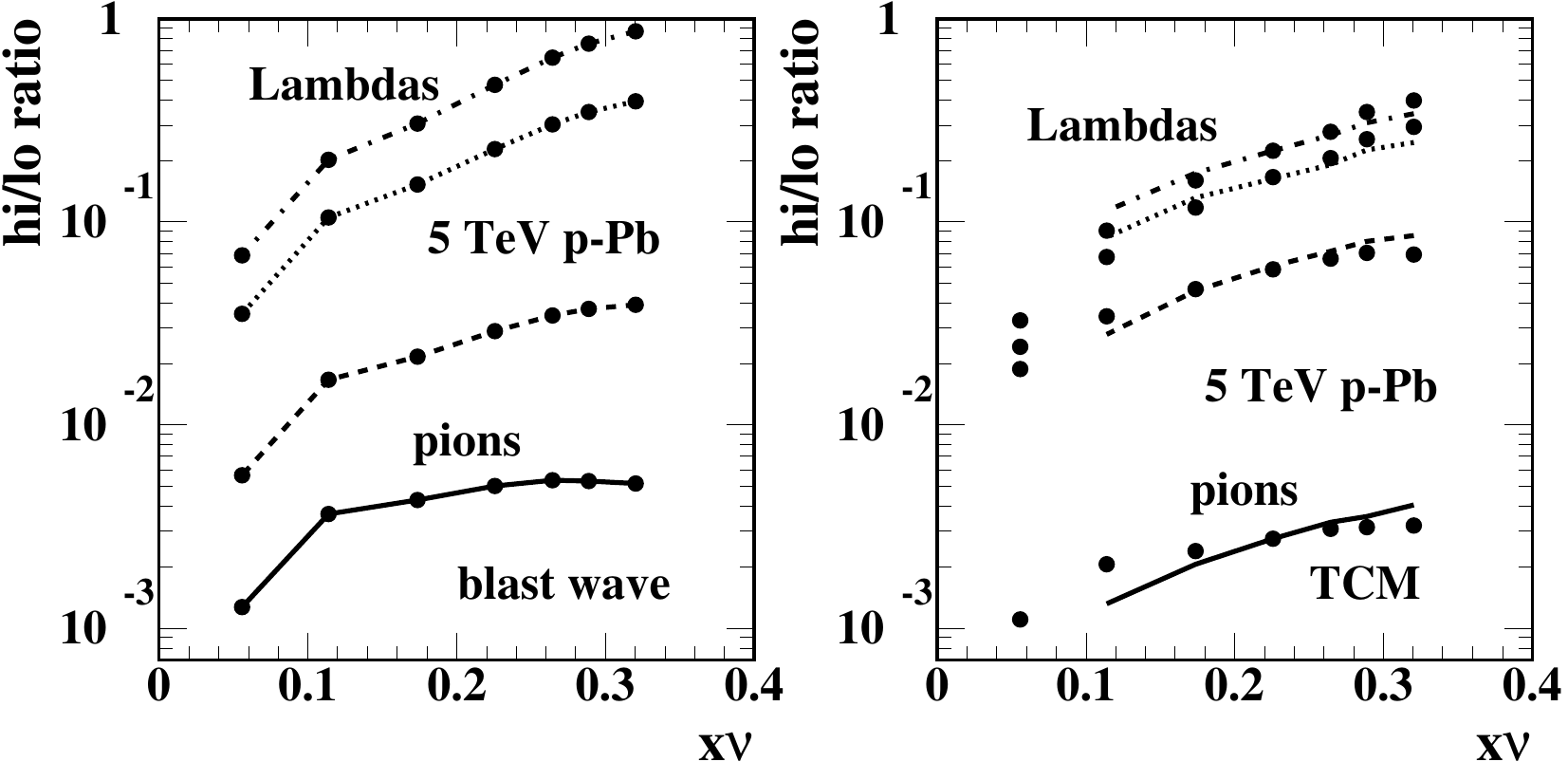}
	\caption{\label{hilorats}
		Left: High/low (hi/lo) ratios derived from integrals of BW model spectra integrated over $p_t < 0.15$ GeV/c (low) and $p_t > 4$ GeV/c (high) for four hadron species.
		Right: High/low ratios for data represented by the variable TCM. The lower limit for the ``high'' integral is 4 GeV/c to match the left panel. The curves represent the $\tilde z_{i}  x(n_s)\nu(n_s)$ trend expected for fixed spectrum hard components. Note the systematic difference between mesons and baryons as found in Ref.~\cite{pidpart2}.
	}  
\end{figure}

Figure~\ref{hilorats} (left) shows high/low (hi/lo) ratios derived from integrals of BW model spectra integrated over $p_t < 0.15$ GeV/c (low) and $p_t > 4$ GeV/c (high) for four hadron species. The lower limit for ``high'' is reduced compared to the 5 GeV/c limit in Fig.~\ref{hilowtrends} to accommodate the restricted \yt\ range of the BW model spectra.

Figure~\ref{hilorats} (right) shows high/low ratios for data (points) represented by the variable TCM. The lower limit for ``high'' is here 4 GeV/c to match the left panel. The curves are equivalent to solid curves in Fig.~\ref{hilowtrends} representing the $\tilde z_{i}  x(n_s)\nu(n_s)$ trend expected for fixed spectrum hard components. The curves are scaled vertically to pass through data points corresponding to event class $n = 4$.

While there is some similarity in  the general shapes of BW and data high/low \nch\ trends, there are large differences in mean values which are due in  part to the very different \pt\ intervals imposed on BW model fits. BW ratios for baryons are a factor 3-4 greater than those for data/TCM. The BW fit interval for baryons is confined to higher \pt\ and extends well beyond hard-component modes. In contrast, the BW ratio trend for neutral kaons falls significantly below that for data/TCM, and the fit interval extends on \pt\ from just above the hard-component mode down to zero. The BW ratios for pions are comparable to those for data/TCM, but the \pt\ fit interval is negligible in comparison to the other cases.

\subsection{Differential spectrum curvatures}

While comparisons of high/low ratios as above do provide some quantitative information on spectrum structure they are still not definitive. A more-differential approach is called for. One possibility is logarithmic derivatives. In Ref.~\cite{pidpart2} Sec.~IV E a logarithmic derivative was applied to investigate conjectures in Ref.~\cite{aliceppbpid} concerning power-law dependences of integrated \pt\ intervals on charge density $\bar \rho_0$. It was demonstrated there that the approximate power-law trends on $\bar \rho_0$ of baryon/meson spectrum ratios resulted from transport of peaked spectrum hard components (for baryons) to higher \yt\ with increasing \nch, the trends corresponding quantitatively to the Gaussian (+ exponential tail) shapes on \yt\ of spectrum hard components. In the present context direct analysis of \yt\ spectra via logarithmic derivative provides the required information.

The logarithmic derivative applied to spectra $\bar \rho_0(y_t)$ can be illustrated using the TCM hard-component model [i.e.\ $\bar \rho_0(y_t) \rightarrow \hat H_0(y_t)$, see Ref.~\cite{pidpart2} Eq.~(8)]:
\bea \label{first}
-\frac{d\ln[\bar \rho_0(y_t)]}{dy_t}&\rightarrow&  \frac{y_t - \bar y_t}{\sigma^2_{y_t}}~~\text{near the mode}
\\ \nonumber
&\approx& q
 ~~\text{well above the mode}.
\eea
The second derivative leads to
\bea \label{second}
-\frac{d^2\ln[\bar \rho_0(y_t)]}{dy_t^2}&\rightarrow& 1/\sigma_{y_t}^2~~\text{near the mode}
\\ \nonumber
&\approx& 0~~\text{well above the mode}
\eea
The second derivative functions as a measure of {\em local curvature} or rate of change of local slope of a spectrum. 

\ppb\ spectra can be characterized in general as follows (based on TCM structure): At low \pt\ spectra vary as $\hat S_0(m_t)$ -- approximately as a Boltzmann exponential on $m_{ti}$ with $m_{ti} \rightarrow m_i \cosh(y_{ti})$ for hadron species $i$. At high \pt\ spectra approximate a power law on \pt\ $\sim p_t^n$ (with $p_t \sim m_0 \sinh(y_t)$) or exponential on \yt. Based on  those characteristics one expects for Eq.~(\ref{second})

\bea \label{limits}
-\frac{d^2\ln[\bar \rho_0(y_t)]}{dy_t^2}&\sim& \cosh(y_t)~~\text{at low $p_t$}
\\ \nonumber
&\sim& 0~~~\text{at high \pt}.
\eea
Note that $\hat S_0(y_t)$ (a L\'evy distribution on \mt) itself goes to a power law on \mt\ at higher \yt. Its logarithmic second derivative should thus also fall to zero at higher \yt.

\begin{figure}[h]
	\includegraphics[width=1.65in]{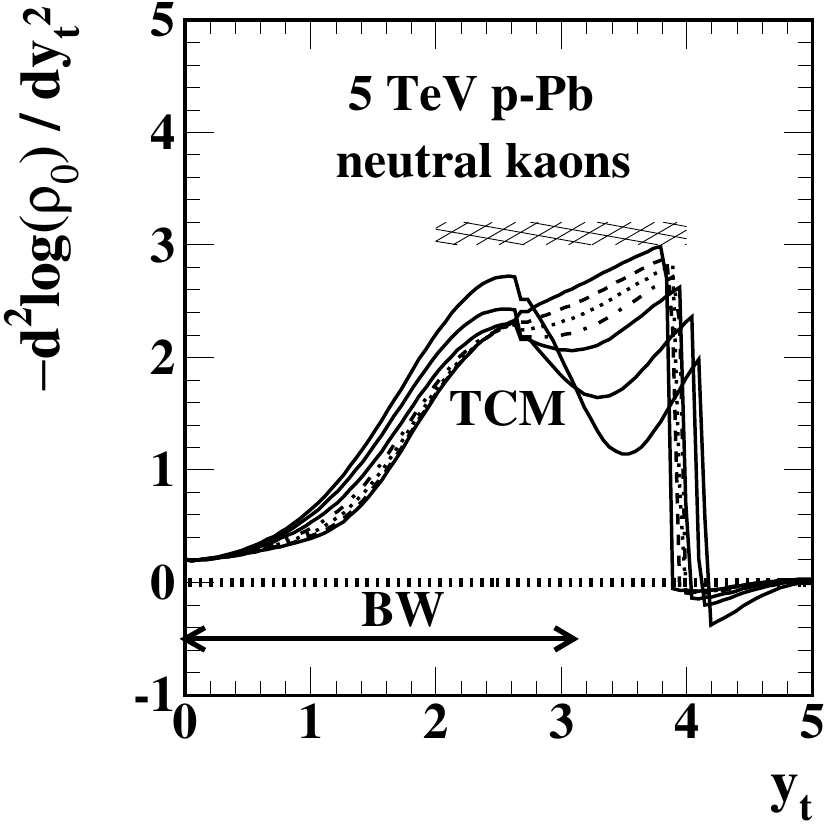}
\includegraphics[width=1.65in]{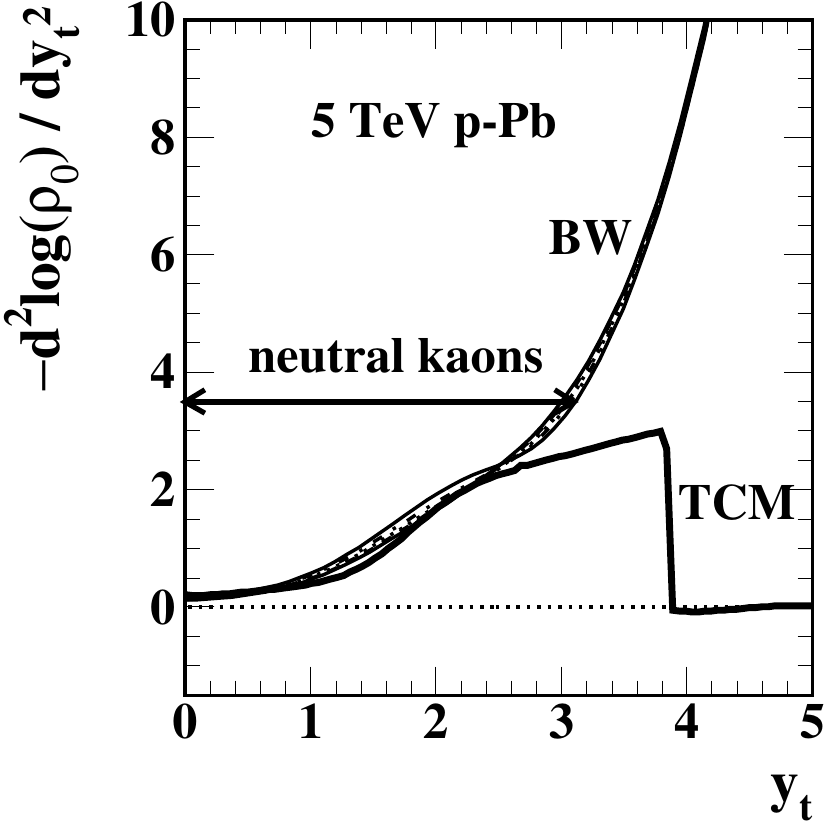}
	\caption{\label{logderiv}
	Left: Logarithmic second derivative Eq.~(\ref{second}) applied to variable TCM spectra accurately representing data for neutral kaons. Line styles proceed down from most-central as solid, dashed, dotted, dash-dotted. The hatched band is limiting case $1 / \sigma_{y_t}^2$ for Eq.~(\ref{second}).
	Right: Logarithmic second derivative Eq.~(\ref{second}) applied to BW spectra with  the same line-style sequence. The bold solid curve is the $n=1$ most-central curve from the left panel.
	} 
\end{figure}

Figure~\ref{logderiv} (left) shows Eq.~(\ref{second}) (left side) applied to  neutral kaon spectra from 5 TeV \ppb\ collisions as represented by the variable TCM. The kaon data are preferred because of the large \yt\ acceptance. The line styles proceeding down from most-central are solid, dashed, dotted and dash-dotted followed by solid for the remainder. The trends at low \yt\ correspond to $\cosh(y_t)$ and at high \yt\ drop to zero as anticipated by Eq.~(\ref{limits}). The hatched band corresponds to $1/\sigma^2_{y_t}$ for neutral kaons (see Table~\ref{pidparamsxx}). The trend expected for data hard components alone is constant (within the hatched band) terminating near \yt\ = 4 (where  the hard component transitions from Gaussian to exponential) dropping toward zero above that point. Variation between \yt\ = 2 and 4 corresponds to relative contributions of soft and hard components at a given \yt\ varying with \ppb\ centrality ($\sim$ product $x\nu$).

Figure~\ref{logderiv} (right) shows Eq.~(\ref{second}) (left side) applied to BW spectra (curves of several line styles) as reported in Ref.~\cite{aliceppbpid} and as in Fig.~\ref{hydrokalam} (a). The result for the variable TCM representing most-central ($n = 1$) \ppb\ data (left panel) is included as a reference. The strong deviation between BW and data/TCM is evident above \yt\ = 2.5 ($p_t \approx 0.8$ GeV/c). Variation of BW model parameters with \nch\ leads to rather small spectrum shape variations (compared to the left panel) below \yt\ = 3 where $\chi^2$ fits should matter. What follows is a procedure to examine the effect of varying each of  three BW parameters in turn.

\subsection{BW model response to individual parameters}

\begin{figure}[h]
	\includegraphics[width=1.65in]{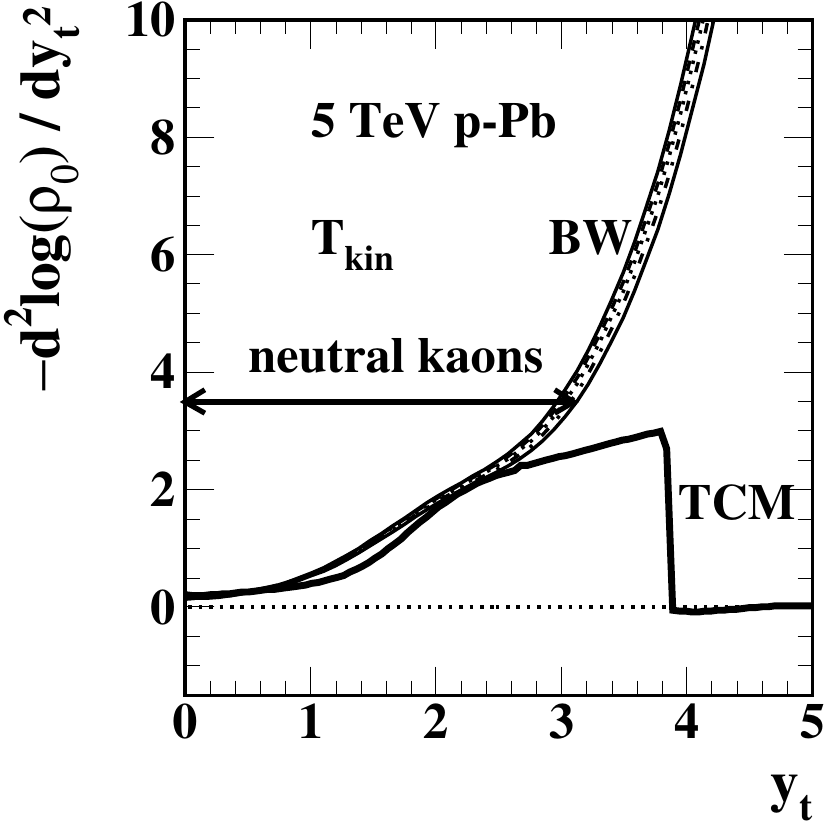}
\includegraphics[width=1.65in]{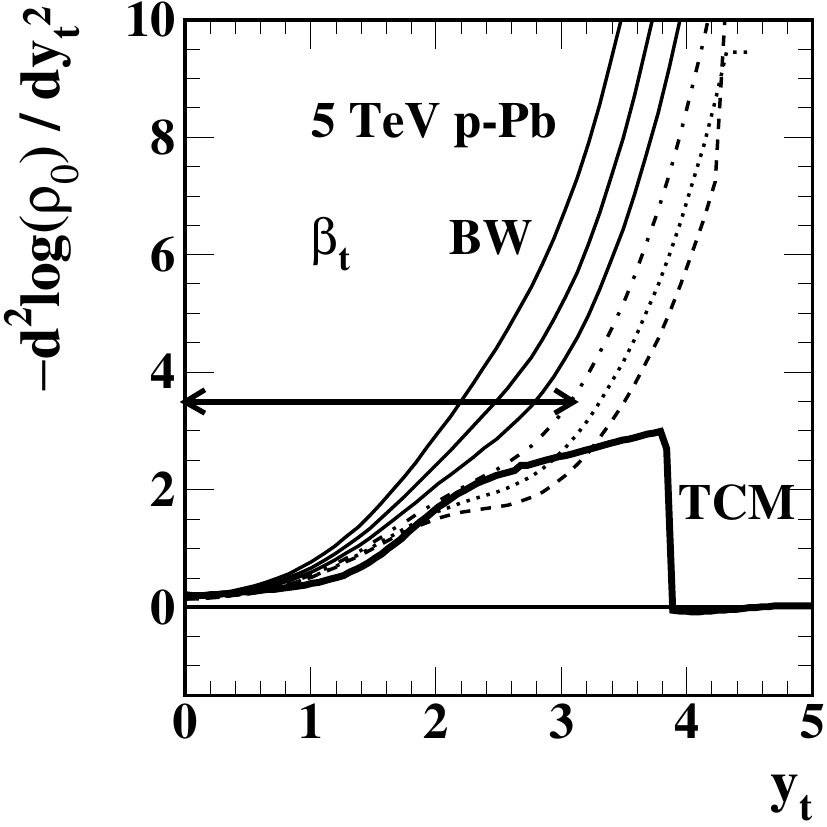}
	\put(-139,105) {\bf (a)}
	\put(-18,105) {\bf (b)}\\
	\includegraphics[width=1.65in]{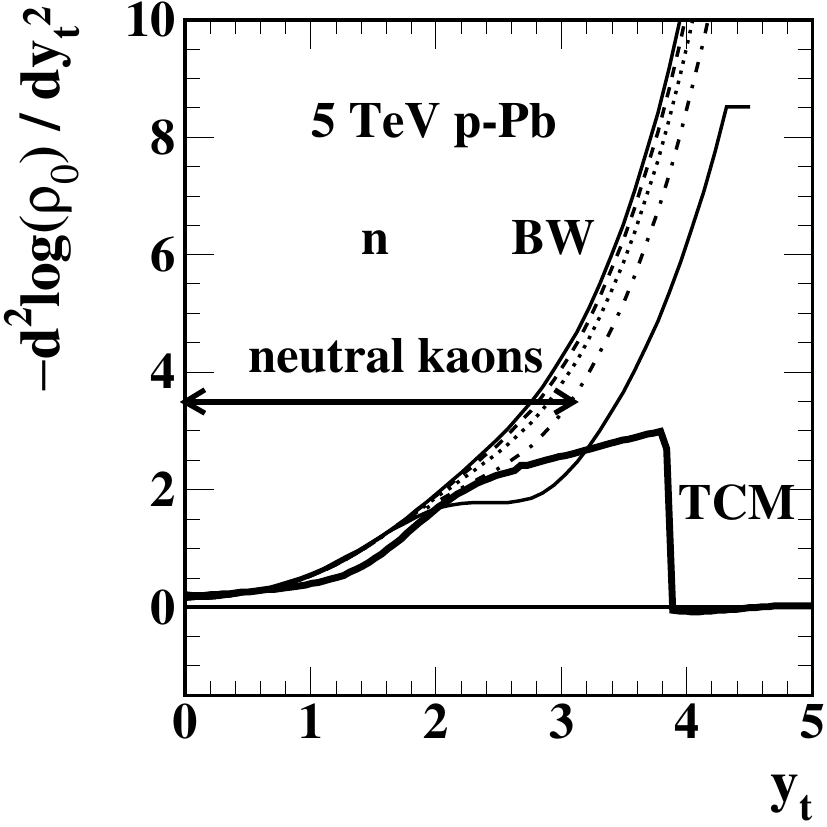}
	\includegraphics[width=1.65in]{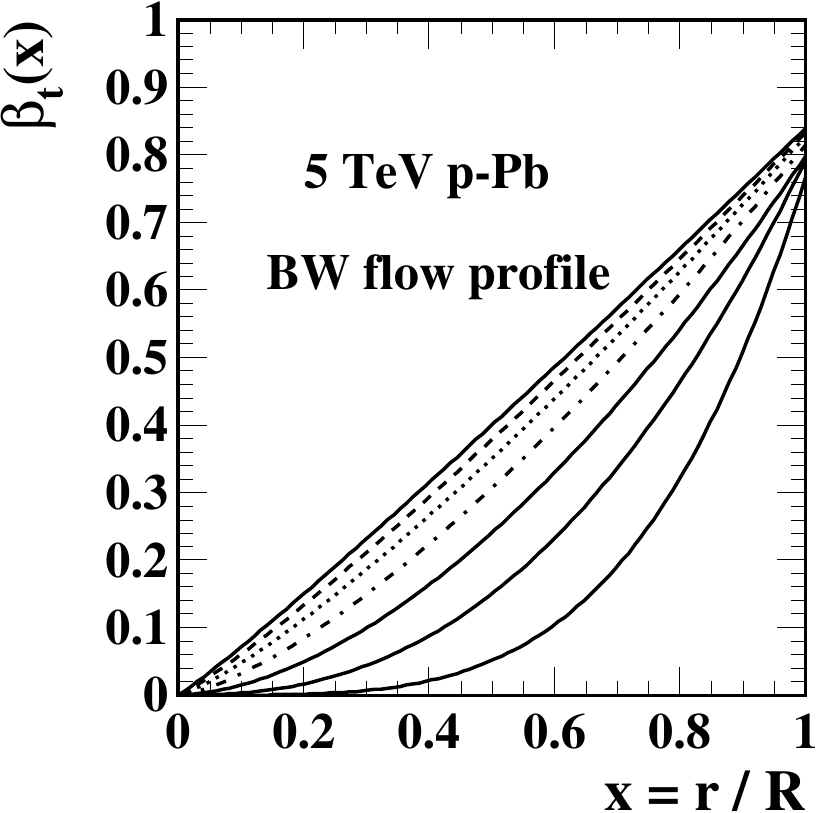}
	\put(-139,105) {\bf (c)}
	\put(-20,102) {\bf (d)}
	\caption{\label{logdetails}
	BW second derivatives with only one parameter varying: (a) $T_{kin}$ varies, (b) $\bar \beta_t$ varies, (c) profile parameter $n$ varies. In each case other BW parameters are maintained at centrality $n = 4$ values and the TCM reference for $n = 1$ is included. (d) BW $\beta_t(r)$ profiles for seven \ppb\ centralities.
	} 
\end{figure}

Figure~\ref{logdetails} (a) shows the effect of varying ``kinetic freezeout'' temperature $T_{kin}$ over the values reported in Ref.~\cite{aliceppbpid} while holding the other two parameters fixed at their values for middle centrality $n = 4$. The result is relatively little variation of BW spectrum shapes.

Figure~\ref{logdetails} (b) shows the effect of varying mean radial speed $\bar \beta_t$ over its reported values while holding the other two parameters fixed. The result is large variation of curvatures, spectra being shifted strongly to {\em higher} \yt\ with increasing centrality across the entire \yt\ interval.

Figure~\ref{logdetails} (c) shows the effect of varying flow-profile parameter $n$ over its reported values (per Table~\ref{flowparams}) while holding the other two parameters fixed. The  result is moderate shift of spectra to {\em lower} \yt\ with increasing centrality, but only above \yt\ = 2. Figure~\ref{logdetails} (d) shows the effect of varying parameter $n$ on the ``flow profile'' or variation of radial speed $\beta_t(x)$ with fractional radius $x = r/R$. 

Based on Fig.~\ref{logdetails} (a) and Z-scores  in the next section the BW model as applied to neutral kaons is not relevant above $y_t \approx 2.8$ ($p_t \approx 1.15$ GeV/c). Below that point $T_{kin}$ does not significantly affect the model compared to the other two BW parameters (acts mainly as an overall scale factor that is seen as not relevant). $\bar \beta_t$ affects the model over a broad interval but $n$ is controlling only above \yt\ = 1.8 ($p_t \approx 0.4$ GeV/c), and the two parameters produce strongly opposing variations in spectrum shape. In effect, the BW model as applied to neutral kaons is a two-parameter model function that attempts to describe data over a  limited \pt\ interval. Based on the above results physical interpretation of model parameters {\em directly} via comparisons with data trends seems problematic.

\begin{figure}[h]
	\includegraphics[width=1.65in]{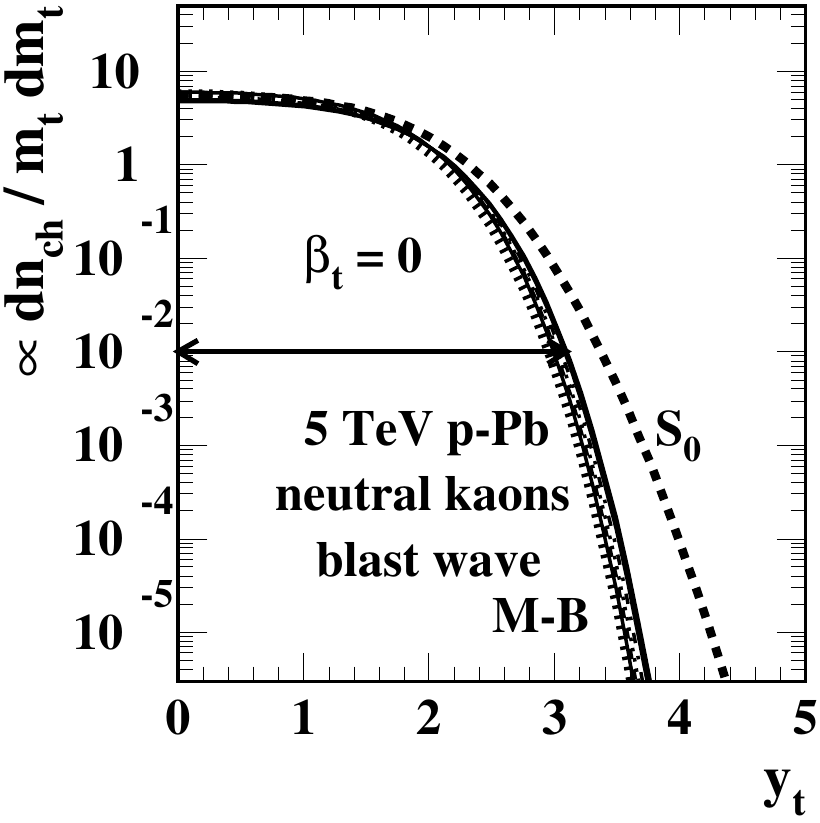}
	\includegraphics[width=1.65in]{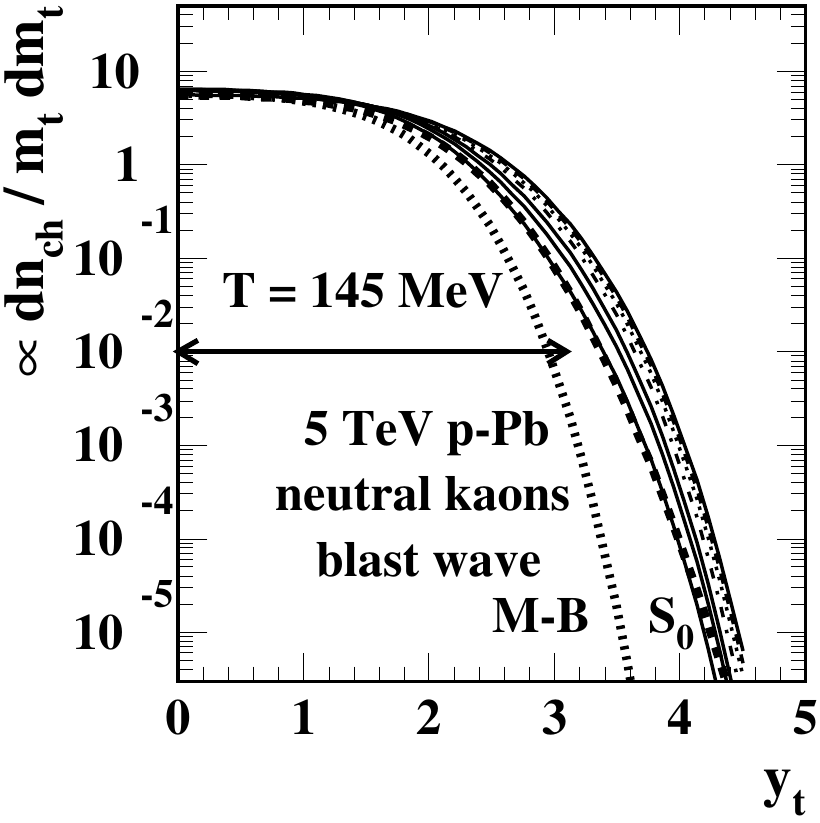}
	\caption{\label{betatkin}
	Left: BW model spectra with $\bar \beta_t = 0$ (various line styles). $\hat S_0(y_t)$ (bold dashed) corresponds to $T_i = 200$ MeV for kaons. The reference Boltzmann distribution (bold dotted) corresponds to $T = 145$ MeV.
	Right: BW model spectra with fixed $T_{kin} = 145$ MeV (= TCM $T$ value for unidentified hadrons) with $\bar \beta_t$ and $n$ varying as in Table~\ref{flowparams}.
	}  
\end{figure}

Figure~\ref{betatkin} (left) shows the effect of turning off radial flow ($\bar \beta_t = 0$). The shape change corresponds to faster fall-off with decreasing $T_{kin}$ value (and {\em increasing} \ppb\ centrality). The bold dotted curve is the Boltzmann (exponential on $m_t$) limit of $\hat S_0(y_t)$ (i.e.\ if L\'evy exponent $n \rightarrow \infty$) with $T = 145$ MeV. BW model spectra approach that limit for central \ppb\ collisions and $T_{kin} \approx 145$ MeV.

Figure~\ref{betatkin} (right) shows the effect of holding temperature $T_{kin}$ fixed at 145 MeV while the other two parameters remain consistent with Table~\ref{flowparams}. The BW model fitted to the most-peripheral \ppb\ data (with $\bar \beta_t \approx 0.25$) corresponds closely to TCM model $\hat S_0(y_t)$ (bold dashed).

Note that $\hat S_0(y_t)$ for the TCM (bold dashed) in the left panel is {\em defined} to match data spectra as $n_{ch} \rightarrow 0$. As such it has no jet contribution and represents {\em zero particle density}. The BW model with no jet component and zero flow contribution should have the best opportunity to describe spectrum data there {\em given BW model assumptions}. But the BW model strongly deviates from $\hat S_0(y_t)$ above \yt\ = 2 ($p_t \approx 0.5$ GeV/c). BW curves with $\bar \beta_t = 0$ and $T_{kin} \approx 145$ MeV (central \ppb\ in Table~\ref{flowparams}) instead closely approximate a Boltzmann exponential (bold dotted) on transverse mass \mt\ with slope parameter $T = 145$ MeV scaled to coincide with $\hat S_0(y_t)$ at low \yt.

The power-law tail on $\hat S_0(y_t)$ (as a L\'evy distribution~\cite{levywilk}) can be interpreted to represent inhomogeneity of the emitting system, i.e.\ the L\'evy model parameter $n$ can be expressed in the form  $1/n \sim \sigma^2_T / \bar T^2$~\cite{wilklevy}. The inferred width $\sigma_T$ may be interpreted in terms of incomplete equilibration or in terms of $k_t$ broadening within a longitudinal splitting cascade. The result in the right panel suggests that a substantial part of the $\bar \beta_t$ contribution simply accommodates the L\'evy shape of soft component $\hat S_0(y_t)$ common to all A-B collision systems and has nothing to do with radial flow {\em per se}. That explains why low-\nch\ \pp\ and peripheral A-B collision data require $\bar \beta_t \approx 0.25$~\cite{alicepppid} for Boltzmann-based BW models.

\section{BW $\bf vs$ TCM Spectrum fit quality} \label{quality}

This section quantitatively evaluates spectrum fit quality for the BW model compared to the TCM as applied to 5 TeV \ppb\ collisions. Three goodness-of-fit measures are compared: (a) Z-scores, (b) $\chi^2$ and (c) data/model spectrum ratios. Fit quality based on statistical errors is compared with that based on systematic errors, both as published with spectrum data in association with Ref.~\cite{aliceppbpid}.

\subsection{Goodness-of-fit measures}

A standard measure of data description or fit quality for a model is the Z-score (for an $i^{th}$ observation)~\cite{zscore}
\bea \label{zscore}
Z_i &=&  \frac{O_i - M_i}{\sigma_i},
\eea
where $O_i$ is an observation (datum), $M_i$ is a model prediction and $\sigma_i$ is the uncertainty (``error'') for the corresponding observation. In some presentations $M_i \rightarrow E_i$ (expectation) and $\sigma_i \rightarrow \sqrt{E_i}$ (assuming Poisson statistics for $O_i$). For an acceptable model one expects $Z_i$  to have an r.m.s.\ value near 1. The $\chi^2$ statistic is simply related:
\bea \label{chisq}
\chi^2 &=&\sum_{i=1}^N Z_i^2
\eea
for $N$ data points. An acceptable model should yield $\chi^2 \sim \nu = N - D$ where $D$ is the number of degrees of freedom of the model (e.g.\ number of model parameters). 

There is a tendency to evaluate model quality in terms of data/model {\em ratios} -- e.g.\ Fig.~7 of Ref.~\cite{aliceppbpid}. The relation between data/model ratios and Z-scores is
\bea \label{suppress}
\frac{\text{data}}{\text{model}} - 1 &\approx& \text{Z-score} \times \frac{\text{error}}{\text{data}},
\eea
with error/model (exact) $\rightarrow $ error/data (approximate). The error/data ratio may vary by orders of magnitude between different particle types and collision systems, and even across \yt\ intervals (see Figs.~\ref{bwsyser1} and \ref{bwsyser2}). Whereas the correct test of model validity is the relation of Z-scores to 1 (or $\chi^2$ to $\nu$) the interpretation of data/model ratios relative to 1 is problematic and typically quite misleading. The error/data ratios on the right are typically $\ll 1$ thereby suppressing, in data/model ratios on the left, what may be very significant data-model deviations.

\subsection{Fit quality based on statistical uncertainties} \label{qualstats}

\begin{figure}[h]
	\includegraphics[width=3.3in]{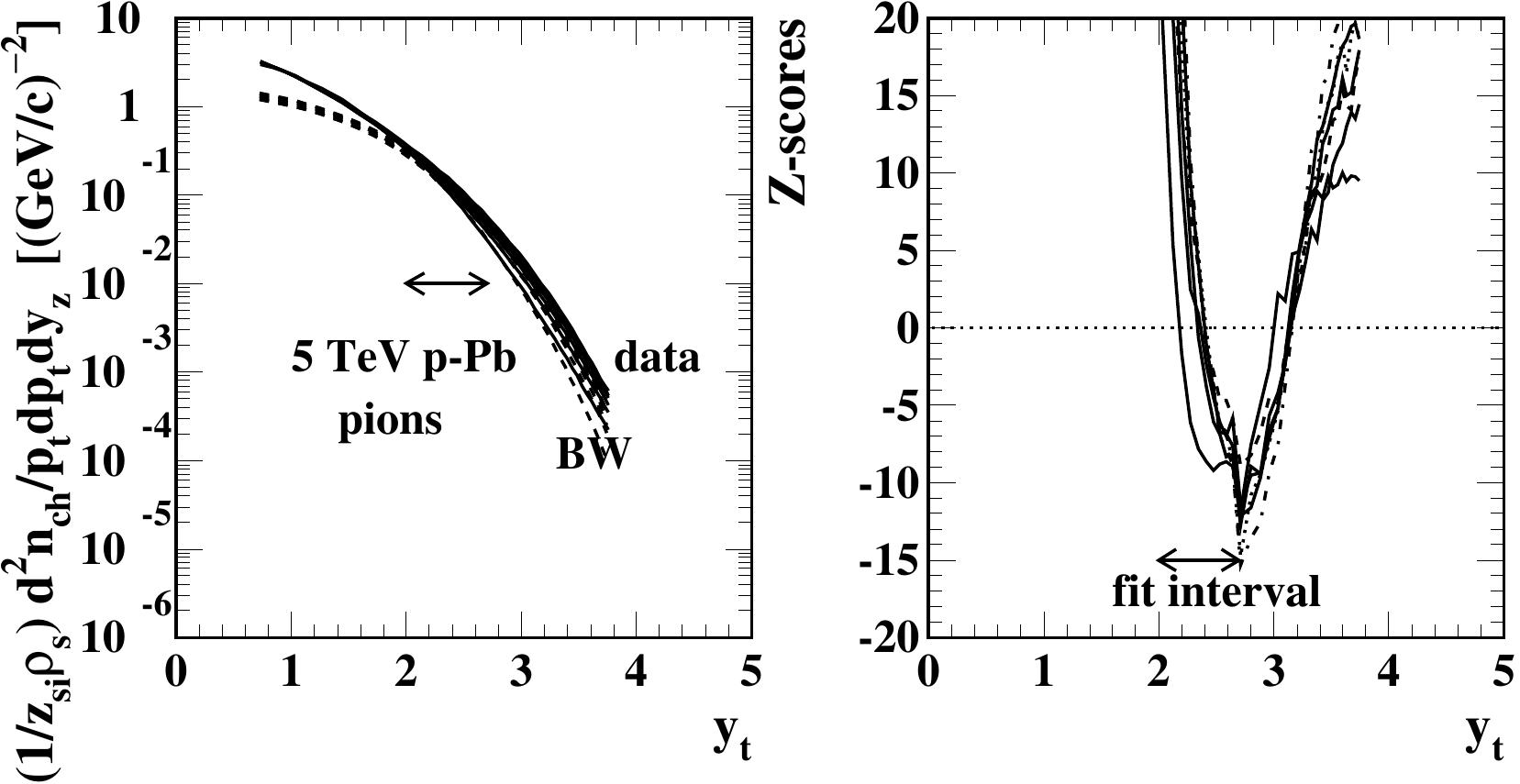}
	\put(-142,105) {\bf (a)}
	\put(-23,105) {\bf (b)}\\
	\includegraphics[width=3.3in]{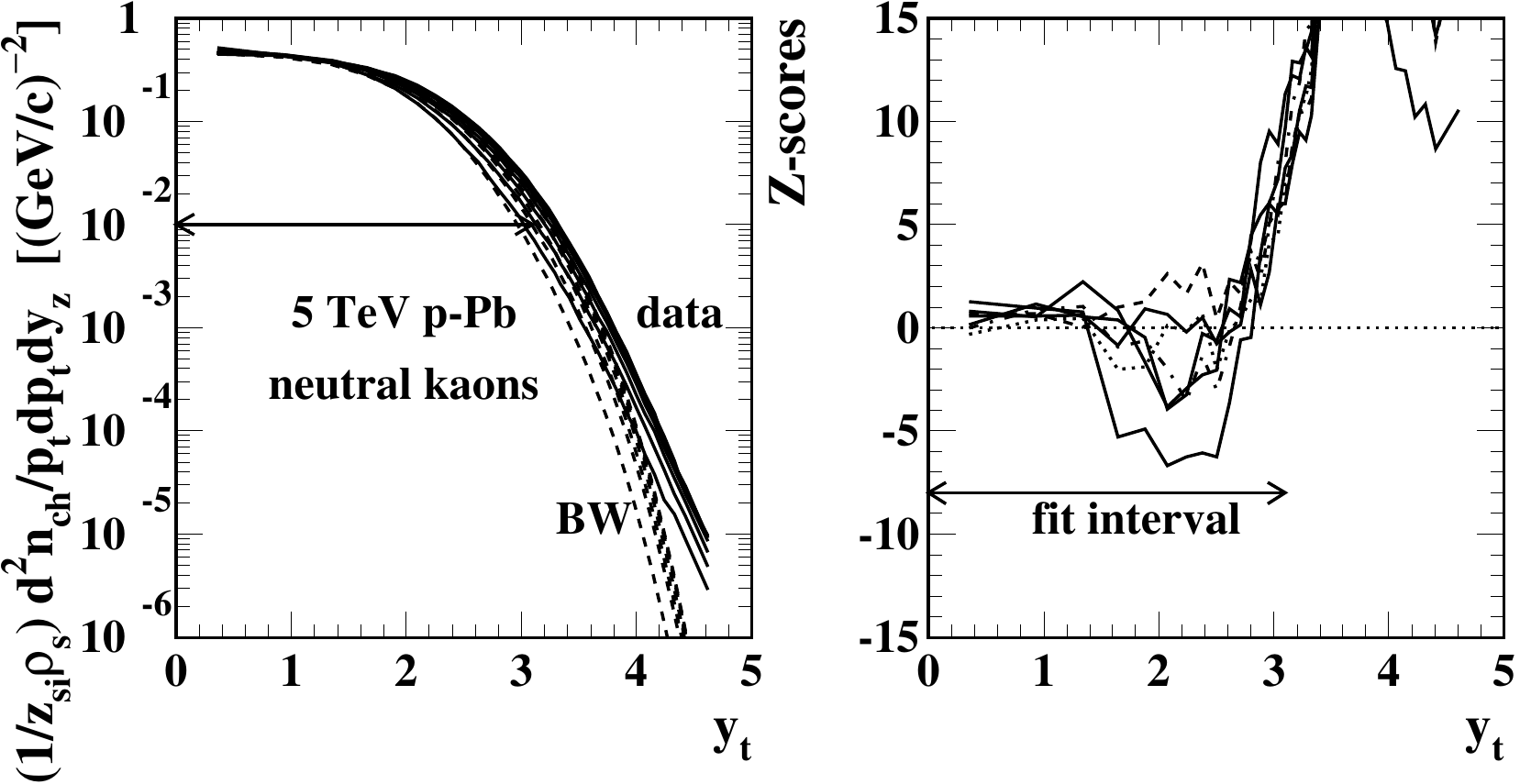}
	\put(-142,105) {\bf (c)}
	\put(-23,105) {\bf (d)}\\
	\caption{\label{bwzscore1}
	Data \pt\ spectra (solid) for pions (a) and neutral kaons (c) and corresponding BW model curves (dashed). Data spectra are rescaled to quantity $X_i(y_t)$ as defined in Eq.~(\ref{eq4}). BW curves are scaled to match the data within fit intervals (arrows). Panels (b,d) show corresponding Z-scores based on {\em statistical} errors published with the spectrum data.
	}  
\end{figure}

\begin{figure}[h]
	\includegraphics[width=3.3in]{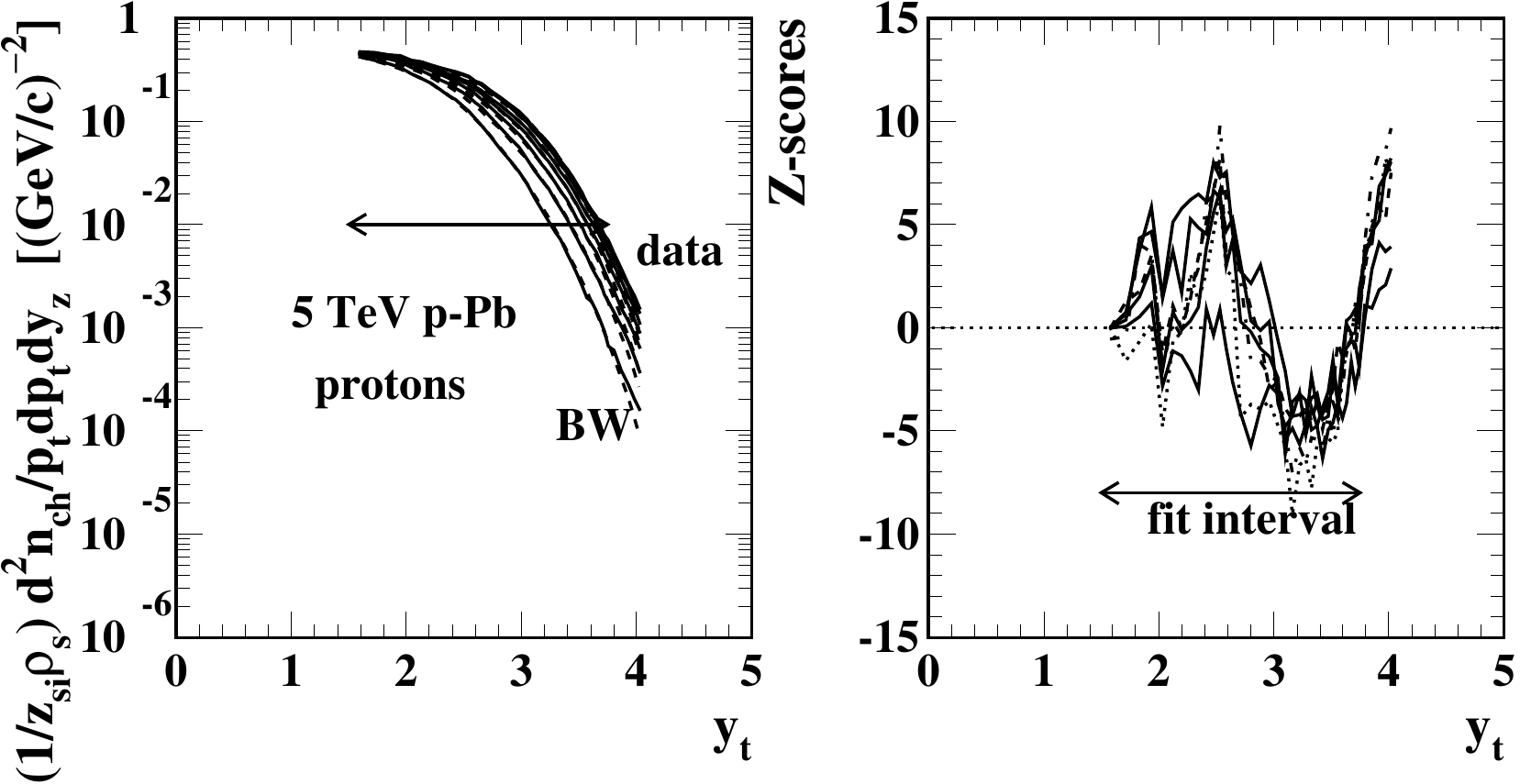}
	\put(-142,105) {\bf (a)}
	\put(-23,105) {\bf (b)}\\
	\includegraphics[width=3.3in]{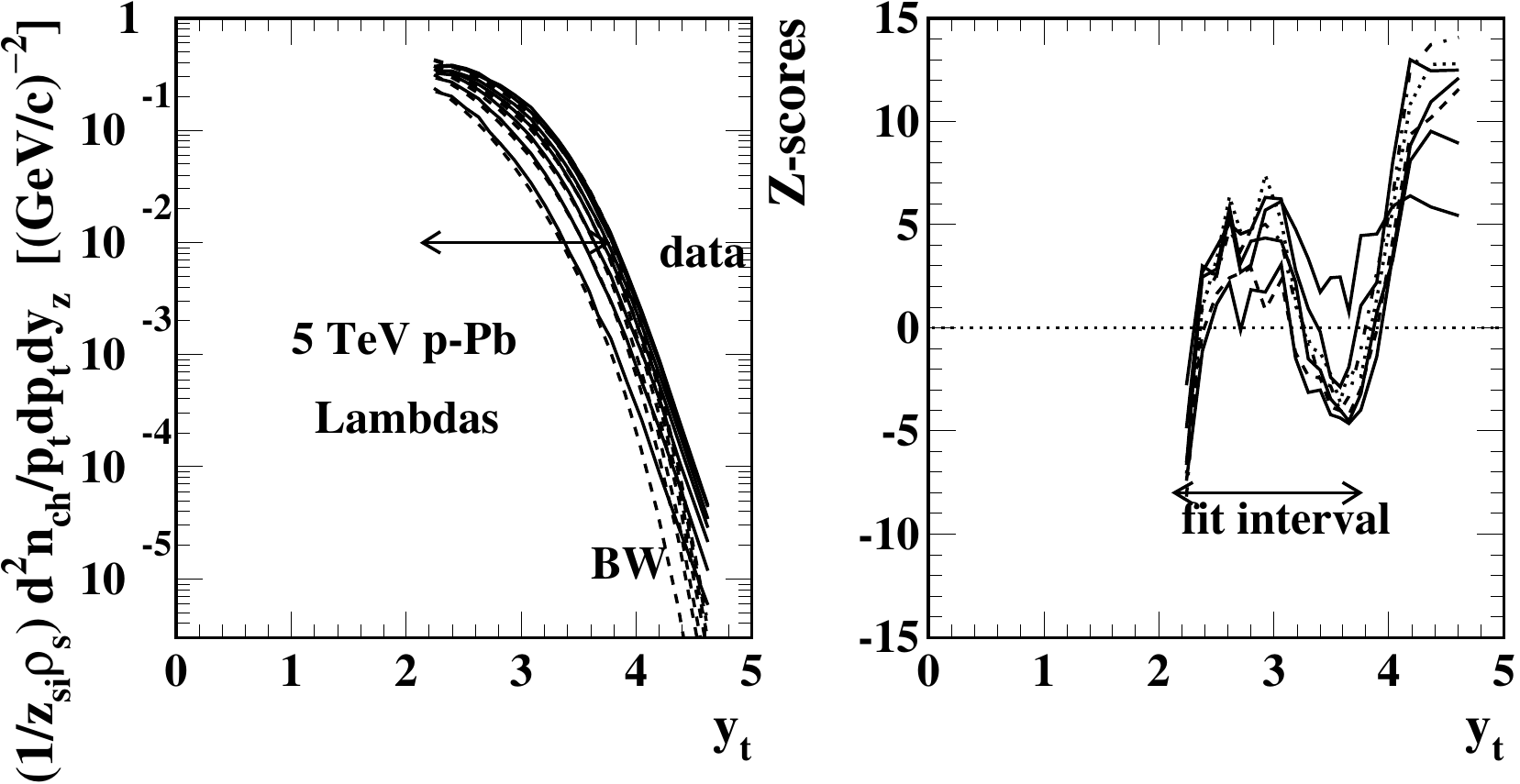}
	\put(-142,105) {\bf (c)}
	\put(-35,105) {\bf (d)}\\
	\caption{\label{bwzscore2}
	Data \pt\ spectra for protons (a) and Lambdas (c)  (solid) and corresponding BW model curves (dashed). Data spectra are rescaled to quantity $X_i(y_t)$ as defined in Eq.~(\ref{eq4}). BW curves are scaled to match the data within fit intervals (arrows). Panels (b,d) show corresponding Z-scores based on statistical errors published with the spectrum data.
	}  
\end{figure}

Figures~\ref{bwzscore1} and \ref{bwzscore2} (a,c) compare BW model fits (dashed) to \ppb\ spectrum data (solid). The spectrum data are scaled as shown in the $y$-axis labels and as reported in Refs.~\cite{pidpart1,pidpart2}. The BW curves correspond to the fitted parameters in Table~\ref{flowparams} and are scaled to best match the data, consistent with the fitting procedure.

Figures~\ref{bwzscore1} and \ref{bwzscore2} (b,d) show Z-scores for BW model vs spectrum data using statistical uncertainties that accompany the published data. It is evident that even within \yt\ intervals chosen to produce an acceptable fit~\cite{aliceppbpid} the Z-scores are consistently substantially greater than 1. Results for BW fits can be compared with results from a TCM analysis of the same data reported in Refs.~\cite{pidpart1,pidpart2}.

\begin{figure}[h]
	\includegraphics[width=3.3in]{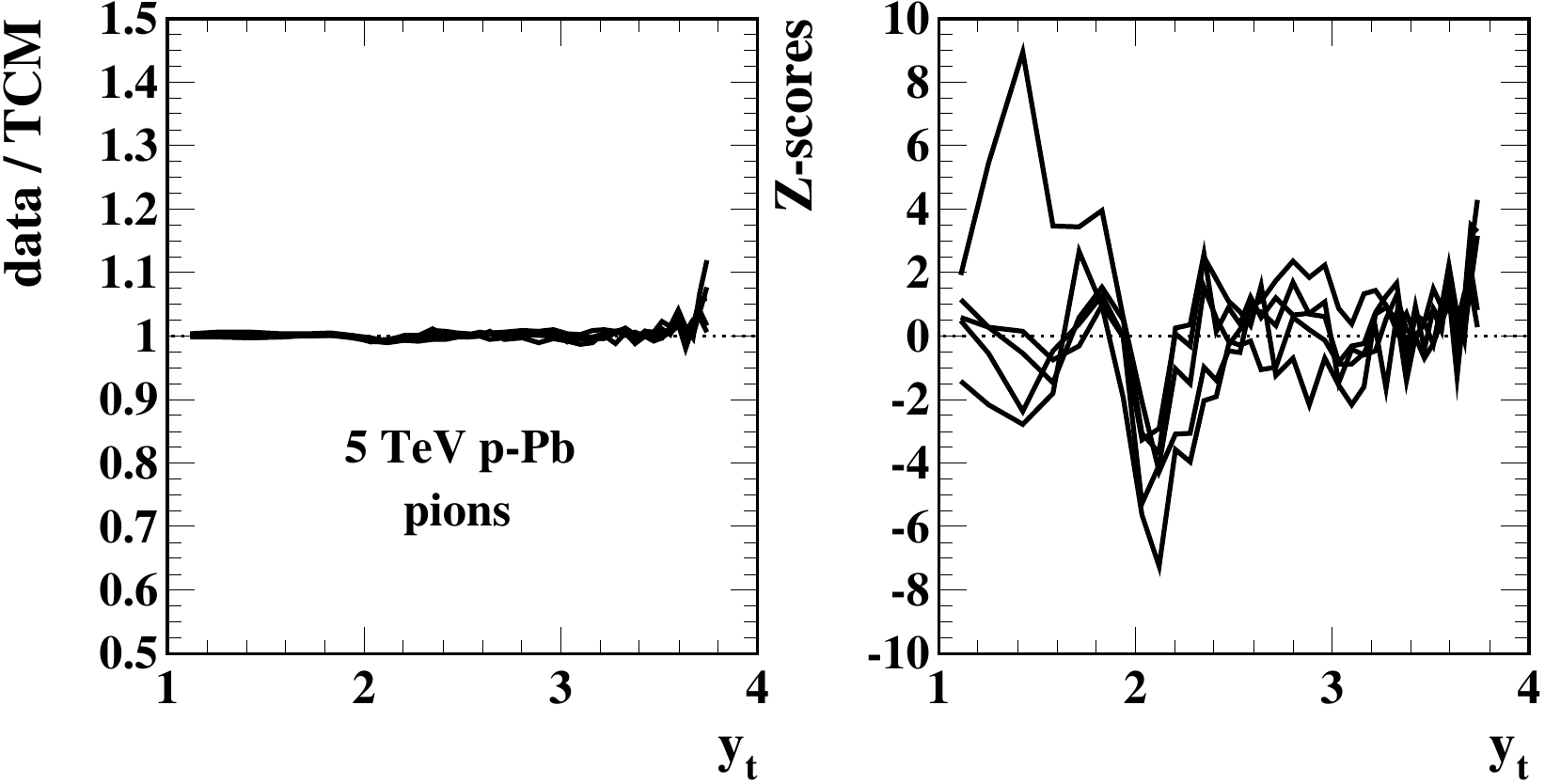}
	\put(-142,105) {\bf (a)}
	\put(-23,105) {\bf (b)}\\
	\includegraphics[width=3.3in]{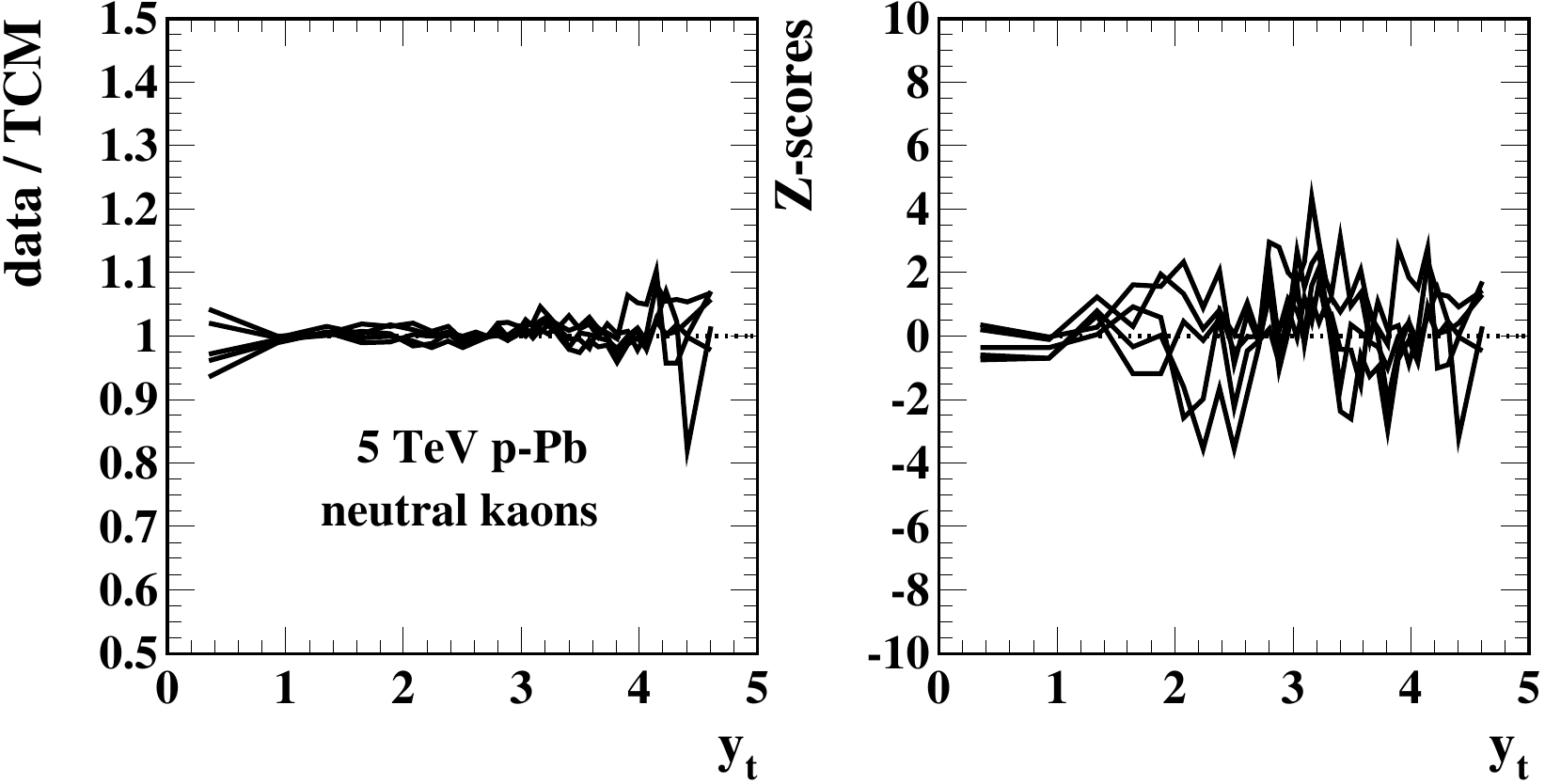}
	\put(-142,105) {\bf (c)}
	\put(-23,105) {\bf (d)}\\
	\caption{\label{zsmeson}
		Left: Data/TCM spectrum ratios for pions (a) and neutral kaons (c).
		Right: 	Z-scores for  pions (b) and neutral kaons (d). The model is the {\em variable} TCM defined in Ref.~\cite{pidpart2}.
	}   
\end{figure}

\begin{figure}[h]
	\includegraphics[width=3.3in]{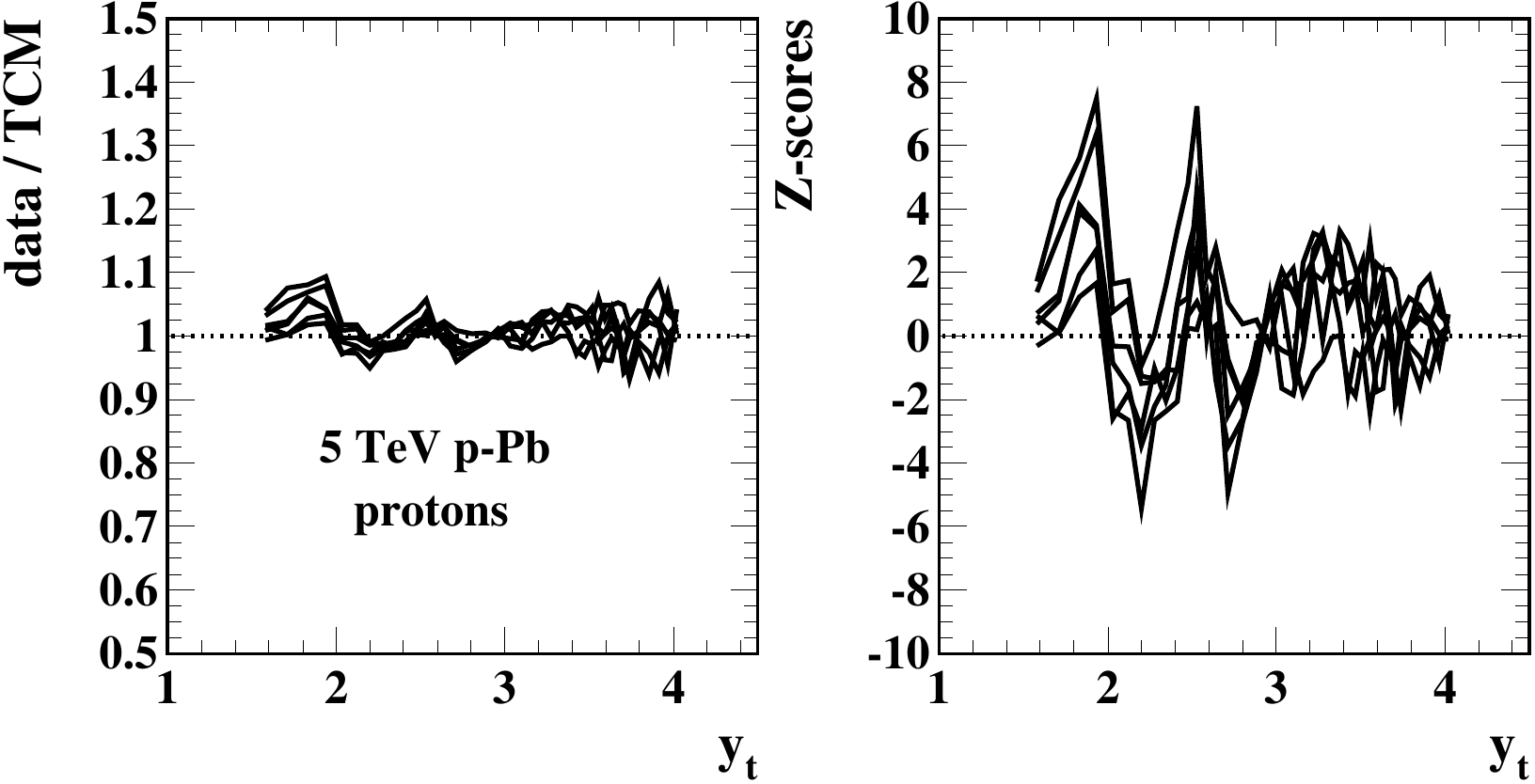}
	\put(-142,105) {\bf (a)}
	\put(-23,105) {\bf (b)}\\
	\includegraphics[width=3.3in]{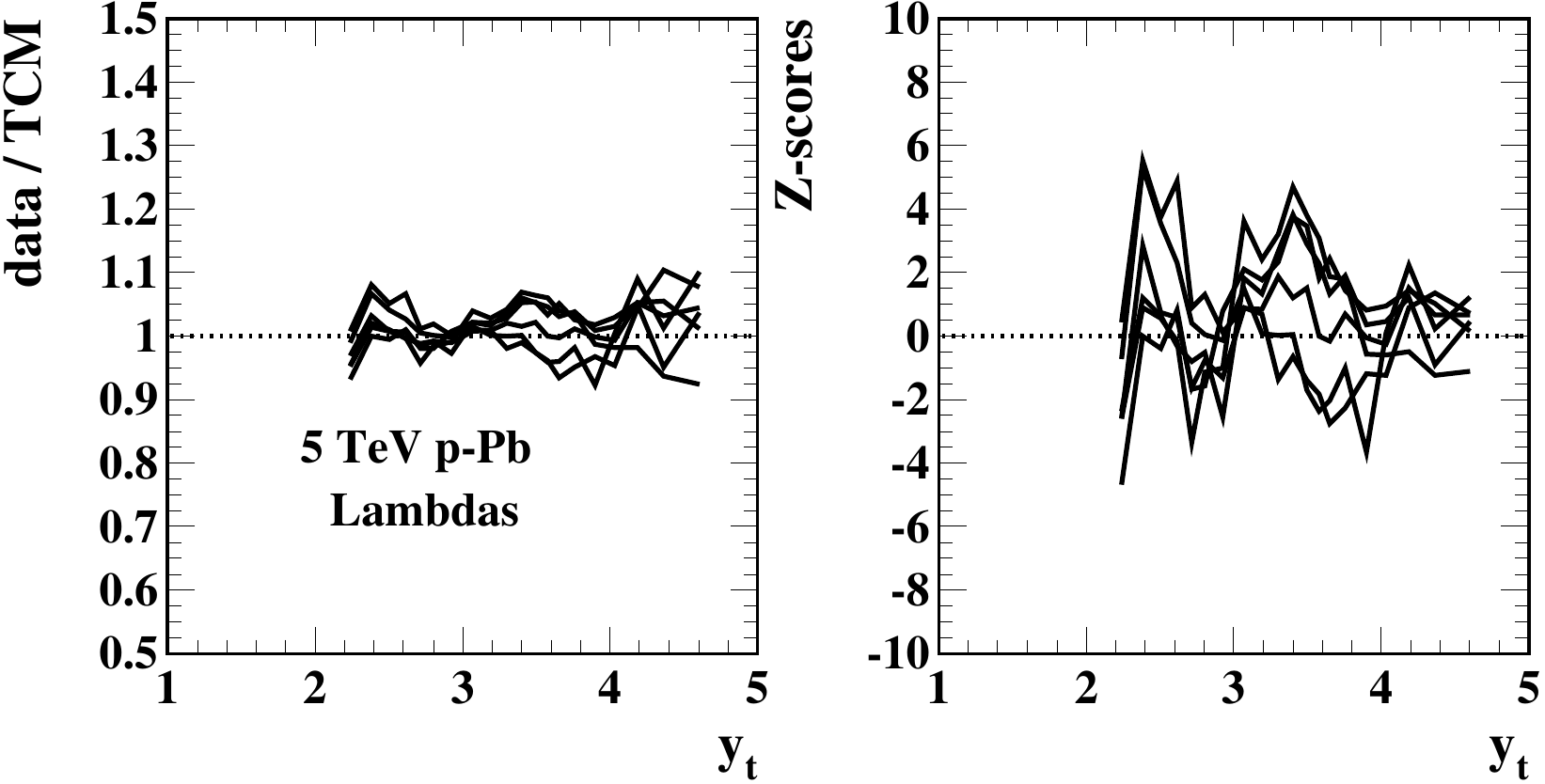}
	\put(-142,105) {\bf (c)}
	\put(-23,105) {\bf (d)}\\
	\caption{\label{zsbaryon}
		Left: Data/TCM spectrum ratios for protons (a) and Lambdas (c).
Right: 	Z-scores for  protons (b) and Lambdas (d).  The model is the {\em variable} TCM defined in Ref.~\cite{pidpart2}.
	} 
\end{figure}

Figures~\ref{zsmeson} and \ref{zsbaryon} compare data/TCM ratios (left panels) to Z-scores (right panels), for mesons (pions and kaons) in Fig.~\ref{zsmeson} and for baryons (protons and Lambdas) in Fig.~\ref{zsbaryon}. Based solely on data/model ratios it would seem that the TCM description for pion data is much better than that for Lambda data. Yet a comparison of corresponding Z-scores shows that the data descriptions are actually of comparable quality. While Z-scores within some \yt\ intervals are consistent with statistical deviations (i.e.\ r.m.s.\  consistent with 1) prominent localized excursions are apparent -- for instance \yt\ near 2 for pions and \yt\ within 2-3 for protons. Given that the significant data-TCM deviations are consistent across \ppb\ \nch\ classes and highly localized on \yt\ it is unlikely that the deviations result from model issues. It is reasonable to conjecture that such structures are inevitable within the complex procedure of processing particle data. These results suggest that the TCM describes \ppb\ PID spectrum data within statistical uncertainties except for local anomalies that may be due to detector effects. 

\subsection{Fit quality based on systematic uncertainties} \label{sysuncertain}

The BW-fit $\chi^2/\text{ndf}$ values appearing in Table~\ref{flowparams}, as reported in Ref.~\cite{aliceppbpid}, are notable for two reasons: they are (a) consistently {\em low} compared to the expected r.m.s.\ value $\approx 1$ for acceptable fits and (b) apparently at odds with the Z-scores shown in Figs.~\ref{bwzscore1} and \ref{bwzscore2} that are consistent with $\chi^2/\text{ndf}$ values values substantially greater than 10. Those results suggest that $\chi^2/\text{ndf}$ values reported in Ref.~\cite{aliceppbpid} were obtained assuming {\em systematic} rather than statistical uncertainties $\sigma_i$ in Eq.~(\ref{zscore}).

\begin{figure}[h]
	\includegraphics[width=3.3in]{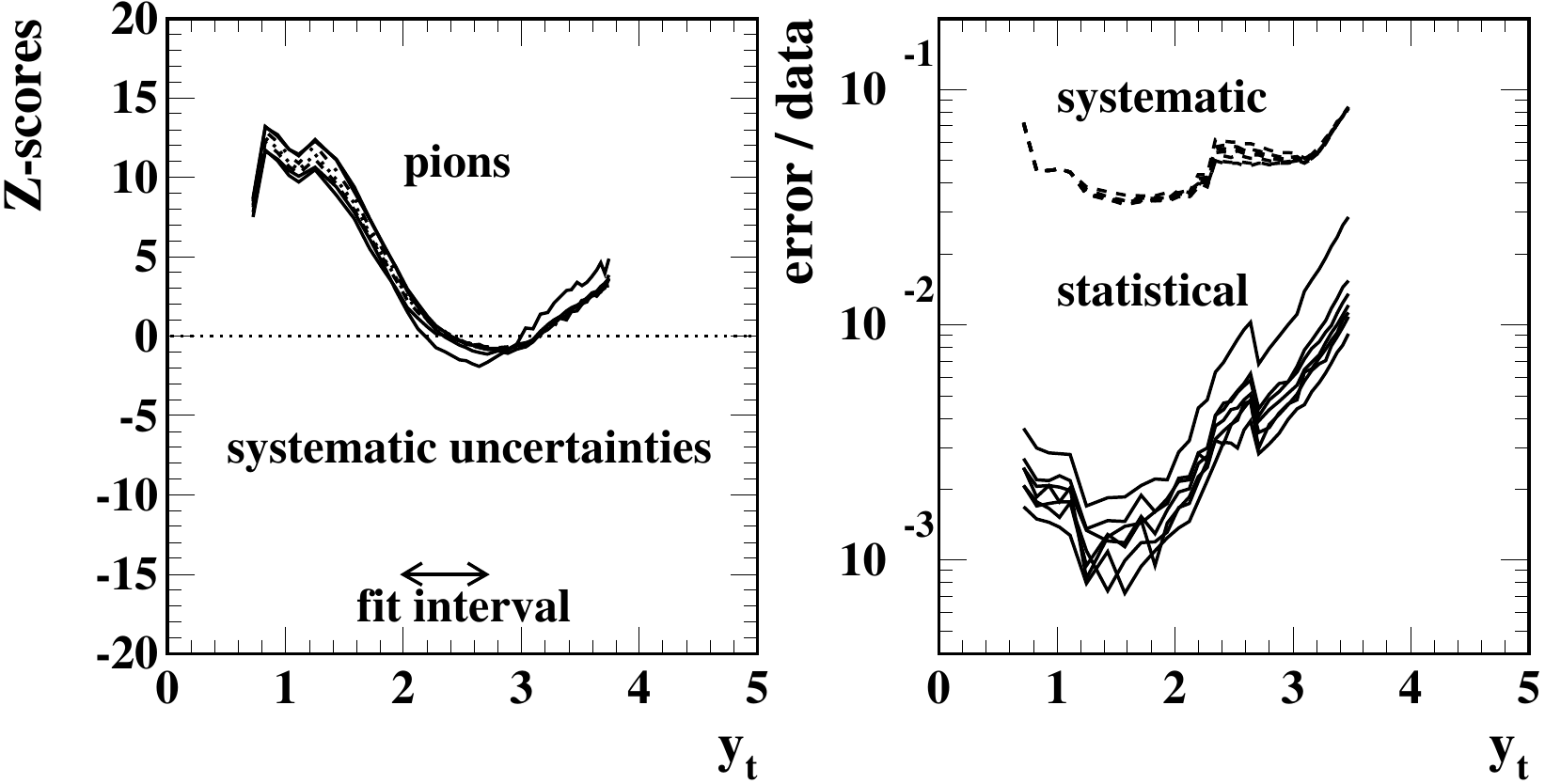}
	\put(-142,105) {\bf (a)}
	\put(-23,105) {\bf (b)}\\
	\includegraphics[width=3.3in]{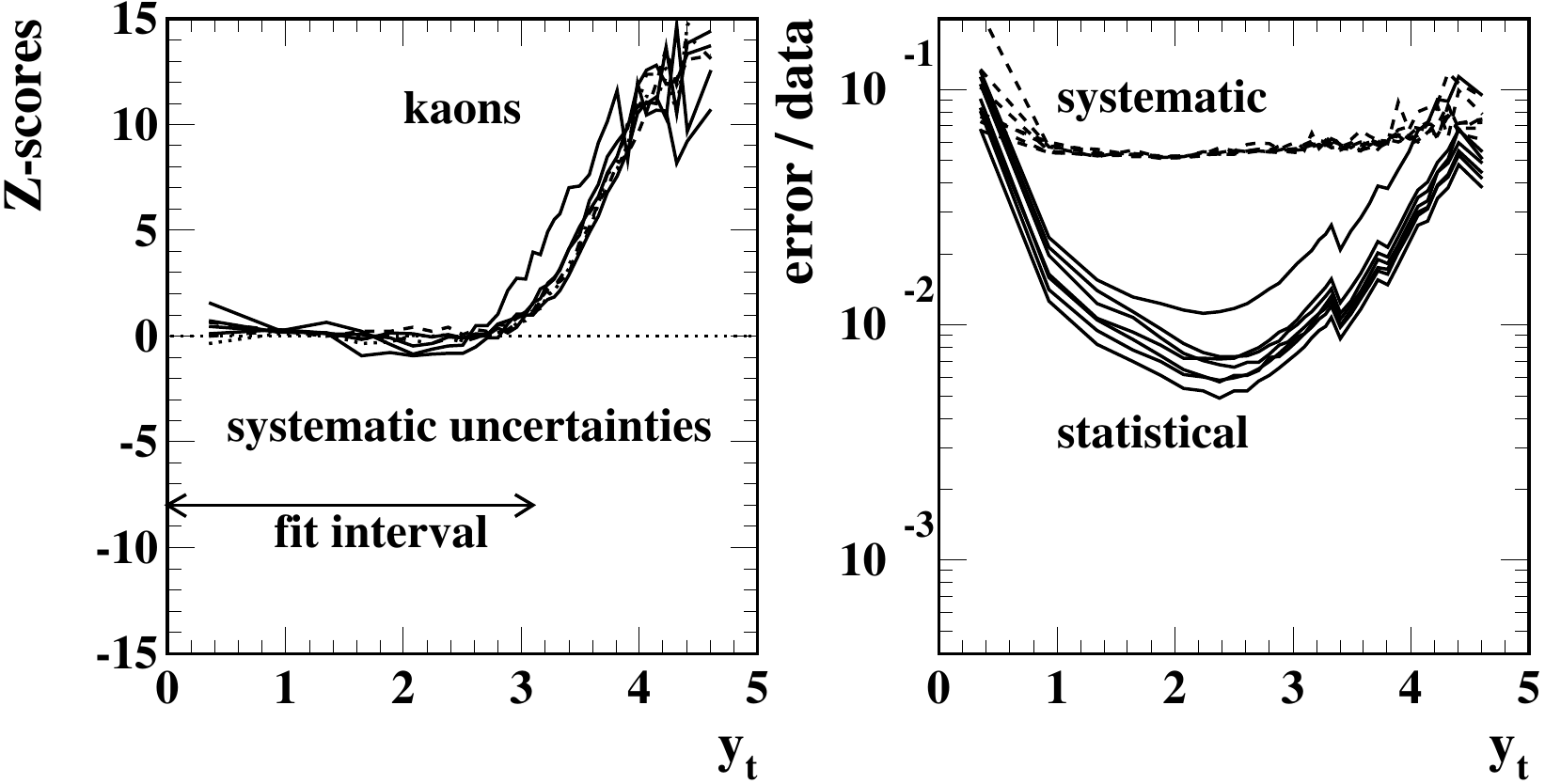}
	\put(-142,82) {\bf (c)}
	\put(-23,72) {\bf (d)}
	\caption{\label{bwsyser1}
	BW model Z-scores for pions (a) and neutral kaons (c) based on {\em systematic} uncertainties or errors published with spectrum data. Panels (b,d) show error/data ratios for statistical errors (solid) and systematic errors (dashed).
	}  
\end{figure}

\begin{figure}[h]
	\includegraphics[width=3.3in]{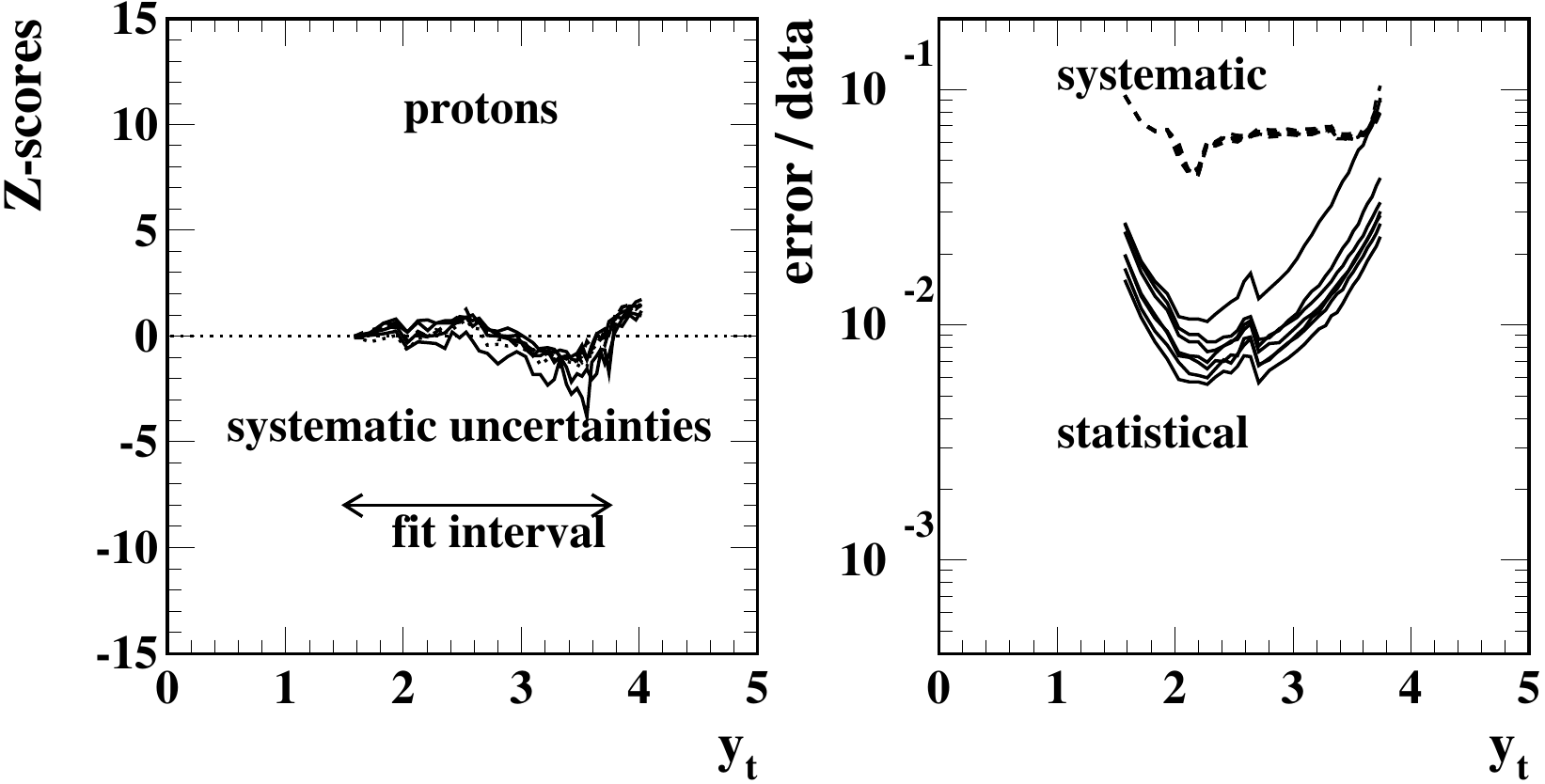}
	\put(-142,105) {\bf (a)}
	\put(-23,105) {\bf (b)}\\
	\includegraphics[width=3.3in]{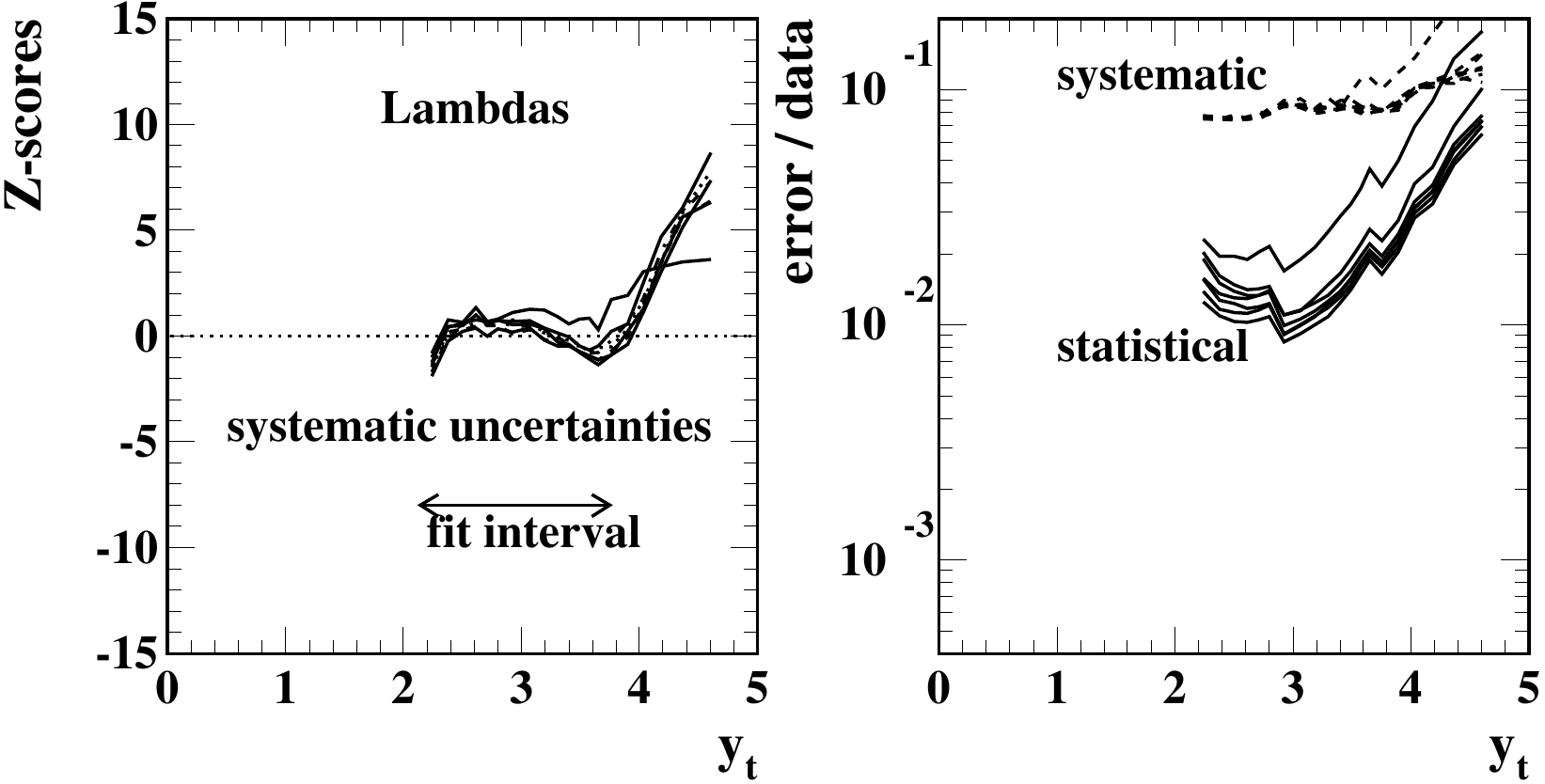}
	\put(-142,105) {\bf (c)}
	\put(-23,72) {\bf (d)}
	\caption{\label{bwsyser2}
	BW model Z-scores for protons (a) and Lambdas (c) based on {\em systematic} uncertainties or errors published with spectrum data. Panels (b,d) show error/data ratios for statistical errors (solid) and systematic errors (dashed).
	} 
\end{figure}

Figures~\ref{bwsyser1} and \ref{bwsyser2} (a,c) show BW-fit Z-scores recalculated using systematic errors included with published \ppb\ spectra from Ref.~\cite{aliceppbpid}. Vertical scales are the same as those in Figs.~\ref{bwzscore1} and \ref{bwzscore2} (a,c) for comparison. Those results suggest that systematic uncertainties were in fact used to calculate  $\chi^2/\text{ndf}$ values appearing in Table~\ref{flowparams}. 

Figures~\ref{bwsyser1} and \ref{bwsyser2} (b,d) compare published \ppb\ systematic uncertainties (dashed) with statistical uncertainties (solid) in ratio to corresponding spectrum data values. Systematic uncertainties are approximately constant on \yt\ with magnitudes 5-10\% of data values. The numbers are consistent with {\em total} systematic uncertainties presented in Tables~3 and 4 of Ref.~\cite{aliceppbpid}. The systematic uncertainties tend to exceed statistical uncertainties by a factor five or more, especially within fit intervals employed for BW fits, implying that reported $\chi^2/\text{ndf}$ values are at least a factor 25 less than what would be obtained using statistical uncertainties. That description responds to items (a) and (b) at the beginning of this subsection.

Appropriate uncertainty estimates are of central importance for realistic assessment of model quality. The most basic estimate (where data result from counting discrete elements -- e.g.\ particles) is the statistical uncertainty corresponding to Poisson statistics. Corresponding Z-scores then show all {\em statistically significant} data-model deviations no matter what their origin. Various sources of systematic uncertainty, whether correlated with collision event class (\nch) or particle properties (\yt, hadron species) or not, may be relevant to correct assessment of model quality but should be considered only relative to Z-scores based on statistical uncertainties.

More generally the $\chi^2$ statistic is not very  effective for model testing. $\chi^2$ is intended to measure a significant difference between model and data but contains much less information than the differential Z-scores it represents as in Eq.~(\ref{chisq}). Z-scores based on statistical errors clearly show not only what data-model differences are statistically significant but also how such deviations are correlated, e.g.\ across \yt\ bins and \nch\ classes. In comparing two models the one with the greater $\chi^2$ may even be preferred because of  the detailed structure of  Z-scores.

\section{Blast wave model evolution} \label{bwmodelx}

In order to interpret BW model fit results it is useful to review early theoretical developments that led to the current general form applied to ultrarelativistic A-B collisions and some significant variations in the structure of the model over several decades.

\subsection{Early BW model development} \label{earlybw}

The blast-wave model arguably has its origins in the collision theories of Landau (1950s, e.g.\ Ref.~\cite{landau}) and Hagedorn (1960s, e.g.\ Ref.~\cite{haged}). Hadron production is based in the first case on hydrodynamic evolution of a hadronic fluid and in the second case on emission from rapidly-moving ``fireballs.''  In either case ``collectivity'' -- collective motion of a continuous hadronic fluid or discrete collection of objects (fireballs) in configuration space -- is a central feature. Detected hadrons that emerge are thereby {\em correlated} in momentum space, such correlations interpreted to indicate a form of collectivity.  

The correct relation between the measured particle momentum distribution in the center-of-momentum (CM) or laboratory (lab, for symmetric colliders) frame and the hadron emission distribution in the local, comoving or boost frame was determined by Cooper and Frye~\cite{cooper} as
\bea \label{cfeq}
E\frac{d^3N}{dp^3} &=& \frac{d^3N}{m_t dm_t dy_z d\phi_p} = \int_\sigma f(x,p,u) p^\mu d^3\sigma_\mu,
\eea
where $u(x)$ is the velocity field of an emitting fluid in the lab frame, $p^\mu$ is the particle four-momentum measured in the lab frame, $\sigma$ is the ``freezeout'' hypersurface ($\sim$ transition from small to large mean free path), and $d^3\sigma_\mu$ is a differential volume four-vector. A limiting case for $f(x,p,u)$ in the local or boost frame is isotropic emission, according to Bose or Fermi statistics or (as an approximation) the Boltzmann distribution. That formalism was employed to describe spherical expansion within nucleus-nucleus collisions in Ref.~\cite{siemens} wherein the term ``blast wave'' was introduced in connection with A-B collisions. More complex expansion geometries (i.e.\ {\em nucleon} flows) were encountered within the Bevalac  program~\cite{bevalac}.

\subsection{BW model for ultrarelativistic A-B collisions}

For the Bevalac and alternating gradient synchrotron (AGS) nucleus-nucleus programs, where the collision energy per nucleon is of order the nucleon mass-energy, fractional deviations from spherical expansion (e.g.\ squeeze-out, side-splash) were measured via the sphericity tensor or related measures~\cite{sphericity,daniel}. For A-B collisions at the super proton synchrotron (SPS), RHIC or LHC, collision energies are much greater than the nucleon mass-energy. As a consequence the velocity field represented by $u(x)$ above is highly asymmetric -- strongly elongated along the collision axis. Reference~\cite{ssflow} reported a version of Eq.~(\ref{cfeq}) adapted to that context and applied to SPS S-S collisions at $\sqrt{s_{NN}} \approx 19$ GeV. It is useful to examine details of the derivation and  accompanying arguments.

It is assumed that within a local (co-moving or boost) frame hadron emission is locally thermal and isotropic (i.e.\ within a {\em transparent} environment): ``...the invariant distribution function [in the CM or lab frame]...we assume to be an isotropic thermal distribution boosted by the local fluid velocity $u^\mu$, and we approximate the respective Bose and Fermi distributions by the Boltzmann distribution.'' Equation~(\ref{cfeq}) is modified by
\bea
f(x,p,u) &\rightarrow & [g /(2\pi)^3] \exp[- (u^\nu p_\nu - \mu) / T],
\eea
and potential $\mu$ is omitted below (see comment below on ``normalization'').
Configuration space is represented by cylindrical coordinates with a hypersurface defined by
\bea \label{sigmaeq}
\sigma_\mu &=& [t(\eta_s),r \hat e(\phi_x),z(\eta_s)]
\\ \nonumber
&=& [\tau \cosh(\eta_s),r \hat e(\phi_x),\tau \sinh(\eta_s)]
\eea
with $\zeta \rightarrow \eta_s$ and
\bea \label{d3sigma}
d^3\sigma_\mu &=& \left[\frac{\partial z}{\partial \eta_s},0,0,\frac{\partial t}{\partial \eta_s}\right]d\eta_s r dr d\phi_x
\\ \nonumber
&\rightarrow&  \tau [\cosh(\eta_s),0,0,\sinh(\eta_s)] d\eta_s r dr d\phi_x.
\eea
Space-time rapidity is $\eta_s = (1/2)\ln[(t + z)/(t-z)] = \ln[(t + z)/\tau]$.
Quantities $u_\mu$ and $p_\mu$ are as defined in Eqs.~(\ref{umu}) and (\ref{pmu}). Whereas Ref.~\cite{ssflow} invokes ``flow  angles'' $\rho$ and $\eta$, equivalent quantities are presented here by $\eta_t$ and $\eta_z$ respectively.  The conventional assumption $\eta_z \rightarrow \eta_s$ is consistent with Bjorken expansion.

Given those definitions Eq.~(\ref{cfeq}) then becomes
\bea \label{ssfloweq}
\frac{d^3N}{m_t dm_t dy_zd\phi_p} &=& \frac{g}{(2\pi)^3} m_t 
\\ \nonumber
&& \hspace{-.7in} \times\int_{-H_s}^{H_s} d\eta_s \left[ \cosh(y_z)\frac{\partial z}{\partial \eta_s} -  \sinh(y_z) \frac{\partial t}{\partial \eta_s}  \right]
\\ \nonumber
&& \hspace{-.7in} \times \int_0^R rdr \exp\left\{ - m_t \cosh[\eta_t(r)] \cosh(y_z - \eta_z)/T \right\}
\\ \nonumber
&& \hspace{-.7in} \times  \int_0^{2\pi} d\phi_x \exp\{p_t \sinh[\eta_t(r)] \cos(\phi_x - \phi_p)/T\},
\eea
where $\eta_s \in [-H_s,H_s]$ delimits the particle source.
The quantity in square brackets then simplifies to
\bea \label{brackets}
\cosh(y_z)\frac{\partial z}{\partial \eta_s} -  \sinh(y_z) \frac{\partial t}{\partial \eta_s} &\rightarrow& \tau \cosh(y_z - \eta_s).
\eea
The integral over $\phi_x$ reduces to $2 \pi I_0[p_t \sinh(\eta_t)/T]$. Using Eq.~(\ref{brackets}) and integrating both sides over $y_z$ leads to integrand $K_1[m_t \cosh(\eta_t)/T]$ in the third line. ``We obtain the transverse mass spectrum $dn/m_Tdm_T$ by  integrating over rapidity [$y_z$] using the modified Bessel function $K_1$....'' The integral over $\eta_s$ then simply leads to factor $2 Z_t \rightarrow 2H_s$ consistent with the interval over space-time rapidity $\zeta \rightarrow \eta_s$ assumed in Eq.~(14) of Ref.~\cite{ssflow}. The result is Eq.~(\ref{bweq}) above except $\rho \rightarrow \eta_t = \tanh^{-1}(\beta_t)$, with $\beta_t(r) = \beta_s (r/R)^n$. That expression is used by Ref.~\cite{aliceppbpid} to obtain the BW model fits summarized in Sec.~\ref{bwmodel}.

As noted, the absolute scale of the BW model in relation to spectrum data is not seen as significant (in contrast to the TCM). Reference~\cite{ssflow} concludes: ``...we have shown that a thermal model is perfectly possible for S+S collisions despite (or because of?)  the similarity of S+S and pp spectra.  The data force us to include resonance decays and longitudinal flow while they make no decisive statement about the existence of transverse flow.''




\subsection{Alternative BW model applications}

It is informative to examine some alternative routes to BW model functions and their applications to a variety of collision data. The examples below relate to the experimental programs at the SPS, RHIC and LHC. Further details on BW model derivations and applications are provided in App.~\ref{derive}.

\subsubsection{Tom\'asik, Wiedemann and Heinz~\cite{tomasik}}

The study in Ref.~\cite{tomasik} combines BW spectrum analysis of single-particle spectra and Bose-Einstein correlation (BEC) analysis of two-particle correlations to infer space-time properties of the particle source (emission region, freezeout state). The models (BW + BEC) were applied to NA49 data from 158A GeV \pbpb\ collisions. The particle source is modeled by an ``emission function'' $S(x,p)d^4x$ that includes the factor $p^\mu d^3 \sigma_\mu$ from Eq.~(\ref{cfeq}) plus densities on $r$, $\eta_s$ and $\tau$ that explicitly define the integrated space-time (source) volume.

Combining Eqs.~(3.1) and (3.10) from Ref.~\cite{tomasik}
\bea \label{tomasikeq}
E\frac{d^3N}{dp^3} &=& \int S(x,p)d^4x 
\\ \nonumber
&= & \frac{1}{(2\pi)^3} \int  d\eta_s r dr d\phi_x G(r) H(\eta_s)T(\tau)
\\ \nonumber
&& \times m_t\, \tau \cosh(y_z - \eta_s) \exp[- (u^\nu p_\nu - \mu) / T],
\eea
where $H(\eta_s)$ and $T(\tau)$ are Gaussians on $\eta_s$ and $\tau$ respectively, the latter normalized to unity. $p_\mu = [m_t \cosh(y_z),p_t,0,m_t \sinh(y_z)]$. $G(r)$ is modeled as a Gaussian or square (flat) distribution on $r$. In contrast, Ref.~\cite{ssflow} implicitly assumes flat distributions on $r$ and $\eta_s$ and a delta function on $\tau$.
Velocity field $u(x)$ is initially defined in terms of longitudinal $\eta_l \rightarrow \eta_z$ and transverse $\eta_t$ ``expansion rapidities'' (boosts) and  $u(x)$ is then as in Eq.~(\ref{umu}). $\eta_l$ is then identified with space-time rapidity $\eta \rightarrow \eta_s$ (i.e.\ Bjorken expansion). In Ref.~\cite{ssflow} transverse speed is parametrized by $\beta_t(r) = \beta_s(r/R)^n$ but  Ref.~\cite{tomasik} expresses transverse {\em boost} as $\eta_t(r) = \eta_f (r/r_{rms})$. 

\subsubsection{Florkowski and Broniowski~\cite{wojciechs}}

Reference~\cite{wojciechs} briefly summarizes the history of the BW model and then presents an alternative route to the model as it appears in Eq.~(\ref{bweq}). Symbols used in Ref.~\cite{wojciechs} are related to the standard set employed in this study.

Space-time rapidity is $\alpha_\parallel \rightarrow \eta_s$ (what correlates $t$ and $z$ given proper time $\tau$). $\zeta$ is a parameter that correlates proper time $\tau$ and radius $\rho \rightarrow r$. That system is assumed boost invariant over all $\eta_s$ in contrast to \cite{ssflow}. Thus, instead of $[t(\zeta),z(\zeta)]$ with $\zeta \rightarrow \eta_s$ space-time rapidity there is $[\tau(\zeta),r(\zeta)]$ with $\zeta \in [0,1]$ simply a correlation parameter.
$\sigma_\mu(x)$ is formally as in the second line of Eq.~(\ref{sigmaeq}), but the differential volume element $d^3\sigma_\mu$ (with $\rho \rightarrow r$) is
\bea \label{wodsig}
\left[ \frac{dr}{d\zeta} \cosh(\eta_s), \frac{d\tau}{d\zeta}\hat e(\phi_x),\frac{dr}{d\zeta} \sinh(\eta_s)   \right] d\zeta \tau r d\eta_s d\phi_x.
\eea
$u(x) = \cosh(\eta_t)[\cosh(\eta_z),\tanh(\eta_t)\hat e(\phi_x),\sinh(\eta_z)]$ is the velocity field (corrected from Ref.~\cite{wojciechs}) with $\alpha_\perp \rightarrow \eta_t$ and in this case $\alpha_\| \rightarrow \eta_z$. Longitudinal flow is represented by $v_z \rightarrow \beta_z = \tanh(\eta_z)$, and transverse flow by $\beta_t(\zeta) = \tanh[\eta_t(\zeta)]$. $\beta_t$ is correlated with $\tau$ and $r$ by parameter $\zeta$ as alternative to explicit $\beta_t(r) = \beta_s(r/R)^n$ in Ref.~\cite{ssflow}. $p_\mu$ and $u^\mu p_\mu$ are as in Eqs.~(\ref{pmu}) and (\ref{upmu}).

The transverse component $d\tau/d\zeta$ in Eq.~(\ref{wodsig}) would introduce additional complexity compared to Eq.~(\ref{ssfloweq}) and is assumed  zero in Ref.~\cite{wojciechs} ``to achieve the simplest possible form of the model.'' $d^3\sigma_\mu$ then has the form of Eq.~(\ref{d3sigma}) (second line) except $d\zeta \rho(\zeta) d\rho /d\zeta \rightarrow r dr$. Equation~(28) of Ref.~\cite{wojciechs} corresponds to Eq.~(\ref{ssfloweq}) with $g \rightarrow 1$, but its Eq.~(29) assumes that $\alpha_\perp(\zeta) \rightarrow \eta_t(\zeta)$ is constant.

\subsubsection{Rath, Khuntia, Sahoo and Cleymans~\cite{cleymans}}

The BW model invoked in Ref.~\cite{cleymans} is nominally that of Ref.~\cite{ssflow} but differs substantially as follows.  The differential volume element is given as
\bea
d^3\sigma_\mu &=& \tau [\cosh(\eta_s),0,0,[-]\sinh(\eta_s)] d\eta_s r dr d\phi_x
\eea
but the bracketed minus sign is incorrect -- compare Eq.~(\ref{d3sigma}) (second line). Quantities $u_\mu$ and $p_\mu$ are as defined in Eqs.~(\ref{umu}) and (\ref{pmu}). Quantity $\eta \rightarrow \eta_s$ is space-time rapidity. It is notable that ``Bjorken correlation in rapidity'' is interpreted as $(y = \eta) \rightarrow (y_z = \eta_s)$. In that case the expression in Eq.~(\ref{brackets}) above goes to $\tau$ and the integration over $\eta_s$ in Eq.~(\ref{ssfloweq}) is trivial. Function $K_1(x)$ could not arise based on that assumption. It would be replaced by an exponential with the same arguments.

Explicitly, $\eta_s$ is space-time rapidity parameterizing $\sigma_\mu(x)$ and is usually assumed equivalent to longitudinal boost $\eta_z$ in velocity field $u_\mu(x)$ (Bjorken expansion), both of those being flows  in configuration space. The assumption $y_z = \eta_s$ equates longitudinal rapidity $y_z$ of detected particles in momentum space with space-time rapidity or boost of a particle source in configuration space. Equation~(14) of Ref.~\cite{cleymans} appears to be missing a combined factor $p_t m_t$ or $m_t^2$ compared to Eq.~(\ref{ssfloweq}). The analysis is based on taking $n = 1$ for ``flow profile'' $\beta_t(r) \propto (r/R)^n$.

\subsubsection{Ray and Jentsch~\cite{lanny}}

Reference~\cite{lanny} uses the BW model as formulated in Ref.~\cite{tomasik} to study the effect of event-wise fluctuations within A-A collisions and their manifestations in single-particle spectra on transverse rapidity $y_t = \ln[(p_t + m_t)/m_0]$ and on two-particle correlation space $(y_{t1},y_{t2})$. The single-particle distribution in Ref.~\cite{lanny} is derived from Ref.~\cite{tomasik} [see Eq.~(\ref{tomasikeq}) above] but it is assumed that $y \rightarrow y_z \approx 0$. That is a reasonable assumption for data near midrapidity and for the higher energies at the RHIC and LHC in contrast to Ref.~\cite{ssflow} relying on integration over $y_z$ to obtain Eq.~(\ref{bweq}) for description of SPS S-S data at $\sqrt{s_{NN}} \approx 19$ GeV. The expression on the right in Eq.~(\ref{brackets}) then becomes $\tau \cosh(\eta_s)$ (with $\tau_0 \rightarrow \tau$). In Eq.~(\ref{ssfloweq}) $\cosh(y_z - \eta_z) \rightarrow \cosh(\eta_z)$ in the exponential.

\subsubsection{Lao, Liu and Ma~\cite{lao}}

Reference~\cite{lao} invokes the BW model as reported in Eq.~(18) of Ref.~\cite{lanny} to produce its Eq.~(1). A ``blast-wave model with fluctuations'' is attributed to Ref.~\cite{tomasik} but the word ``fluctuations'' does not appear in that document. Reference~\cite{lanny} does describe a BW model with fluctuations but their Eq.~(18), adopted from Ref.~\cite{tomasik} and the basis for Eq.~(1) of Ref.~\cite{lao}, appears before the subject of fluctuations (in $\beta = 1/T$ and in radial boost $\eta_t$) is addressed. In Ref.~\cite{lao} $y \rightarrow y_z$ remains variable whereas $\cosh(y_z - \eta_z) \rightarrow \cosh(y_z)\cosh(\eta_z)$ is assumed. The derivation then proceeds to ``single  source emission'' wherein $\eta_s \rightarrow 0$ and source distribution $H(\eta_s) \rightarrow 1$. The particle source is thereby modeled effectively as a thin disk at $\eta_s = 0$. As a result $K_1[m_t \cosh(\eta_t)/T]$ in Eq.~(\ref{bweq}) $\rightarrow \exp[- m_t \cosh(y_z) \cosh(\eta_t)/T]$ in Eq.~(4) of Ref.~\cite{lao}. The $\beta_t(r)$ flow-profile parameter $n_0 \rightarrow n$ is held fixed at 2. It is further remarked that ``$n_0$ is not a sensitive quantity. It does not matter if $n_0 = 1$ or $n_0 = 2$.'' But refer to Fig.~\ref{logdetails} (c) for the strong effect of parameter $n$.

Equation~(4) of \cite{lao} is then applied to a broad array of LHC A-B collision data. The study notes  ``...one can see that we have used the species-dependent parameters to fit the spectra... [i.e.\ BW fits to individual hadron species]. This is not the usual way to use the blast-wave model, which fits various spectra simultaneously. We have examined the simultaneous fit of the model for various spectra and know that narrow and different $p_T$ ranges have to be used for different particles [e.g.\ Ref.~\cite{aliceppbpid}]. We do not think that the simultaneous fitting of various spectra can be better in wide $p_T$ ranges. Instead, we may use the individual fit for different spectra and obtain better fits.'' 

Even with fits to individual hadron species $\chi^2/\text{ndf}$ numbers are often much greater than 1. Reference~\cite{lao} comments that large $\chi^2/\text{ndf}$ values ``...indicates that the fitting was qualitative and approximately acceptable, and the large dispersion between the curve and data exists.'' One observation is  notable in the context of the present study: ``However, the disadvantage of the [BW] model is also obvious. In some cases, one[-]component model cannot fit the data well. In fact, it is only applicable to the low-$p_T$ region, but not to the high-$p_T$ region. This problem is not only a disadvantage in the blast-wave model...but also a disadvantage in all thermal models.''

These examples serve to illustrate several BW issues: Assumptions vary concerning the structure of the flow field and emission volume and how they should be represented. Concerning derivation of a specific BW model function the basic Cooper-Frye context is accepted. However, significant details of derivations vary. Different symbol choices by authors for basic kinematic quantities complicate interpretation. There is also substantial variation concerning BW model application to data: e.g.\ a common fit to all hadron species as opposed to individual fits to species.  There is the question what \pt\ fit intervals should be used, as determined by what criteria? It is not clear how fits to data might be used to clarify such ambiguities. As a matter of interpretation is there any circumstance in the comparison of model to data wherein the BW model (which version?) might be falsified?

\subsubsection{Tsallis-BW models~\cite{tsallisbw}}

A basic assumption of the BW models above is that any deviation from a {\em Boltzmann} exponential (say below $m_t \approx 3$ GeV/c) must arise from a moving particle source as in fluid flow~\cite{ssflow}. That assumption fails for two reasons: (a) The Boltzmann exponential is an idealization quite unlikely in real situations, especially in relation to parton splitting cascades resulting in heterogeneous systems (parton showers) as hadron sources. (b) The strong contribution from minimum-bias jet fragments (from large-angle-scattered partons) extends down to 0.5 GeV/c with a peak near 1 GeV/c. That system is described accurately by the TCM as demonstrated for 5 TeV \ppb\ collisions in Secs.~\ref{spectrumtcm} and \ref{qualstats} and Refs.~\cite{ppbpid,pidpart1,pidpart2}.

Incorporation of the {\em Tsallis} spectrum model ($q$-exponential, Tsallis statistics) as a replacement for the Boltzmann exponential within the BW model formulation~\cite{tsallisbw} is, in effect, an attempt to deal with (a) but relies on variation of Tsallis model parameters with \mbox {A-B} centrality which is never observed for soft component $\hat S_0(m_t)$ of the TCM. As a monolithic spectrum model the Tsallis model has, {\em in principle}, no capacity to accommodate item (b) above. But model parameters $T$ and $q$ are varied as an attempt to accommodate strong increase of the jet contribution with centrality. Model fits in Ref.~\cite{tsallisbw} are restricted to the range $p_t \in [0.5,3]$ GeV/c. 

The main difference from the Boltzmann BW versions is that $\bar \beta_t$ for \pp\ and peripheral A-B collisions is now consistent with zero as one might expect, since the Tsallis model is a variant of soft-component model $\hat S_0(m_t)$ that describes all A-B soft components within statistical uncertainties. Ironically, the {\em decrease} with increasing centrality of Tsallis parameter $q - 1$ (corresponding to increase of TCM parameter $n$ indicating softening of the $\hat S_0(m_t)$ power-law tail) may actually respond in part to ``jet quenching.'' Although it cautions that ``...the physical interpretation of this [Tsallis] statistical model in the context of high energy nuclear collisions remains to be fully understood'' Ref.~\cite{tsallisbw} interprets the $q - 1$ trend as ``...indicating an evolution from a highly non-equilibrated system in p+p collisions toward an almost thermalized system in central Au+Au.'' There is no acknowledgement of a jet contribution to spectra assumed (erroneously) as confined to ``high-$p_T$ particles.'' The possibility that parameter $\bar \beta_t$ and $T_{kin}$ values, as inferred from a BW model effectively falsified by data, may not be physically interpretable is also not considered.

\section{Systematic uncertainties} \label{sys}

Systematic uncertainties for BW model fits to spectra relate to the primary PID spectra, various BW model versions and fit methods and the validity of any BW model version for part or all of the spectrum \pt\ acceptance.

\subsection{Systematic uncertainties from Ref.~\cite{aliceppbpid}}

Regarding PID spectra themselves Ref.~\cite{aliceppbpid} states that ``The main sources of systematic uncertainties for the analysis of charged and neutral particles are summarized in Tables 3 and 4, respectively. The study of systematic uncertainties was repeated for the different multiplicity bins in order to separate the sources of uncertainty which are dependent on multiplicity and uncorrelated across different bins (depicted as shaded boxes in the figures).'' From that and other language in the text  there are apparently three types of spectrum uncertainties reported: statistical, \nch-dependent and total denoted by solid bars, shaded boxes and unshaded boxes respectively. Captions for Figs.~2, 3, 4 and 5 include ``The empty boxes show the total systematic uncertainty; the shaded boxes indicate the contribution uncorrelated across multiplicity bins....'' 

Spectrum data reported in Ref.~\cite{aliceppbpid} apparently include statistical and {\em total} uncertainties, the latter consistent with ``Total'' in Tables 3 and 4 of Ref.~\cite{aliceppbpid}. There is no distinction in those tables among systematic errors that are \yt\ dependent or not and \nch\ dependent (correlated vs uncorrelated) or not, and there is no estimation of what might contribute to {\em point-to-point} (on \yt) uncertainties. See Figs.~\ref{zsmeson} and \ref{zsbaryon} (right) for statistically-significant systematic {\em errors}  that are correlated across multiplicity \nch\ bins but uncorrelated (i.e.\ vary strongly) across \yt\ bins.

Properly-estimated systematic uncertainties relevant to model descriptions would ideally be consistent with actual data-model discrepancies in the event of a valid model. The results in Figs.~\ref{zsmeson} and \ref{zsbaryon} indicate that actual data-model systematic {\em errors} are rather small, highly-localized on \yt\ and closely-correlated on \nch. As noted above, TCM data descriptions are not obtained by fits to individual spectra. The Z-scores in those figures suggest that the systematic errors presented in Tables 3 and 4 of Ref.~\cite{aliceppbpid} greatly {\em overestimate} the uncertainties relevant to data description by an appropriate model.

\subsection{Sensitivity to BW fit ranges}

A particular issue for BW model fits is the \pt\ or \yt\ intervals over which fits to data are imposed. Fit ranges are determined by ``the available data at low $p_T$'' (i.e.\ the effective particle acceptance lower cutoff) and ``the agreement with the data at high $p_t$, justified considering that the assumptions underlying the blast-wave model {\em are not expected to be valid at high $p_T$}'' [emphasis added]. That policy begs the question where, if anywhere, is the BW model valid and by what criteria? Parameter systematic uncertainties are assigned empirically based on range variations: ``The [fit] results are reported in Tab.~5 and Fig.~6. Variations of the fit range[s] lead to large shifts ($\sim$ 10\%) of the fit results (correlated across centralities)....'' What criteria are invoked to determine such variations?

Concerning utility of BW fits there is also the statement ``It has be be kept in mind, however, that the actual values of the fit parameters depend substantially on the fit range. In spite of this limitations [sic], the blast-wave model still provides a handy way to compare the transverse momentum distributions and their evolution in different collision systems.'' The statement apparently refers to BW fits to {\em individual} hadron species over the {\em entire} \yt\ acceptance as they appear in Fig.~1 of Ref.~\cite{aliceppbpid}. The quality of those data descriptions is not reported, and the parameter values are presumably meaningless based on the argument for limiting fit ranges above (``...not expected to be valid at high $p_T$''). As to convenient data representation the variable-TCM description in Fig.~\ref{tcmdata} above describes all PID spectrum data within their statistical uncertainties  {\em without relying on fits to individual spectra} and based on physically meaningful parameters.

\subsection{Systematic vs statistical errors and Z-scores}

As noted in Sec.~\ref{quality}, Z-scores provide a superior {\em differential} measure of fit quality compared to the $\chi^2$ statistic. Spectrum data associated with Ref.~\cite{aliceppbpid} include both statistical and systematic uncertainties. As noted in Sec.~\ref{sysuncertain}  $\chi^2$ values presented in Table 5 of Ref.~\cite{aliceppbpid} appear based on total systematic uncertainties rather than statistical uncertainties. That is one reason to question interpretation of such $\chi^2$ values. Another is the practice to limit model-fit \pt\ intervals to what will provide a ``good'' fit, perhaps as determined by resulting $\chi^2$ values.

An example of Z-score usage as standard practice for particle physics analysis can be found in Ref.~\cite{alephff} reporting properties of hadronic Z decays from the LEP. In its Fig.~1 appears ``the difference between the distributions of the QCD models and the [\ee] data in units of the data error,'' which is just the Z-score defined in Eq.~(\ref{zscore}). The ``data error'' here is ``combined statistical and systematic errors.'' Although that seems to correspond formally to $\chi^2$ values in Ref.~\cite{aliceppbpid} the two uncertainty types have approximately the same magnitudes for the \ee\  spectrum data in Fig.~\ref{eedata} below, in contrast to the factor $\approx$ 10 difference for spectrum data reported in Ref.~\cite{aliceppbpid}.

Part of systematic uncertainty estimation relates to the {\em interpretability} of analysis results, but applies in this case to {\em model} uncertainty estimations themselves as opposed to data. If a statistic is employed to determine data-model agreement/disagreement do its resulting values provide a reliable evaluation of model validity? In the case of $\chi^2$ values in Table 5 of Ref.~\cite{aliceppbpid} it is likely that the values reported are misleading, hence uncertain.

It can be argued that Z-scores should be calculated only with statistical uncertainties, which are usually unambiguous in a context where discrete objects are counted. Z-scores based solely on statistical uncertainties directly reveal systematic {\em errors} (as opposed to uncertainties) in both data and model. {\em Data} systematic errors are apparent for instance in Fig.~\ref{zsbaryon} (b) (large narrow excursions) since the applied model (TCM) has no capacity to describe, and the collision process itself is unlikely to generate, such narrow structures. On the other hand, large and smooth deviations apparent in Fig.~\ref{bwzscore1} (right) unambiguously reveal {\em model} systematic errors. A more effective venue for estimation of {\em systematic} model and data {\em errors} is differential Z-scores based on statistical uncertainties that are unambiguous in their definition.

\section{Discussion: two cultures}  \label{disc}

The blast-wave model represents one of two ``cultures'' relating to the dynamics of high-energy nuclear collisions, associated with a hydrodynamics- or hadronic-fluid-based description (flows) on the one hand and a particle or high-energy physics (partons) version of QCD on the other. As noted in Sec.~\ref{earlybw} the former has its origins in the nuclear theory of Landau~\cite{landau} (1950s) and elements of the fireball description of Hagedorn~\cite{haged} (1960s) which substantially predate the experimental discovery of quarks and development of QCD theory (1970s) at which point it became evident that parton cascades or showers play a major role in high-energy collision dynamics.

\subsection{Parton splitting cascades or showers}

One of the most popular Monte Carlo models of elementary collisions is PYTHIA~\cite{pythia1,pythia2}: ``The [PYTHIA] program is designed to simulate the physics processes that can occur in collisions between high-energy particles, e.g.\ at the LHC collider at CERN. 
... A combination of perturbative results and models for semihard and soft physics ... are combined to trace the evolution towards complex [e.g.\ hadronic] final states.'' 
Initial developments that led to the PYTHIA Monte Carlo -- ``reproducing the work of Field and Feynman, and extending it to the analytical model developed in Lund'' --  were motivated by an early model of jet formation reported in Refs.~\cite{field1,field2,field3}. ``The `jets' observed in both cases [\ee\ and \pp\ collisions] are thought to arise from {\em quarks that fragment or cascade into a collection of hadrons moving in roughly the direction of the original quark} [emphasis added]''~\cite{field3}.

Concerning early development of detailed mechanisms ``Both of these [early jet models other than PYTHIA] were based on the concept of independent fragmentation, wherein each of the $q$, ${\bar q}$ and $g$ jets are assumed to fragment symmetrically around a jet axis defined by the direction of the respective parton in the CM frame of the event. In Lund, instead another picture had been developed, string fragmentation. Here the connecting colour field is approximated by a massless relativistic string, with gluons represented by pointlike momentum-carrying `kinks'.'' ``PYTHIA lacked the parton showers that gave the other two programs [ISAJET and FIELDAJET] realistic jet shapes.''
``It was clear that parton showers would play a key role in order to produce multijet final states...''~\cite{pythia2}.

Reference~\cite{pythia2} notes that ``...the soft-gluon emission pattern around a $q{\bar q}g$ topology could be viewed as a sum of radiation off two independent dipoles...mimicking the nonperturbative string picture. ...this offered a starting point to formulate a [parton] shower as a successive branching [cascade] of dipoles, an idea that today is a standard choice for most shower algorithms....'' In Ref.~\cite{gosta} a Lund Dipole Cascade model, implemented as the DIPSY Monte Carlo, is described: The model ``gives a good description of inclusive pp and ep cross-sections (including diffraction), and a fair description of exclusive final states (min.\ bias and underlying event). The model can also be applied to reactions with nuclei,
with some early results available.'' In particular the model is able to describe $\eta$ charge densities within $\eta \in [-2.5,5]$ for 0.9 and 7 TeV \ppbar\ collisions at the LHC.  In summary, various Monte Carlo models based on parton splitting cascades or showers are able to describe the hadronic final states of elementary (\ee, \pp) collisions in considerable detail.

\subsection{Hadron gas in thermodynamic equilibrium}

An interesting alternative approach to \ee\ collision data is reported in Ref.~\cite{statmodel}. The study acknowledges that \ee\ collisions (as an example) include both hard processes (parton showers) described by perturbative QCD and soft processes (parton fragmentation to hadron jets) that must be described phenomenologically, with the Jetset Monte Carlo (precursor to PYTHIA) as one example.  Reference~\cite{statmodel} comments: ``The main unsatisfactory feature of these [Monte Carlo] models...is the large number of free parameters required in order to correctly reproduced experimental data. As a consequence, those models have a rather poor predictive power.''

The analysis assumes that ``each jet [of an \ee\ hadronic final state] represents an independent phase in complete thermodynamic equilibrium....'' That assumption is taken to imply that ``one can describe a jet as an object defined by thermodynamic and mechanical quantities such as temperature, volume....'' The basis for analysis is ``...the canonical partition functions of systems with internal symmetries.'' The analysis is applied to final-state abundances of an array of hadron species. The study concludes that ``...this model is able to fit impressively well the average multiplicities of light hadrons....'' The model parameters are temperature $T$, volume $V$ and $\gamma_s$ related to strange-quark chemical equilibrium.

While that analysis is informative and technically competent it can be argued that ``an independent phase in complete thermodynamic equilibrium'' is not demonstrated by model agreement with hadron species abundances. In discussing the statistical properties of hadrons emerging from ``freezeout'' of A-A collisions at the SPS Ref.~\cite{stock} warns that ``...this apparent `thermal' equilibrium is a result of the decay process, the nature of which lies well beyond the statistical model which `merely' captures the apparent statistical order, prevailing right after decay. ... The observed equilibrium is, thus, {\em not achieved by inelastic transmutation of the various hadronic species densities, in final hadron gas rescattering cascades}, i.e. not by hadron rescattering approaching a dynamical equilibrium [emphasis added].'' One may argue by analogy that the partons within a jet do not rescatter so as to achieve thermodynamic equilibrium prior to hadronization. It is the hadronization process itself, as a quantum transition following a least-action principle, that leads to approximate agreement with the description in Ref.~\cite{statmodel}.

The array of species abundances among jet fragments from \ee\ collisions (what is addressed by Ref.~\cite{statmodel}) is only part of the information carried by the fragment distribution from a jet ensemble. Single-particle momentum distributions and {\em multiparticle correlations} (on angle~\cite{porter2,porter3} and \pt\ or \yt~\cite{ytyt}), as well as {\em local} correlations among charge, flavor and baryon number, clearly indicate that jets are not featureless equilibrated gases. In summary, while thermal equilibrium of a phase might imply a certain data profile, observation of such a data profile by experiment does not demonstrate thermal equilibrium.

\subsection{Hadron emission from a flowing source}

Within a parton cascade the initial momentum of a single leading particle is distributed by splitting to a shower of lower-momentum particles. Transfer of momentum to final-state hadrons can be described as {\em top-down}.
In contrast, within a fluid-dynamics picture of high-energy collisions the initial projectile energy is to some extent ``stopped'' in the CM leading to high energy densities and pressure gradients that drive fluid flow. The resulting flow field, convoluted with a locally-isotropic Boltzmann distribution (Cooper-Frye formalism~\cite{cooper}), is expected to describe final-state momenta of emitted hadrons. Transfer of momentum/energy can be described as {\em bottom-up}.


Reference~\cite{wong} discusses A-A collision dynamics at the LHC in the context of the Landau hydrodynamic  model of high-energy nuclear collisions:  ``Consider first the case of central AA collisions with A $\gg$ 1 such that nucleons of one nucleus collide with a large numbers of nucleons of the other nucleus and the {\em whole energy content} is used in particle production. This is the case of `{\em full stopping}' [emphasis added].
...
Common to both Landau hydrodynamics and Bjorken hydrodynamics is the basic assumption that in the [conjectured] dense hot matter produced in high-energy heavy-ion collisions, the density of the quanta of the medium 
is so high that a state of local thermal equilibrium can be maintained through out.'' The large densities imply corresponding large density {\em gradients} that drive flows. In such models much of the hadron momentum originates solely from pressure-induced flows combined with thermal emission. Any small jet contribution is assumed restricted to limited \pt\ intervals.

That is the context in which the BW model has been applied to particle data at the SPS and higher energies. Reference~\cite{ssflow} asserts ``In this paper we want to develop a phenomenological model for the hadronic matter by starting out with thermalization as the basic assumption and adding more features as they are dictated by the analysis of the measured hadronic spectra.'' Referring to {\em nucleon} flows at the Bevalac is the statement ``Though this directed flow [at the Bevalac] is of different nature as our collective expansion flow [at the SPS], it suggests the relevance of hydrodynamics also at higher energies.'' 

But there immediately appears evidence of conflict between such assumptions and what had emerged from high-energy (particle) physics study of elementary collisions. Reference~\cite{ssflow} first admits that ``Our current understanding of QCD results basically from high energy experiments with small collision systems [e.g.\ \pp\ collisions] suffering hard interactions which are relatively easy to analyse.'' But given the similarity of \ss\ and \pp\ spectra at SPS energies Ref.~\cite{ssflow} then argues that  ``It would be too impulsive to deduce from the apparent similarity of pp and S+S spectra that the collective interpretation [i.e.\ flows in \ss] is wrong, since a) pp is by no means an elementary collision system which we understand in sufficient detail to serve as the antipode [opposite] of a collective system and b) only a few of the observed features of the [hadron] spectra can be fully reproduced by these two radically different philosophies which share only a small set of common principles as local energy-momentum conservation, relativistic space-time picture, etc.''

One may assume that ``radically different philosophies'' refers to ``high energy experiments'' and resulting parton-cascade descriptions (e.g.\ PYTHIA) on the one hand and a hydrodynamic approach implemented via a BW model on the other. If ``pp is by no means an elementary collision system'' then what is --- \ee? Point a) above essentially excludes \pp\ collisions as a {\em reference} system for the study of A-B collisions although that has subsequently been common practice at the RHIC and LHC.

Concerning point b) is the statement ``For other observables, which are not directly tied to the dynamics, e.g. strangeness, we already see a big enhancement compared to [minimum-bias] pp, thus indicating fundamental differences between both [S-S and \pp] collision systems.'' But that conclusion can be strongly questioned in the context of measured {\em jet production} in \pp\ and for instance \ppb\ collisions as reported in Refs.~\cite{ppbpid,pidpart1,pidpart2}, in which minimum-bias jets are found to make strong contributions to strangeness and baryon production. 
The ``fundamental differences'' between \ss\ and \pp\ are more likely simply due to increased jet production in \ss\ collisions due to {\em multiple} \nn\ binary collisions (and hence increased jet production) {\em per nucleon participant} with otherwise no significant deviation from \pp\ observations. 

In summary, descriptions of hadron emission from a flowing source, e.g.\ the BW model, compete directly with data descriptions in terms of parton cascades -- within projectile nucleons leading to (soft) hadron distributions along the beam axis or within (hard) jets resulting from large-angle scattering of partons from projectile nucleons. 

\subsection{BW model compared with elementary collisions}

It is informative to consider some examples of ``elementary'' collisions in comparison to model elements from the TCM. In particular, the $\sqrt{s_{NN}} = 19.4$ \ss\ spectrum data invoked in Ref.~\cite{ssflow} are compared  with 17 GeV (SPS) \pp\ data and with the soft-component model function that describes $\sqrt{s} = 200$ GeV \pp\ spectra~\cite{ppprd}. These comparisons address in part the question: what is an elementary collision and how should it be modeled?

\begin{figure}[h]
	\includegraphics[width=1.68in]{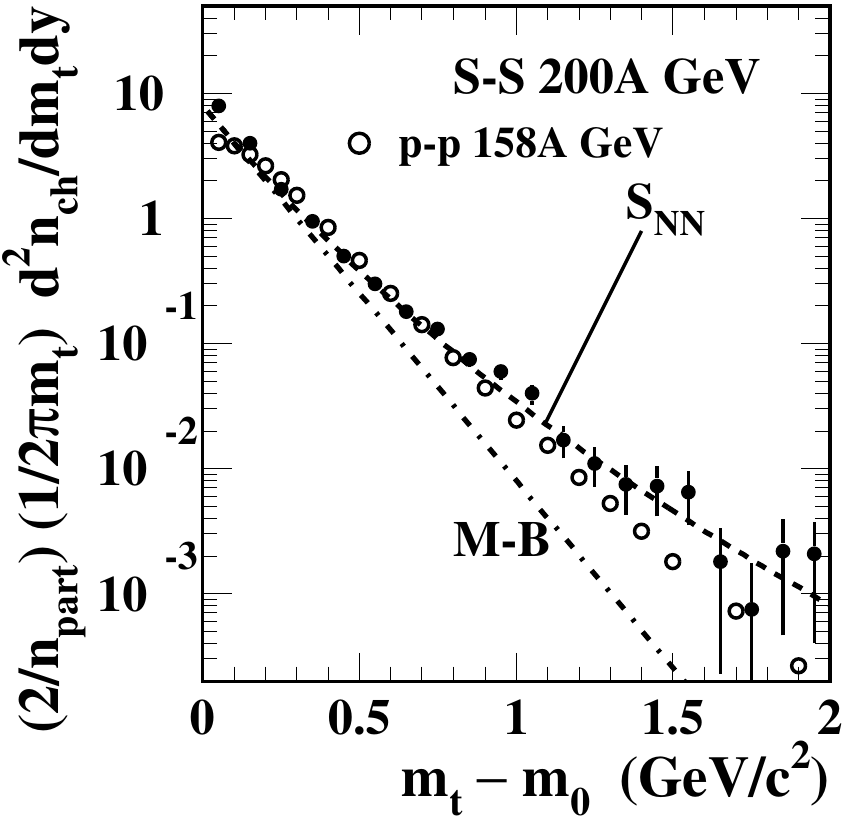}
	\includegraphics[width=1.63in]{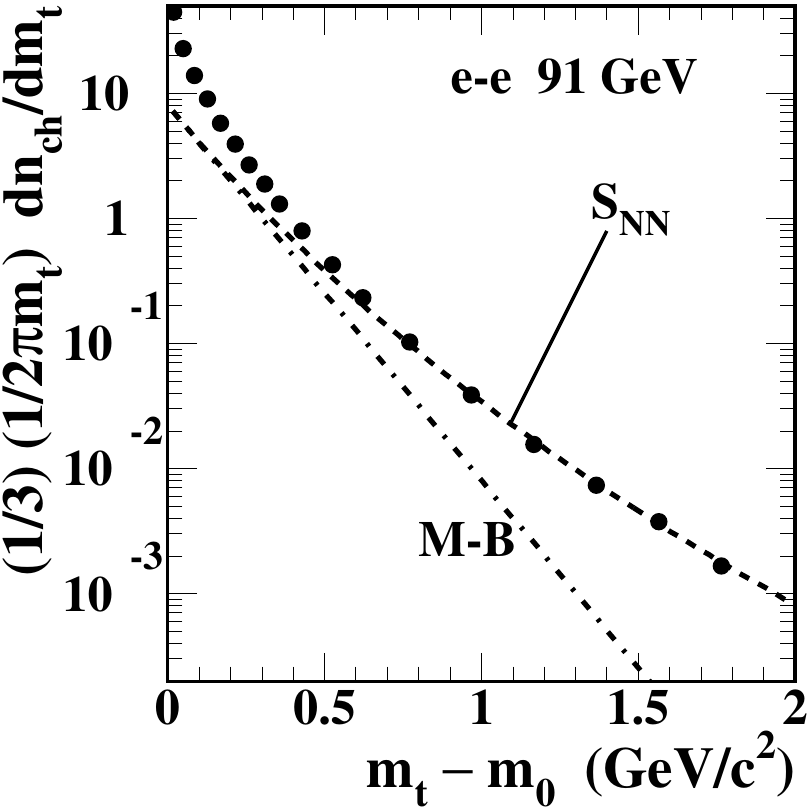}
	\caption{\label{spsdata}
		Left panel: $m_t$ spectra from 0-2\% central \ss\ collisions at 200A GeV (solid points, $\sqrt{s_{NN}} = 19.4$ GeV)~\cite{na35ss} and from \pp\ collisions at $\sqrt{s} = 17.3$ GeV (open circles)~\cite{na49pp}. The dashed curve is L\'evy soft component $\hat S_0(m_t)$ from 200 GeV \pp\ collisions. The dash-dotted curve is the Boltzmann limiting case of $S_{NN}$.
		Right panel: $m_t$ spectrum (points) from \ee collisions at $\sqrt{s} = 91$ GeV~\cite{alephff}. The curves are duplicated from the left panel. The \ee\ data are rescaled as in the axis label to match the $S_{NN}$ curve at larger $m_t$.
	}  
\end{figure}

Figure~\ref{spsdata} (left) shows a pion spectrum on $m_t$ (solid dots, $2\times\pi^-$) for SPS fixed-target 0-2\% central {\ss} collisions at 200A GeV ($\sqrt{s_{NN}} = 19.4$ GeV)~\cite{na35ss}. A hydro-model (BW)  analysis of those data was used to infer radial flow with mean transverse speed $\langle \beta_t \rangle \rightarrow \bar \beta_t \sim 0.25$~\cite{ssflow} (see Fig.~\ref{betatkin}, right, and associated text). The curve labeled $S_{NN}$ is the soft component for $\sqrt{s} =$ 200 GeV NSD \pp\ collisions from the RHIC~\cite{ppprd}. The line labeled M-B is the Boltzmann (exponential) limit of L\'evy distribution $S_{NN}$ with $T \approx 145$ MeV. A $\pi^+ + \pi^-$ spectrum from \pp\ collisions at 158A GeV (open circles, $\sqrt{s}=17.3$ GeV) is included for comparison~\cite{na49pp}. 

Fig.~\ref{spsdata} (right) shows an $m_t$ spectrum (points) from LEP \ee collisions at $\sqrt{s} \approx 91$ GeV~\cite{alephff}. \ee\ hadronic events from Z$_0$ decays are dominated by two-jet events (see Fig.~13 of Ref.~\cite{alephff}).  The spectrum, derived from a sphericity analysis of \qqbar\ dijets, reveals the jet fragment momentum distribution transverse to the thrust (dijet) axis. The LEP $dn_{ch}/p_tdp_t$ data were rescaled to overlap  the \pp\ soft component $S_{NN}$ (dashed curve), and hence the SPS \ss\ spectrum. 
The shape of the $m_t$ spectrum from 91 GeV Z$_0$ decays  is consistent with the soft component of \pp\ spectra at 200 GeV and the spectrum from central 19 GeV \ss\ collisions.  Commonality of the soft-component shape (L\'evy distribution) across energies and collision systems suggests that the TCM soft component is a universal feature of fragmentation for any leading particle -- be it parton or hadron. 

While the comparison in Fig.~\ref{spsdata} (right) provides a hint of the relation among dijet formation in \ee\ collisions, the hadronic final state of A-B nuclear collisions and the BW model a more direct comparison can be made.

\begin{figure}[h]
	\includegraphics[width=1.65in]{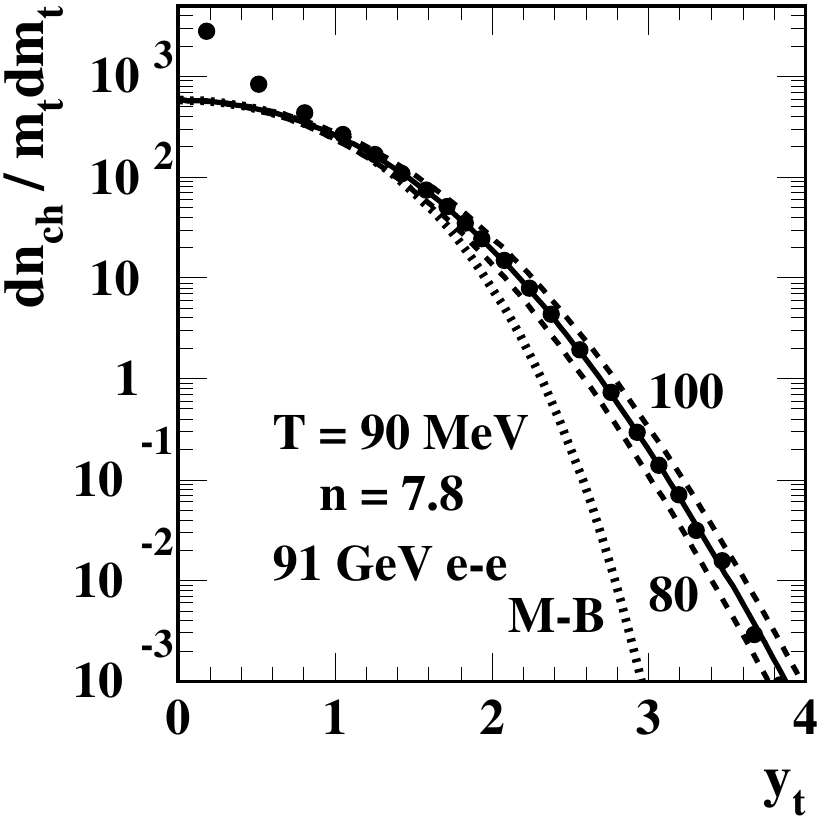}
	\includegraphics[width=1.65in]{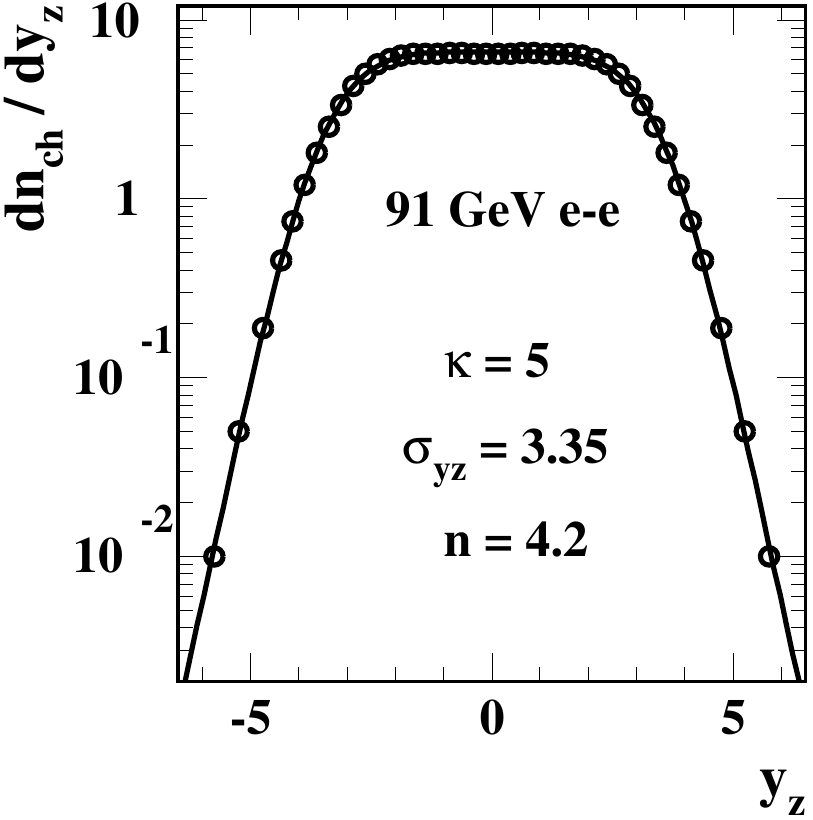}
	\caption{\label{eedata}
		Left: The hadron distribution on \yt\ (perpendicular to the dijet axis) from 91 GeV \ee\ collisions~\cite{alephff}. The solid curve is a L\'evy distribution [i.e.\ $\hat S_0(y_t)$] with parameters ($T,n$) noted in the figure. The dashed curves correspond to variation of $T$ by 10 MeV. Dotted curve M-B is the exponential (on \mt) limit for $n \rightarrow \infty$ and $T = 90$ MeV.
		Right: Hadron distribution on \yz\ (parallel to the dijet axis) from 91 GeV \ee\ $\rightarrow$ \qqbar\ collisions. The curve is Eq.~(\ref{gcurve}) explained in the text.
	} 
\end{figure}

Figure~\ref{eedata} (left) shows the \ee\ \pt\ spectrum from Fig.~\ref{spsdata} (right) replotted on \yt\ (derived from measured \pt\ assuming all hadrons are pions as for \yz\ below). The points are published $dn_{ch}/dp_t$ data (Fig.~18a of Ref.~\cite{alephff}) divided by \pt\ with no other scaling. According to Ref.~\cite{alephff} this is the ``charged particle momentum component transverse to the sphericity [dijet] axis and projected into the event plane.''  The solid curve is a L\'evy distribution on \mt\ [i.e.\ $\hat S_0(m_t)$] with slope parameter $T = 90$ MeV and L\'evy exponent $n = 7.8$. Dashed curves show the effect of varying $T$ by 10 MeV. For comparison, soft-component $\hat S_0(y_t)$ for pions in Fig.~\ref{tcmdata} (a) has $T = 145$ MeV and $n = 8.5$. The \ee\ data are described within their published uncertainties except for the lowest three points  at 0.02, 0.07 and 0.13 GeV/c.

Figure~\ref{eedata} (right) shows a hadron distribution on \yz\ from $Z$ decays reported by Ref.~\cite{alephff}. $y_z$  ($y_s$ in Fig.~16a of Ref.~\cite{alephff} with $s$ for ``sphericity'') is derived from measured $p_L \rightarrow p_z$ along the event sphericity (dijet) axis and is based on assigning the pion mass to all hadrons. The solid curve is a generalized $q$-Gaussian distribution
\bea \label{gcurve}
G(y_z) &=& \frac{A}{[1 + (|y_z|/\sigma_{y_z})^\kappa/n]^{n}}
\\ \nonumber
&\sim & A \exp_q[-(|y_z| /\sigma_{y_z})^\kappa],
\eea
with $\kappa = 5$, $\sigma_{y_z} = 3.35$, $n = 4.2$ and $A = 6.6$.
The  distribution differs from a Gaussian in two ways: (a) the parameter value $\kappa =$ 5 rather than 2 leads to a more rectangular distribution and (b) the parameter value $n =$ 4.2 controls the distribution tails. The $q$ in ``$q$-Gaussian'' refers to the relation $q - 1 = 1/n$. For $n \rightarrow \infty$ (and $\kappa = 2$) the distribution reverts to a standard Gaussian form. The kinematic limits of the data distribution are $y_z \approx \pm 6.5$ (for pions) corresponding to $\sqrt{s}/ 2 \approx 46$ GeV. The ``1/e'' points at $\pm 3.35$ correspond to 2 GeV/c.

Several issues are addressed by Figs.~\ref{betatkin}, \ref{spsdata} and \ref{eedata}:

Figure~\ref{betatkin} demonstrates the relation between the commonly-applied BW model and (a) the Boltzmann distribution (agreement expected for $\bar \beta_t = 0$ by definition of BW model) and (b) $\hat S_0(y_t)$ inferred from spectrum data as the zero-density limit but requiring $\bar \beta_t \approx 0.25$ from the BW model to approximate the L\'evy distribution.

Figure~\ref{spsdata} demonstrates that the soft-component model (L\'evy distribution) $\hat S_0(y_t)$ ($S_{NN}$ in those figures) describes spectra from SPS \pp\ and S-S collisions and (for higher \pt) LEP \ee\ collisions, all of  which deviate strongly from a Boltzmann-distribution exponential for reasons related to known QCD processes. The small difference between SPS {\em central} S-S spectra and \pp\ spectra may be due  an expected jet contribution at $\sqrt{s_{NN}} \approx 19$ GeV~\cite{jetspec2}. The power-law tail of soft component $\hat S_0(y_t)$ inferred from 200 GeV \pp\ collisions is apparently able to approximate the jet contribution to the 19 GeV S-S spectrum, consistent with collision-energy dependence of the \pp\ soft component (see Ref.~\cite{alicetomspec}, Fig.~12 right).

Figure~\ref{eedata} reveals a direct connection between the TCM description of hadron-hadron collisions and \ee\ collisions from LEP~\cite{alephff}. 
The left panel shows the \pt\ spectrum from \ee\ collisions (solid dots) consistent in shape with soft component $\hat S_0(y_t)$ (solid curve) from \pp\ and \ppb\ collisions. While the soft component for pions has $T \approx 145$ MeV that for \ee\ collisions has the substantially lower value $T \approx 90$ MeV. Whereas the parton parents of final-state hadrons from A-B collisions experience Fermi motion within projectile nucleons and $k_t$ contributions within a parton cascade the \qqbar\ pair in a \ee\ collision has {\em zero} transverse momentum in the CM frame leading to a ``cooler'' hadron fragment spectrum. On the other hand $n = 7.8$ for \ee\ collisions represents a  ``harder'' power-law tail for that spectrum than  the $n = 8.5$ value for \pp\ collisions. Both observations are consistent with $\hat S_0(y_t)$, as the description {\em inferred from A-B data}, reflecting hadron production via parton splitting cascades within a conventional QCD context. The resulting L\'evy data trend is {\em inconsistent} with the Boltzmann exponential trend assumed for the BW model.

Figure~\ref{eedata} (right) shows the complementary dijet fragment distribution on \yz\ (from Ref.~\cite{alephff} Fig.~16a) that can be compared with similar distributions on pseudorapidity from hadronic A-B collisions. Within a hydrodynamic model context such distributions are attributed within hadronic collisions to longitudinal flow or Bjorken expansion. Yet the very similar fragment distribution from \ee\ collisions may be attributed to parton splitting cascades leading to dijets. In the first case longitudinal hadron momentum is attributed to pressure gradients, in the second case to a parton parent in the cascade carrying a fraction $x$ of the leading-parton momentum. One description (flows) says that a particle has been pushed while another (partons) says that it has been ``pulled.''

For both longitudinal and transverse momentum distributions from \ee\ collisions the trend at higher momentum is a power law controlled by $n = 4.2$ for \pz\ and $n = 7.8$ for \pt, indicated by straight-line trends in Figure~\ref{eedata} on \yt\ or \yz. A similar power-law trend is observed for \pt\ spectrum hard components [e.g.\ Fig.~\ref{kalam} (b,d)] associated quantitatively in that case with large-angle-scattered partons fragmenting to jets~\cite{fragevo}.

Detailed study of \pt\ spectra from 5 and 13 TeV \pp\ collisions~\cite{tomnewppspec} demonstrates the two-component (projectile-proton dissociation and scattered parton fragmentation to dijets) nature of hadron production {\em consistent with basic QCD} in such ``elementary'' collisions. Statistical analysis of two models applied to the same data with intent to demonstrate thermalization and/or flows in \pp\ collisions~\cite{tommodeltests} are falsified via Z-score tests as in Sec.~\ref{quality}.
One might expect that ``elementary'' collisions (\ee, \pp\ and even \pa\ systems) should be well understood after fifty years of theoretical and experimental QCD-related research whereas larger A-B collision systems might remain ambiguous. However, there seems to be a persistent tendency to assume that A-A collisions comprise the fixed reference and ``elementary'' collisions are seen as not elementary and still returning unanticipated results.

\section{Summary}\label{summ}

Hadron \pt\ spectrum analysis is a principal component of data interpretation relating to high-energy nuclear collisions. A major element of spectrum analysis is fitting  models to spectrum data both as a form of data compression (reducing spectra to a few, possibly interpretable, parameters) and as a test of specific physical model predictions.
Several models available for spectrum description are comprised of simple functions: (a) the Boltzmann exponential on transverse mass $m_t$, (b) the Tsallis model (``Tsallis statistics'') on \mt\ or \pt, (c) the blast-wave (BW) model on \mt\ and (d) the two-component (soft+hard) spectrum model (TCM). Also available are several complex Monte Carlo models (e.g.\ PYTHIA).

The BW model in particular, formulated to describe the effect on hadron momenta in the center-of-momentum frame of moving (flowing) particle sources, was intended to facilitate discovery of flows in A-A collisions as one manifestation of QGP formation. Early results from the relativistic heavy ion collider (RHIC) seemed to demonstrate flow manifestations in spectra.

More recently, the BW model as applied to \pa\ and even \pp\ collisions has been interpreted to support the same conclusions for small systems, although matter and energy densities achievable in such systems are {\em a priori} unlikely to produce a QGP. Such results cast doubt on the legitimacy of the BW model applied to collision data.

The present study of BW model applications to identified-hadron (PID) spectra from 5 TeV \ppb\ collisions is a response to that situation. This study is intended to assess basic formulation(s) of the BW model and its (their) underlying assumptions, to determine data fit quality in relation to possible model falsification and to compare the BW model to alternative models that may provide superior data description and lead to different (and perhaps more credible) physical interpretations.

The general procedure includes the following steps: 
(a) visually compare BW model and TCM to data in basic semilog plots on transverse rapidity $y_t$ where data trends have a particularly simple structure,
(b) compare data and model {\em shapes} (what is emphasized in BW model applications) using model-independent measures,
(c) assess data-model fit quality differentially and {\em quantitatively} using Z-scores, and
(d) review BW model evolution across several decades via a number of published examples.

The specific results are as follows:
(a) As established in a previous study the TCM describes spectra for four hadron species within point-to-point uncertainties over the entire \pt\ acceptance and the TCM is {\em not} fitted to individual spectra. In contrast, the BW model fails to describe spectra except over limited \pt\ intervals {\em defined by data-model agreement}. That failure is most obvious for the case of pion spectra.
(b) Data-model comparisons using model-independent methods (e.g.\ logarithmic derivatives) demonstrate qualitative differences between BW model and data. On that basis the model is falsified.
(c) Application of the Z-score statistic, a standard measure of data-model agreement, demonstrates that the BW model is excluded as a spectrum model for this collision system, whereas the TCM is an acceptable model as established in a previous study.
(d) While the general structure of the BW model is accepted, specific applications may be quite different in detail (what parameters are allowed to vary over what ranges, even basic algebraic structure) leading to difficulty in interpreting results.

If the BW model is accepted as a valid physical data model, albeit only within artificially-restricted \pt\ intervals (with doubtful justification), it is commonly concluded that the great majority of hadron production arises from a moving source exhibiting ``collective'' motion. However, the present study establishes that for the \ppb\ collision system at 5 TeV the BW model is falsified by PID spectrum data. Previous studies have arrived at the same conclusion for 5 and 13 TeV \pp\ collisions.

In contrast, the TCM does provide a statistically-acceptable data description for {\em all} available PID spectrum data from 5 TeV \ppb\ collisions. From that result the following conclusions may be drawn: 

The TCM soft component is consistent with {\em nuclear transparency} (1976) wherein a projectile nucleon may transit a target nucleus intact, albeit ``excited'' (wounded nucleon model), with later dissociation to hadron fragments {\em outside} the collision space-time volume. That picture is consistent with the soft-component model {\em inferred from data} -- a Boltzmann exponential with power-law tail indicating a heterogeneous source (e.g.\ parton shower).

The TCM hard component is quantitatively consistent with {\em minimum-bias dijet production} as represented by measured jet spectra and parton fragmentation functions. In that case final-state hadrons do indeed emerge from moving particle sources which however happen to be scattered partons as expected from basic QCD theory.

\begin{appendix}

\section{BW model derivation} \label{derive}

Formulation of the BW model for single-particle momentum spectra requires evaluation of the integral~\cite{cooper}
\bea \label{cfeqq}
E\frac{d^3N}{dp^3} &=& \frac{g}{(2\pi)^3} \int_\sigma e^{-(u^\mu p_\mu) / T} p^\nu d^3\sigma_\nu,
\eea
where a factor $\exp(\mu/T)$ is omitted.
Evaluation requires determination of quantities  $u^\mu(x)$, $p_\mu$, $\sigma_\mu$ and  $d^3\sigma_\mu$, where  $u^\mu(x)$ is a fluid velocity field on space-time $x$ in the CM or lab frame, $p_\mu$ is particle four-momentum in the lab frame and $\sigma(r,\phi_x,\eta_s)$ is the volume (hypersurface) from which detected particles emerge (freeze out). The Boltzmann exponential assumes particle emission according to an isotropic thermal distribution in the local frame. 

\subsection{Basic four-vectors} \label{bwtheory}

The basic four-vectors required for evaluating Eq.~(\ref{cfeqq}) are defined using a set of self-consistent symbols that overlap as much as possible the different conventions utilized in an assortment of relevant publications.

\subsubsection{Space-time rapidity}

The position four-vector in cylindrical coordinates preferred for large $\sqrt{s}$ is
\bea
x_\mu &=& [t,x,y,z]
\\ \nonumber
&=& [\tau \cosh(\eta_s),r\hat e(\phi_x),\tau \sinh(\eta_s)],
\eea
where $\hat e(\phi_x)$ is a unit vector, with space-time rapidity
\bea
\eta_s &\equiv & \frac{1}{2} \ln\left( \frac{t + z}{t - z} \right) = \ln\left( \frac{t + z}{\tau} \right) 
\eea
where $t = \tau \cosh(\eta_s)$, $z = \tau \sinh(\eta_s)$ and $v_s = z / t = \tanh(\eta_s)$. 
Azimuth angles in configuration and momentum spaces are distinguished by $\varphi$ vs $\phi$ or $\phi_x$ vs $\phi_p$, with the latter notation used below to reduce ambiguity. The space-time parameter $\zeta$ in Ref.~\cite{ssflow} here goes to $\eta_s$.

Configuration-space rapidity $\eta_s$ is distinguished from (a) momentum-space {\em pseudorapidity} $\eta$ with no subscript and (b) longitudinal flow angle or boost $\eta_z$ (in configuration space). Those distinctions are consistently maintained below unless explicitly noted. The approximation $\eta_z(z) \approx \eta_s(z) $ (Bjorken expansion) is often assumed. In what follows the two quantities are maintained distinct unless specifically noted otherwise.

\subsubsection{Fluid (particle source) velocity field} \label{upmuu}

Fluid velocity fields are distributed on configuration space $x$. The following are Lorentz hyperbolic (boost) angles for particle sources moving in the CM frame
\bea \label{etax}
\eta_t(r)  &=& (1/2)\ln\left( \frac{1 + \beta_t}{1 - \beta_t}\right) = \ln[\gamma_t (1 + \beta_t)]
\\ \nonumber
\eta_z(z) &=& (1/2)\ln\left( \frac{1 + \beta_z}{1 - \beta_z}\right) = \ln[\gamma_z (1 + \beta_z)],
\eea
with $\beta_x = \tanh(\eta_x)$, $\gamma_x = \cosh(\eta_x)$ and $\gamma_x^2 = 1 / (1- \beta_x^2)$.
Given those definitions the four-velocity field distributed on configuration space is
\bea \label{umu}
u^\mu(x) &=& \cosh(\eta_t) [\cosh(\eta_z),\tanh(\eta_t) \hat e(\phi_x),\sinh(\eta_z)]
\nonumber \\
&=& \gamma_t [\gamma_z,\beta_t\hat e(\phi_x),\gamma_z \beta_z ]
\\ \nonumber
u^\mu u_\mu 
&=& 1 
\eea
In Refs.~\cite{ssflow,cleymans}   $\rho \rightarrow \eta_t$ and $\eta \rightarrow \eta_z$ (``boost angles'') are functions of $r$ and $z$ respectively based on assumptions.

\subsubsection{Particle momenta in the CM or lab frame}

Given measured momenta in cylindrical coordinates $(p_t,\theta,\phi)$ (e.g.\ in a solenoidal magnetic field) pseudorapidity $\eta = - \ln\left[ \tan(\theta / 2)\right]$ ($\sim y_z$ rapidity). The three-vector momentum is $\vec p = (p_x,p_y,p_z) = (p_t,\eta,\phi_p)$ with magnitude $p = p_t \cosh(\eta)$ and $p_z = p_t \sinh(\eta)$. Symbol $\eta$ without subscript here always represents pseudorapidity.

Transverse mass \mt\ is defined by $m_t^2 = p_t^2 + m_0^2$. Transverse rapidity is $y_t = \ln[(m_t + p_t) / m_0]$ in a longitudinally comoving frame with $m_t = m_0 \cosh(y_t)$, $p_t = m_0 \sinh(y_t)$. Longitudinal rapidity is $y \rightarrow y_z = \ln[(E+p_z)/m_t]$
with $E = m_t \cosh(y_z)$, $p_z = m_t \sinh(y_z)$.
With those definitions the particle four-momentum is
\bea \label{pmu}
p_\mu 
&=& [E,p_t \hat e(\phi_p),p_z]
\\ \nonumber
&=& \ [m_t \cosh(y_z), p_t\hat e(\phi_p), m_t \sinh(y_z)]
\\ \nonumber 
&=& m_t [\cosh(y_z), \tanh(y_t) \hat e(\phi_p), \sinh(y_z)] 
\\ \nonumber
p^\mu p_\mu
&=& m_0^2
\eea

\subsubsection{Particle energy in the boost frame}

With the velocity field for particle emission and the particle four-momentum defined the product represents particle energy in a conjectured boost frame based on measured particle momentum in the CM or lab frame~\cite{cooper}.
\bea \label{upmu}
u^\mu p_\mu &=& \gamma_t m_t [\cosh(y_z)\cosh(\eta_z) - \sinh(y_z)\sinh(\eta_z)]
\nonumber
\\ 
&& - \gamma_t \beta_t p_t \cos(\phi_x - \phi_p)
\\
\nonumber
&=& \gamma_t [m_t \cosh(y_z - \eta_z) - \beta_t p_t \cos(\phi_x - \phi_p) ]
\\ \nonumber
&\rightarrow& m_t \cosh(y_z) = E~~\text{in the lab frame for no flows}.
\eea

\subsection{Emission hypersurface} \label{hyper}

The freezeout hypersurface $\sigma$ appearing in Eq.~(\ref{cfeqq}) is the configuration-space volume from which particles are emitted (decoupled from the source fluid) and then fly freely to a detector. In Ref.~\cite{cooper} the differential product $p^\mu d\sigma_\mu = (E - \beta p)dx = \bar E d\bar x$ for a 1D system, where bars denote quantities in the local, comoving or boost frame. That follows since $p_\mu = [E,p]$, $\sigma_\mu = [t,x]$, $d\sigma_\mu = [dx,dt]$ and $dt = \beta dx$ if $d\bar t = 0$. In what follows, several examples are presented for defining $d^3\sigma_\mu$ and determining $p^\mu d^3\sigma_\mu$.

\subsubsection{Schnedermann et al.~\cite{ssflow}}

In this study the emission hypersurface (with $\zeta \rightarrow \eta_s$ the space-time rapidity, $\eta_s \in [-H_s, H_s]$ and $r \in [0,R]$) is
\bea
\sigma(r,\phi_x,\eta_s) &\rightarrow& \sigma_\mu = [t(\eta_s),r \hat e(\phi_x),z(\eta_s)].
\eea
The differential freezeout volume is then
\bea  \label{dsigmu}
d^3\sigma_\mu &=& \left[\frac{\partial z}{\partial \eta_s},0,0,\frac{\partial t}{\partial \eta_s}\right]d\eta_s  rdr d\phi_x
\eea
assuming instantaneous freezeout in $r$, and with
\bea
p^\mu d^3\sigma_\mu &=& \left[  E\frac{\partial z}{\partial \eta_s} - p_z \frac{\partial t}{\partial \eta_s} \right] d\eta_s  r dr d\phi_x.
\eea
The velocity field and four-momentum $u^\mu(x)$ and $p^\mu$ are as in App.~\ref{upmuu}, with ``boost angles'' $\eta \rightarrow \eta_z(z)$ and $\rho(r) \rightarrow \eta_t(r)$ and with longitudinal momentum $p_L \rightarrow p_z$.  Azimuth angles are $\varphi \rightarrow \phi_p$ and $\phi \rightarrow \phi_x$. If $z = \tau \cosh(\eta_s)$ and $t = \tau \sinh(\eta_s)$, with $E = m_t \cosh(y_z)$ and $p_z = m_t \sinh(y_z)$, then
\bea \label{pmusigma}
p^\mu d^3\sigma_\mu &\rightarrow& m_t \tau \cosh(y_z - \eta_s)  d\eta_s  r dr d\phi_x.
\eea

\subsubsection{Tomasik et al.~\cite{tomasik}}

In this study the single-particle momentum spectrum is defined in terms of a ``source function'' $S(x,p)$
\bea
E\frac{d^3N}{dp^3} &=&  \int d^4x S(x,p).
\eea
The source function is defined by
\bea \label{sourceeq}
S(x,p) d^4x &\propto& m_t \tau \cosh(y_z - \eta_s) e^{-u^\mu(x) p_\mu / T}
\\ \nonumber
&&\times~ G(r) H(\eta_s)  d\eta_s rdr d\phi_x
\eea
with $y \rightarrow y_z$, $\eta \rightarrow \eta_s$ and $\varphi \rightarrow \phi_x$. The function $H(\eta_s)$ is a Gaussian with width $\Delta \eta_s$. The function $G(r)$ is Gaussian with width $R_G$ or flat over $r \in [0,R_B]$. A unit-normal Gaussian on $\tau$ is assumed integrated to yield mean value $\tau$. The emission hypersurface is defined by
\bea
\sigma_\mu &=& [\tau \cosh(\eta_s),r\hat e(\phi_x),\tau \sinh(\eta_s)]
\eea
with (assuming $d\tau=0$)
\bea \label{sigmamu}
d^3\sigma_\mu &=&  \tau\left[\cosh(\eta_s),0,0, \sinh(\eta_s) \right] d\eta_s r dr d\phi_x.
\eea
Four-vectors $u_\mu$ and $p_\mu$ are as in App.~\ref{upmuu}, with flow angles $\eta_l \rightarrow \eta_z(z)$ and $\eta_t(r)$ as above in Eq.~(\ref{etax}). Factors $m_t \tau \cosh(y_z - \eta_s)$ in Eq.~(\ref{sourceeq}) result if $d^3\sigma_\mu$ defined above is combined with $p_\mu$ as in App.~\ref{upmuu}. Note that here $\eta_t(r) = \tanh(\beta_t) \propto r$ instead of $\beta_t \propto r$ as in Ref.~\cite{ssflow}.

\subsubsection{Florkowski and Broniowski Ref.~\cite{wojciechs}}

In this study the emission hypersurface (with $\alpha_\| \rightarrow \eta_s$ introduced as the space-time rapidity) is
\bea
\sigma_\mu &=& [\tau(\zeta) \cosh(\eta_s),\rho(\zeta) \hat e(\phi_x),\tau(\zeta) \sinh(\eta_s)],~
\eea
where $\zeta \in [0,1]$ is here simply a parameter that correlates proper time $\tau$ and radius $\rho$, not space-time rapidity $\eta_s$.  The associated volume element (with $\phi \rightarrow \phi_x$) is

\bea
d^3 \sigma^\mu &=& \left[ \frac{\partial \rho}{\partial \zeta} \cosh(\eta_s),\frac{d\tau}{d\zeta} \hat e(\phi_x),\frac{d\rho}{d\zeta} \sinh(\eta_s) \right]
\\ \nonumber
&& \times  \tau(\zeta) d\zeta  \rho(\zeta) d\eta_s d\phi_x.
\eea
The velocity field $u_\mu(x)$ [with $\alpha_\perp(\zeta) \rightarrow \eta_t(\zeta)$] is consistent with Eq.~(\ref{umu}) except $\alpha_\| \rightarrow \eta_z$ (boost angle), not $\eta_s$ (space-time rapidity), and an error in factorization has been corrected. Particle momentum $p_\mu$ is consistent with App.~\ref{upmuu} if $m_\perp \rightarrow m_t$, $y \rightarrow y_z$ and $\varphi \rightarrow \phi_p$. In that case
\bea
p^\mu d^3\sigma_u &=& \left[ m_t \cosh(y_z - \eta_s) \frac{\partial \rho}{\partial \zeta} - p_t \cos(\phi_p - \phi_x) \frac{d\tau}{d\zeta} \right]
\nonumber \\
&&\times \rho(\zeta) \tau(\zeta) d\zeta d\eta_s d\phi_x.
\eea
If one assumes $d\tau/d\zeta \rightarrow 0$ as in Ref.~\cite{wojciechs} and $\rho \rightarrow r$ then
\bea
d^3 \sigma_\mu 
&\rightarrow& \tau \left[\cosh(\eta_s),0,0,\sinh(\eta_s) \right] d\eta_s  r dr  d\phi_x
\eea
which is consistent with Eq.~(\ref{sigmamu}), and
\bea
p^\mu d^3\sigma_u &\rightarrow& m_t \tau(\zeta) \cosh(y_z - \eta_s) d\eta_s  r(\zeta)dr d\phi_x.~~
\eea

To clarify, the distinction between space-time rapidity and longitudinal boost angle for  Ref.~\cite{wojciechs} is as follows: $\alpha_\|$ is initially introduced as space-time rapidity ($\rightarrow \eta_s$). But $\alpha_\perp$ is then initially introduced as transverse boost angle ($\rightarrow \eta_t$) with transverse flow $v_r \rightarrow \beta_t = \tanh{\eta_t}$, and $\alpha_\|$ is later reintroduced as longitudinal boost angle $\rightarrow \eta_z$) with longitudinal flow $v_z \rightarrow \beta_z = \tanh(\eta_z)$. The initial assumption is that $\eta_s \approx \eta_z$ with $\alpha_\|$ filling both roles. In this appendix space-time $\eta_s$ and velocity-field $\eta_z$ are maintained distinct unless indicated otherwise.

\subsubsection{Rath et al.~\cite{cleymans}}

In this study $u_\mu$ and $p_\mu$ are as in App.~\ref{upmuu} with $\eta \rightarrow \eta_z$ (longitudinal flow angle), $\rho \rightarrow \eta_t$, $y \rightarrow y_z$, $\phi \rightarrow \phi_p$ and $\phi_r \rightarrow \phi_x$. $d^3\sigma_\mu$ is as in Eq.~(\ref{sigmamu}) except $\eta \rightarrow \eta_s$ (space-time rapidity) and with a minus sign corrected. With those definitions
\bea \label{pmudsigmamu}
p^\mu d^3\sigma_\mu &=& m_t \tau [\cosh(y_z)\cosh(\eta_s) - \sinh(y_z)\sinh(\eta_s)]
\nonumber
\\
&& \times d\eta_s r dr d\phi_x
\\ \nonumber
&=& m_t \tau \cosh(y_z - \eta_s) d\eta_s  r dr d\phi_x
\eea
and
\bea \label{umupmu}
u^\mu p_\mu &=& \gamma_t [m_t \cosh(y_z - \eta_z) - \beta_t p_t \cos(\phi_x - \phi_p) ].~~~~~~
\eea
The ``Bjorken correlation in rapidity'' is specified as $(y \approx \eta) \rightarrow (y_z \approx \eta_s)$ rather than the conventional $\eta_z \approx \eta_s$. That assignment of $y_z$ is incorrect [see Eq.~(\ref{pmudsigmamu})]. 

In summary, the above examples illustrate the differences and similarities among several approaches to the BW model over nearly thirty years, from 1993 to 2020.

\subsection{Single-particle spectra and blast-wave model}

The several definitions of relevant quantities in the previous subsections are  combined to evaluate Eq.~(\ref{cfeqq}) with
\bea
E\frac{d^3N}{dp^3} &=&  \frac{d^3N}{m_t dm_t dy_z d\phi_p}.
\eea
For each case in App.~\ref{hyper} Eq.~(\ref{cfeqq}) is evaluated with the relevant quantities and consistent symbols adopted herein. Distinction between longitudinal flow angle $\eta_z$ and space-time rapidity $\eta_s$ is maintained unless stated otherwise. But just as $\beta_t(r) = \tanh^{-1}(\eta_t) \sim r$ is often assumed $\beta_z(z) = \tanh^{-1}(\eta_z) \sim z$ may also be assumed in the form of Bjorken expansion $\eta_z(z) \approx \eta_s = \ln[(t + z)/\tau]$.

\subsubsection{Schnedermann et al.~\cite{ssflow}}

The formulation in Eq.~(14) of Ref.~\cite{ssflow} is modified by $\zeta \rightarrow \eta_s$, ${\cal Z} \rightarrow H_s$, $y \rightarrow y_z$, $\rho \rightarrow \eta_t$, $\eta \rightarrow \eta_z$ and invoking the expression in Eq.~(\ref{pmusigma}), in which case
\bea  \label{sshintegrals}
E\frac{d^3N}{dp^3}  &=& \frac{g}{(2\pi)^3} m_t  \tau \int_{-H_s}^{H_s} d\eta_s \cosh(y_z - \eta_s)
\\ \nonumber
&& \hspace{-.5in} \times \int_0^R rdr \exp\left[ - m_t \cosh(\eta_t) \cosh(y_z - \eta_z)/T \right]
\\ \nonumber
&& \hspace{-.5in} \times  \int_0^{2\pi} d\phi_x \exp[p_t \sinh(\eta_t) \cos(\phi_x - \phi_p)/T].
\eea
For the integrations, integrals over $\phi_p$ and $\phi_x$ give factor $(2\pi)^2 I_0\left[ p_t \sinh(\eta_t) /T \right]$. Then $\cosh(y_z - \eta_s) \rightarrow \cosh(y_z - \eta_z) \cosh(\eta_z - \eta_s)$, omitting the odd term $\sinh(y_z - \eta_z)\sinh(\eta_z - \eta_s)$ that will not contribute to the $y_z$ integral. Both sides are integrated over longitudinal rapidity $y_z$ leading to factor $2 K_1[m_t \cosh(\eta_t)/ T]$. 
The integral over $\eta_s$ then yields a constant no matter what the relation between $\eta_s$ and $\eta_z$, although equality (Bjorken expansion) is usually assumed in which case the integral over $\eta_s \rightarrow 2 H_s$ ($= 2Z_t$ in Ref.~\cite{ssflow}).
The result, assuming $\eta_t(r) \sim r$, is
\bea \label{sseq2}
\frac{dN}{m_t dm_t} &\approx& \frac{g}{\pi} m_t \tau 2H_s \times
\\ \nonumber 
&& \hspace{-.5in} \int_0^R r dr K_1[m_t \cosh(\eta_t)/ T] I_0\left[ p_t \sinh(\eta_t) /T  \right]
\eea
near midrapidity ($y _z \approx 0$) which agrees with Eq.~(14) of Ref.~\cite{ssflow} except for the factor $\tau$. $\beta_r(r) \rightarrow \beta_t(r) = \beta_s(r/R)^n$ with $\beta_t = \tanh(\eta_t)$ is assumed with $n = 2$. It is observed that ``the form of the profile [i.e.\ value of $n$] is not important for the [data] analysis.'' But results in Fig.~\ref{logdetails} (c) demonstrate that parameter $n$ does strongly affect the BW  model above $y_t \approx 1.8$ ($p_t \approx 0.4$ GeV/c).

\subsubsection{Tom\'asik et al.~\cite{tomasik}}

In this study explicit model functions on $\eta_s$, $r$ and $\tau$ are introduced to define the emission space-time volume. The integrals in Eq.~(\ref{sshintegrals}) then become (with $\eta \rightarrow \eta_s$ and omitting factor $e^{\mu / T}$)
\bea
E\frac{d^3N}{dp^3}  &=& \frac{1}{(2\pi)^3} m_t  \tau \int_{-\infty}^\infty d\eta_s H(\eta_s) \cosh(y_z - \eta_s)
 \nonumber \\
&& \hspace{-.5in} \times \int_0^\infty rdr G(r) \exp\left[ - m_t \cosh(\eta_t) \cosh(y_z - \eta_z)/T \right]
 \nonumber \\
&& \hspace{-.5in} \times  \int_0^{2\pi} d\phi_x \exp[p_t \sinh(\eta_t) \cos(\phi_x - \phi_p)/T],
\eea
where an additional integral over $\tau$ is simply represented by mean value $\tau$ above. It is noted that the longitudinal rapidity or flow angle $\eta_l(x) \rightarrow \eta_z(x)$ is identified with space-time rapidity $\eta_s$ such that longitudinal speed $\beta_z = \tanh(\eta_z) \rightarrow \tanh{\eta_s} = z/t$. That is analogous to a common transverse prescription $\beta_t(r) = \beta_s(r/R)^n$, in this case with $n = 1$. However, in this study the adopted {\em transverse} relation is $\eta_t(x) = \eta_f(r/r_{rms})$. The integrals over $\phi_p$ and $\phi_x$ should give the same result as in the previous case. However, if there is no integration over $y_z$ the integrals over $\eta_s$ and $r$ may require numerical techniques.

\subsubsection{Florkowski and Broniowski Ref.~\cite{wojciechs}}

In this study the single-particle spectrum is expressed as (omitting factor $e^{\mu / T}$ and including $d\phi_p$ at left)
\bea
\frac{d^3N}{d^2p_t dy_zd\phi_p} &=&\frac{1}{(2\pi)^3} \int_0^{2\pi} d\phi_x \int_{-\infty}^\infty d\eta_s \int_0^1 d\zeta\, \rho(\zeta) \tau(\zeta)
 \nonumber \\
&& \hspace{-.7in} \times \left[ m_t \cosh(y_z - \eta_s) \frac{d\rho}{d\zeta} - p_t \cos(\phi_x - \phi_p) \frac{d\tau}{d\zeta}  \right]
\\ \nonumber
&& \hspace{-.7in} \times \exp[ - m_t \cosh(\eta_t) \cosh(y_z - \eta_z)/T]
\\ \nonumber
&& \hspace{-.7in}\times \exp\left[p_t \sinh(\eta_t) \cos(\phi_x - \phi_p)/T \right].
\eea
The final result assumes that $d\tau / d\zeta \rightarrow 0$, in which case $d\zeta \rho(\zeta)\tau(\zeta) d\rho / d\zeta \rightarrow \tau(r)rdr$ with presumably $r \in [0,R]$ for some $R$.  The integrals over $\varphi \rightarrow \phi_p$ and $\phi \rightarrow \phi_x$ should then give the same result as in the previous cases. If $\eta_z \rightarrow \eta_s$ (assumed Bjorken expansion) then the integral over $\alpha_\| \rightarrow \eta_s$ gives $2K_1(m_t \cosh(\eta_t)/T)$ as above. The further assumption is made that $\alpha_\perp(\zeta) \rightarrow \eta_t(r)$ is constant, in which case the integral over $\zeta \rightarrow r$ gives $\bar \tau R^2/2$. If that assumption is dropped the result is
\bea \label{floreq}
\frac{d^2N}{m_tdm_t dy_z} &=&\frac{1}{\pi}   m_t
\\ \nonumber 
&& \hspace{-.5in}  \times \int_0^R \tau(r) r dr K_1[m_t \cosh(\eta_t)/ T] I_0\left[ p_t \sinh(\eta_t) /T  \right],
\eea
which can be compared with Eq.~(\ref{sseq2}).

\subsubsection{Rath et al.~\cite{cleymans}} 

Reference~\cite{cleymans} assumes $u_\mu$, $p_\mu$ and $d^3\sigma_\mu$ as in App.~\ref{bwtheory} and Eq.~(\ref{sigmamu}) with $\eta_z \rightarrow \eta_s$ (space-time rapidity), arriving at the single-particle distribution
\bea \label{cleymanseq}
\frac{d^2N}{dp_T dy} &\propto &  \int_0^{R_0} \hspace{-.15in} r dr K_1\left( \frac{m_T \cosh \rho}{T_{kin}} \right) I_0 \left( \frac{p_T \sinh \rho}{T_{kin}} \right)~~~~~~~
\eea
given the usual $\rho \rightarrow\eta_t $ and $y \rightarrow y_z$. It is further assumed that $y = \eta \rightarrow y_z = \eta_s$ reflects Bjorken expansion which is incorrect, and given the prior assumption $\eta_z \rightarrow \eta_s$ this further assumption is unnecessary. The ``flow profile'' function $\beta_t(r)$ is as for Eq.~(\ref{sseq2}) but with $n = 1$. A substantial problem for Eq.~(\ref{cleymanseq}) is missing factors $p_t m_t$ or $m_t^2$ compared for instance with Eq.~(\ref{floreq}).

\subsubsection{Lao, Liu and Ma \cite{lao}}

In this study the single-particle momentum distribution is derived from Eq.~(3.1) in Ref.~\cite{tomasik} via Eq.~(18) in Ref.~\cite{lanny} (omitting factor $e^{\mu / T}$ and $\hbar \rightarrow 1$) with e.g.\ $\eta_{smax} \rightarrow H_s$ as
\bea \label{laoeq}
E\frac{d^3N}{dp^3}  &=& \frac{1}{(2\pi)^2} m_t  \tau \int_{-H_s}^{H_s} d\eta_s H(\eta_s) \cosh(y_z - \eta_s)
\nonumber \\
&& \hspace{-.5in} \times \int_0^\infty rdr G(r) \exp\left[ - m_t \cosh(\eta_t) \cosh(y_z - \eta_z)/T \right]
\nonumber \\
&& \hspace{-.5in} \times  I_0\left[ p_t \sinh(\eta_t) /T  \right].
\eea
In Ref.~\cite{lanny} $y_z = 0$ is assumed (near midrapidity) and therefore does not appear in its Eq.~(18). In Ref.~\cite{lao} $y_z$ is retained but $\cosh(y_z - \eta_s) \rightarrow \cosh(y_z)\cosh(\eta_s)$ with the $ \sinh(y_z)\sinh(\eta_s)$ term omitted. The expression is further simplified by taking $G(r) \rightarrow 1$ (with $r \in [0,R]$) and $H(\eta_s) \rightarrow 1$ in which case Eq.~(\ref{laoeq}) is reduced to Eq.~(\ref{sshintegrals}) (after integration over $\phi_x$). In that case integration over $\eta_s$ should result in Bessel function $K_1\left[ m_t \cosh(\eta_t)/T \right]$. However, for their Eq.~(4) it is decided to set $\eta_s = 0$ (``single source emission''), and the Bessel function must be replaced by $\cosh(y_z) \exp[-m_t \cosh(y_z)\cosh(\eta_t)/T]$ which near midrapidity ($y_z \approx 0$) is in effect  a Boltzmann distribution with radially varying T(r). That expression then represents the BW model as it is applied to data spectra.

\end{appendix}



\begin{thebibliography}{99}

\bibitem{ssflow}  E.~Schnedermann, J.~Sollfrank and U.~W.~Heinz,
Phys.\ Rev.\ C {\bf 48}, 2462 (1993).

\bibitem{phenradflow} K.~Adcox \textit{et al.} (PHENIX])
Phys. Rev. C \textbf{69}, 024904 (2004).

\bibitem{starradflow} B.~I.~Abelev \textit{et al.} (STAR),
Phys. Rev. C \textbf{79}, 034909 (2009).
	
\bibitem{aliceppbpid}  B.~B.~Abelev {\it et al.} [ALICE Collaboration],
Phys.\ Lett.\ B {\bf 728}, 25 (2014).

\bibitem{cooper} F.~Cooper and G.~Frye,
Phys.\ Rev.\ D \textbf{10}, 186 (1974).

\bibitem{pidpart1} T.~A.~Trainor,
arXiv:2112.09790 [hep-ph].

\bibitem{pidpart2} T.~A.~Trainor,
arXiv:2112.12330 [hep-ph].

\bibitem{ppbpid}  T.~A.~Trainor,
J.\ Phys.\ G \textbf{47}, no.4, 045104 (2020).

\bibitem{ppprd} J.~Adams {\it et al.}  (STAR Collaboration),
Phys.\ Rev.\  D {\bf 74}, 032006 (2006).

\bibitem{busza} W.~Busza {\em et al.}, Phys.\ Rev.\ Lett.\ {bf 34}, 836 (1975).

\bibitem{stopping} S.~Chapman and M.~Gyulassy,
Phys.\ Rev.\ Lett.\ \textbf{67}, 1210-1213 (1991).

\bibitem{bialas} A.~Bialas, M.~Bleszynski and W.~Czyz, Nucl.\ Phys.\ B {\bf 111}, 461 (1976)

\bibitem{eeprd}   T.~A.~Trainor and D.~T.~Kettler,
Phys.\ Rev.\ D {\bf 74}, 034012 (2006).

\bibitem{hardspec}  T.~A.~Trainor,
Int.\ J.\ Mod.\ Phys.\  E {\bf 17}, 1499 (2008).

\bibitem{fragevo}    T.~A.~Trainor,
Phys.\ Rev.\  C {\bf 80}, 044901 (2009).

\bibitem{jetspec2}  T.~A.~Trainor,
Phys.\ Rev.\ D  {\bf 89}, 094011 (2014).

\bibitem{mbdijets} T.~A.~Trainor,
arXiv:1701.07866.
	
\bibitem{tomglauber}  T.~A.~Trainor,
arXiv:1801.05862.

\bibitem{alicempt}  B.~B.~Abelev {\it et al.}  (ALICE Collaboration),
Phys.\ Lett.\ B {\bf 727}, 371 (2013).

\bibitem{tommpt} T.~A.~Trainor,
arXiv:1708.09412.

\bibitem{aliceglauber} J.~Adam {\it et al.} (ALICE Collaboration),
Phys.\ Rev.\ C {\bf 91}, no. 6, 064905 (2015).

\bibitem{cleymans} R.~Rath, A.~Khuntia, R.~Sahoo and J.~Cleymans,
J.\ Phys.\ G \textbf{47}, no.5, 055111 (2020).

\bibitem{alicetomspec}    T.~A.~Trainor,
J.\ Phys.\ G {\bf 44}, no. 7, 075008 (2017).

\bibitem{levywilk} G.~Wilk and Z.~Wlodarczyk,
Nucl. Phys. B Proc. Suppl. \textbf{75}, no.1-2, 191-193 (1999).

\bibitem{wilklevy} G.~Wilk and Z.~Wlodarczyk,
Phys. Rev. Lett. \textbf{84}, 2770 (2000).

\bibitem{alicepppid} S.~Acharya \textit{et al.} (ALICE),
Eur. Phys. J. C \textbf{80}, no.8, 693 (2020).

\bibitem{zscore} E.~Kreyszig (1979). {\em Advanced Engineering Mathematics} (Fourth ed.), Wiley, p.\ 880, eq. 5. 

\bibitem{landau} L.~D.~Landau, Izv.\ Akad.\ Nauk SSSR {\bf 17}, 51 (1953).

\bibitem{haged}R.~Hagedorn, Nuovo Cimento Suppl.\ {\bf 3} 147 (1965).

\bibitem{haged2} R.~Hagedorn,
Nuovo Cim.\ A \textbf{52}, no.4, 1336-1340 (1967). ORPHAN

\bibitem{siemens} P.~J.~Siemens and J.~O.~Rasmussen, Phys.\ Rev.\ Lett.\ {\bf 42}, 880 (1979).

\bibitem{bevalac} H.~A.~Gustafsson, H.~H.~Gutbrod, B.~Kolb, H.~Lohner, B.~Ludewigt, A.~M.~Poskanzer, T.~Renner, H.~Riedesel, H.~G.~Ritter and A.~Warwick, \textit{et al.}
Phys. Rev. Lett. \textbf{52}, 1590-1593 (1984).

\bibitem{daniel} P.~Danielewicz and G.~Odyniec,
Phys. Lett. B \textbf{157}, 146-150 (1985).

\bibitem{sphericity} C.~Cesarotti, M.~Reece and M.~J.~Strassler,
JHEP \textbf{07}, 215 (2021).

\bibitem{tomasik} B.~Tom\'asik, U.~A.~Wiedemann and U.~W.~Heinz,
Acta Phys. Hung. A \textbf{17}, 105-143 (2003).

\bibitem{wojciechs} W.~Florkowski and W.~Broniowski,
Acta Phys. Polon. B \textbf{35}, 2895-2910 (2004).
	
\bibitem{lanny} R.~L.~Ray and A.~Jentsch,
Phys. Rev. C \textbf{99}, no.2, 024911 (2019).

\bibitem{lao} H.~L.~Lao, F.~H.~Liu and B.~Q.~Ma,
Entropy \textbf{23}, no.7, 803 (2021).

\bibitem{tsallisbw} Z.~Tang, Y.~Xu, L.~Ruan, G.~van Buren, F.~Wang and Z.~Xu,
Phys. Rev. C \textbf{79}, 051901 (2009).

\bibitem{alephff}  D.~Buskulic {\it et al.}  (ALEPH Collaboration),
Z.\ Phys.\  C {\bf 55}, 209 (1992).
	
\bibitem{pythia2} T.~Sj\"ostrand,
Comput. Phys. Commun. \textbf{246}, 106910 (2020).

\bibitem{pythia1} T.~Sj\"ostrand, S.~Mrenna and P.~Z.~Skands,
Comput. Phys. Commun. \textbf{178}, 852-867 (2008).

\bibitem{field1} R.~D.~Field and R.~P.~Feynman,
Phys.\ Rev.\ D \textbf{15}, 2590-2616 (1977).

\bibitem{field2} R.~P.~Feynman, R.~D.~Field and G.~C.~Fox,
Phys. Rev. D \textbf{18}, 3320 (1978).

\bibitem{field3} R.~D.~Field and R.~P.~Feynman,
Nucl. Phys. B \textbf{136}, 1 (1978).

\bibitem{gosta} G.~Gustafson,
Acta Phys. Polon. B \textbf{42}, 2581-2606 (2011).
	
\bibitem{statmodel} F.~Becattini,
Z. Phys. C \textbf{69}, no.3, 485-492 (1996).

\bibitem{stock} R.~Stock,
arXiv:hep-ph/0312039 [hep-ph].

\bibitem{porter2} R.~J.~Porter and T.~A.~Trainor  (STAR Collaboration),
J.\ Phys.\ Conf.\ Ser.\  {\bf 27}, 98 (2005).

\bibitem{porter3}  R.~J.~Porter and T.~A.~Trainor  (STAR Collaboration),
PoS C {\bf FRNC2006}, 004 (2006).

\bibitem{ytyt} M.~Abdallah \textit{et al.} (STAR),
arXiv:2204.11661 [nucl-ex].

\bibitem{wong} C.~Y.~Wong,
EPJ Web Conf. \textbf{7}, 01006 (2010).

\bibitem{na35ss} J.~Baechler {\it et al.}  (NA35 Collaboration),
Phys.\ Rev.\ Lett.\  {\bf 72}, 1419 (1994).

\bibitem{na49pp}  C.~Alt {\it et al.}  (NA49 Collaboration),
Eur.\ Phys.\ J.\  C {\bf 45}, 343 (2006).

\bibitem{tomnewppspec} T.~A.~Trainor,
arXiv:2104.08423 [hep-ph].

\bibitem{tommodeltests} T.~A.~Trainor,
arXiv:2107.10899 [hep-ph].



	
\end{thebibliography}
\end{document}